\RequirePackage[2024-06-01]{latexrelease}
\documentclass[
  showkeys,
  reprint,
  onecolumn,
  preprintnumbers,
  nofootinbib,
  groupedaddress, 
  amsmath,amssymb,
  aps,pra,
]{revtex4-2}

\usepackage[english]{babel}
\usepackage[all]{xy}
\usepackage{mathtools} 
\usepackage{graphicx,cancel} 
\usepackage{ifthen}
\usepackage{mathrsfs}  
\usepackage{pgfmath}
\usepackage[dvipsnames]{xcolor}
\usepackage{tikz}
\usetikzlibrary{calc,arrows}
\usepackage{enumerate,comment}
\usepackage{hyperref}
\usepackage{braket}

\usepackage{mathabx}

\def\CircleArrow{\ensuremath{%
  \rotatebox[origin=c]{90}{$\circlearrowleft$}}}
  
\usepackage{lipsum}
\usepackage{stmaryrd,amsthm} 
\theoremstyle{definition}
\newtheorem{remark*}{Remark}

\newcommand*{\Scale}[2][4]{\scalebox{#1}{\ensuremath{#2}}}

\SetSymbolFont{stmry}{bold}{U}{stmry}{m}{n}

\makeatletter
\renewcommand{\section}{\@startsection{section}{1}{\z@}%
  {-3.5ex \@plus -1ex \@minus -.2ex}%
  {2.3ex \@plus .2ex}%
  {\normalfont\Large\bfseries\centering}} 

\renewcommand{\subsection}{\@startsection{subsection}{2}{\z@}%
  {-3.25ex\@plus -1ex \@minus -.2ex}%
  {1.5ex \@plus .2ex}%
  {\normalfont\large\bfseries}} 
\makeatother
\makeatletter
\def\@hangfrom@section#1#2#3{%
  #1%
  \def\@tempa{#2}%
  \ifx\@tempa\@empty\else
    \textup{#2}%
  \fi
  #3%
}

\def\@hangfroms@section#1#2{#1#2}
\makeatother

\begin{document}

\title{\Large Exactly solved Schr\"{o}dinger equations with time-dependent Hamiltonians}

\author{Michael Warnock}
\email{michael.p.warnock3.civ@us.navy.mil}
\affiliation{Department of Physics, Brown University, Providence, Rhode Island 02912, USA}
\affiliation{Ocean Science, Autonomous Systems, and Ranges, Naval Undersea Warfare Center, Newport, RI, 02840 USA}

\author{Ant\^{o}nio Francisco Neto}
\email{antonio.neto@ufop.edu.br}
\affiliation{DEPRO Escola de Minas UFOP, Campus Morro do Cruzeiro, Ouro Preto, 35400-000, Minas Gerais, Brazil}

\author{Pierre-Louis Giscard}
\email{pierre-louis.giscard@univ-littoral.fr}
\affiliation{Laboratoire de Math\'ematiques Pures et Appliqu\'ees Joseph Liouville, Universit\'e du Littoral C\^{o}te d’Opale, 50 rue Ferdinand Buisson, CS 80699, 62228 Calais, France}

\author{Omid Faizy}
\email{omid.faizy_namarvar@sorbonne-universite.fr}
\affiliation{Laboratoire de Chimie de la Matière Condensée de Paris, UMR CNRS 7574, Sorbonne Université, 4, place Jussieu, 75252, Paris, France}

\author{Christian Joachim}
\email{christian.joachim@cemes.fr}
\affiliation{Univ. Toulouse, CNRS, CEMES, Toulouse, France}

\begin{abstract}
We present the analytical, exact, explicit, and assumption free formulas for the evolution operators corresponding to four instances of time-dependent  Hamiltonians relevant to quantum spin batteries including two stochastic cases. We demonstrate how to recover and go beyond existing expansions and approximations directly from the exact solutions giving, for example, an explicit exact formula for Floquet Hamiltonians at all orders. The exact solutions are obtained through a completely novel combination of three mathematical techniques, the $\star$-algebra, path-sums and Omega calculus, which we briefly overview. These are widely applicable to other non-autonomous differential systems.
\end{abstract}

\maketitle

\section{Introduction}
In this work, we consider several instances of 2-level quantum systems driven by \textit{time-dependent} Hamiltonians, including stochastic ones. 
In all cases, we present the mathematically \textit{exact}, analytical solutions to the associated non-autonomous Schr\"{o}dinger's equations. The solutions are fully \textit{explicit} in that they do not involve unevaluated integrals; do not involve coefficients defined implicitly or recursively; do not involve exotic operations or operators;
and are not perturbative expansions. Rather the solutions will be given as \emph{unconditionally convergent} series involving only sums of ordinary functions with well identified arguments and coefficients. For a given problem, the solution is universally valid across all parameter regimes--including regimes where the Hamiltonian is not periodic, or where all physical parameters are of comparable magnitude. 
The progress we present stems from a  triptych of recent mathematical techniques combined for the first time: 1) the $\star$-algebra, a framework that turns differential problems into linear algebraic ones via an integral operation (the $\star$-product) defined on functions of two variables; 2) the method of path-sums, which exploits the combinatorics of graph-walks to express matrix-valued formal solutions as explicit scalar-valued continued fractions terminating at finite depth; and
3) Omega calculus, a technique used to evaluate the $\star$-operations appearing in the path-sum solution.
Omega calculus plays a role analogous to--though much more general than--Laplace and Fourier transforms in autonomous differential systems. The output of these three stages is the evolution operator solution to Schr\"{o}dinger's equation in unconditionally convergent series form. This mathematical machinery is widely applicable to non-autonomous differential systems. 

The text is organized around four cases of physical interest for the quantum energy storage considered in this volume, pairing each problem statement immediately with its solution and some analysis of it (Sections~\ref{FirstCase}-\ref{FourthCase}). In the second case, which includes the Bloch-Siegert Hamiltonian as a subcase, we detail how to recover and go beyond the known Rabi like (RWA) and exact RWA approximations directly from the exact solution, e.g. deriving a novel explicit, commutator-free formula for effective Hamiltonians at all orders. The techniques presented in extensive details in this case are  universal and more succinctly employed in the other cases as well. Concluding remarks are presented in Section \ref{Conclusion}. Proof details are kept to a minimum in the main text, though some background on the $\star$-algebra, path-sums and Omega calculus is provided in \S\ref{MathsSec}. These sections may be bypassed by readers wishing to access the solutions directly. A fuller account of all the proofs can be found in the appendices. 

\subsection{Models}
The general model we consider in this work is the Schr\"{o}dinger's equation: 
\begin{equation}
\frac{d}{dt}\mathsf{U}(t)=-\frac{i}{\hbar}\, \mathsf{H}(t)\mathsf{U}(t),
\end{equation}
where $\hbar$ is a Planck's constant (we shall henceforth use $\hbar=1$), $\mathsf{U}(t)$ is the evolution operator and $\mathsf{H}(t)$ is the system's Hamiltonian. 
More precisely, we consider
\begin{equation}\label{HamForm}
\mathsf{H}(t) = 
\begin{pmatrix}
S_0 + \epsilon(t) & g f(t)\\
\bar{g} \bar{f}(t)& S_1+ i \Gamma
\end{pmatrix},
\end{equation}
where $\bar{g}$ is the complex conjugate of $g$, $\epsilon(t)$ is function of time which we take to be of the form $\epsilon(t)=e_0 \cos(\omega_0 t)$ (Cases 1 \& 2, \S~\ref{FirstCase}, \ref{SecondCase}) or  random noise (Cases 3 \& 4, \S~\ref{ThirdCase}, \ref{FourthCase}). Similarly, the function $f(t)$ will be taken to be $f(t)=1$ (Cases 1 and 3, \S~\ref{FirstCase}, \ref{ThirdCase}), $f(t)=\cos(\omega t)$ (Cases 2 and 3, \S~\ref{SecondCase},\ref{ThirdCase}) and Gaussian  $f(t)\propto \exp(-(t-t_m)^2/2\sigma^2)$ or modulated Gaussian  $f(t)\propto \exp(-(t-t_m)^2/2\sigma^2) \cos(\omega t)$ (Case 4, \S~\ref{FourthCase}).
Parameter $\Gamma$ is describing the lifetime of the excited state after preparing the quantum system in its initial $S_0$ ground state. 
We make no assumption on the physical parameters $S_0, S_1$, $\Gamma$ and $g$ in the analytical results, that are consequently valid in the entire parameter space. 
By contrast, in physical situations we typically have $|g|\ll|S_0-S_1|$ and in numerical simulations we employ $\Gamma=0$ because considering a finite life time for the excited state would have open another vast range of further discussions.
Finally, the mathematical methods employed to arrive at the solutions are not limited to the cases considered here.

\subsection{Objectives}
Denoting $U_{ij}:=\mathsf{U}_{i,j}$, the first objective is to determine the transition probability between both spin states,
\begin{equation}
P_{|0\rangle\to |1\rangle}(t):=|U_{12}(t)|^2.
\end{equation}
The second objective is to determine the energy stored in the qubit when in state $|\varphi(t)\rangle=\mathsf{U}(t)|\varphi(0)\rangle$. This energy is given by $\Delta E(t) = \big||\langle 0|\varphi(t)\rangle|^2S_0+|\langle 1|\varphi(t)\rangle|^2S_1-S_0\big|$, yielding
\begin{equation}
\Delta E(t) = \Big|\big|U_{11}(t)\varphi_0+U_{12}(t)\varphi_1\big|^2S_0+\big|U_{21}(t)\varphi_0+U_{22}(t)\varphi_1\big|^2S_1-S_0\Big|,
\label{eq:Energy_Stored_Large}
\end{equation}
where $|\varphi(0)\rangle=(\varphi_0,\varphi_1)^T$ is the initial pure state, with $|\varphi_0|^2+|\varphi_1|^2=1$. Remark that if $\varphi_0=1$ and $\varphi_1=0$, then choosing without loss of generality that $S_0=0$ yields \begin{equation}\Delta E(t)=P_{|0\rangle\to |1\rangle}(t)\, S_1,
\label{eq:Energy_Stored_Reduced}
\end{equation}
so that studying $P_{|0\rangle\to |1\rangle}(t)$ and $\Delta E(t)$ is the same task. 
In any case, both $\Delta E(t)$ and $P_{|0\rangle\to |1\rangle}(t)$ are determined by the system's evolution operator $\mathsf{U}$, which we shall therefore aim at obtaining.

\section{\Large Mathematical Background}\label{MathsSec}
\subsection{Introduction to the \texorpdfstring{$\star$}{star}-product}\label{SecStarProd}

We introduce only basic definitions and associated notation. A fully detailed account of the $\star$-product is presented in \cite{ryckebusch2023frechetlie}. Here we shall only consider mathematical objects of the form $A(t,s):=a(t,s)\Theta(t-s)$, where $a(t,s)$ is a smooth function of both $t$ and $s$, termed \textit{time} variables, and
\begin{align}
\Theta(x)=\left\{\begin{array}{l}
1,\;{\rm if}\;x\geq 0,\\
0,\;{\rm otherwise,}
\end{array}
\right.
\end{align} 
is the Heaviside Theta function. Then the $\star$-product of two such objects is
\begin{equation}
\big(A\star B\big)(t,s)=\int_s^t a(t,\tau)b(\tau,s)d\tau~ \Theta(t-s),
\end{equation}
with $B(t,s):=b(t,s)\Theta(t-s)$ and $a$ and $b$ are smooth in both their variables, so the integral is well-defined. The $\star$-product is a type of Volterra composition \cite{Volterra1924} that generalizes the convolution. More precisely, a $\star$-product is a convolution if and only if both $a$ and $b$ depend only on the difference between their variables. Thus, e.g., we have
\begin{align}
e^{a(t-s)}\Theta(t-s) \star e^{b(t-s)}\Theta(t-s)
&=\int_s^t e^{a(t-\tau)}e^{b(\tau-s)}d\tau\,\Theta(t-s),\nonumber\\
&=\frac{e^{a(t-s)}-e^{b(t-s)}}{a-b}\Theta(t-s),
\end{align} while
\begin{align}
e^{-as}\Theta(t-s) \star e^{bt}\Theta(t-s)
&=\int_s^t e^{-a\tau}e^{b\tau}d\tau\,\Theta(t-s),\nonumber\\
&=\frac{e^{(b-a)t}-e^{(b-a)s}}{b-a}\Theta(t-s).
\end{align}
The definition of the $\star$-product extends to matrices comprising smooth functions times Heaviside Theta by linearity. Take a matrix $\mathsf{A}$ with $(\mathsf{A})_{ij}=:A_{ij}=a_{ij}(t,s)\Theta(t-s)$ and similarly for a matrix $\mathsf{B}$ of the same size. Then  
\begin{equation}(\mathsf{A}\star \mathsf{B})_{ij}:=\sum_{k}A_{ik}\star B_{kj}=\sum_k \int_{s}^t a_{ik}(t,\tau)b_{kj}(\tau,s)d\tau\,\Theta,
\end{equation}
where all $a_{ik}$ and $b_{kj}$ functions are smooth of both their variables. Note that here and from now on, we omit the $(t-s)$ arguments of $\Theta(t-s)$ in order to alleviate the notation. The $\star$-product admits a unit, the bivariate Dirac delta distribution $1_\star:=\delta(t-s)$, and so the matricial $\star$-product as a unit too, $\mathsf{Id}_\star:=\mathsf{Id}\,1_\star $, $\mathsf{Id}$ being the identity matrix of appropriate size. 

The $\star$-product turns linear differential systems, including non-autonomous ones, into ordinary linear algebraic systems. In particular, the time-ordered exponential of a time-dependent matrix, which represents the general solution of a linear differential system is a matricial $\star$-resolvent \cite{Giscard2015,GisBon},
\begin{equation}\label{SolForm}
\mathscr{T}e^{\int_s^t \mathsf{H}(\tau)d\tau}\,\Theta=\Theta\,\mathsf{Id}\star\big(\mathsf{Id}_{\star}-\mathsf{H}(t)\Theta\big)^{\star-1},
\end{equation}
where $\mathscr{T}$ denotes the time-ordering operator with $\mathscr{T}\{\mathsf{H}(t)\mathsf{H}(s)\}=\mathsf{H}(t)\mathsf{H}(s)$ if $t\geq s$ and $\mathsf{H}(s)\mathsf{H}(t)$ otherwise. Equivalently, the above results states that the differential system's Green's function $\mathsf{G}$ is a matricial $\star$-resolvent, $\mathsf{G}=\left(\mathsf{Id}_{\star}-\mathsf{H}(t)\Theta\right)^{\star-1}$.
The ordinary Dyson series representation of the solution corresponds to the $\star$-Neumann series representation of the matricial $\star$-resolvent above,
\begin{align}\label{Neustar}
\Theta\mathsf{Id}\star\big(\mathsf{Id}_{\star}-\mathsf{H}(t)\Theta\big)^{\star-1}
&=\Theta\mathsf{Id}\star\sum_{k\geq 0}\left(\mathsf{H}(t)\Theta\right)^{\star k},\nonumber\\
&=\mathsf{Id}+\int_s^t\mathsf{H}(\tau)d\tau+\int_s^t \int_s^{\tau_1}\mathsf{H}(\tau_1)\mathsf{H}(\tau_2)d\tau_2d\tau_1+\cdots,
\end{align}
where we assumed $t\geq s$ for the last equality. This series is guaranteed to converge as long as Hamiltonian entries remain finite at finite times \cite{Giscard2020}.
The matricial $\star$-resolvent at the heart of the solution as expressed above is best evaluated with the method of path-sums, which gives an explicit representation of each of its entries as continued fractions that terminates at finite depth. These fractions live in the $\star$-algebra, i.e., they involve $\star$-products and $\star$-inverses of \textit{scalar} functions, which are then evaluated using Omega calculus.  

\subsection{Method of path-sums}\label{SecPS}
Let us now consider the $\star$-resolvent $\left(\mathsf{Id}_{\star}-\mathsf{H}(t)\Theta\right)^{\star-1}$ appearing in the solution to Schr\"{o}dinger's equation as shown above. Since
\begin{equation}
\big(\mathsf{Id}_{\star}-\mathsf{H}(t)\Theta\big)^{\star-1}=\sum_{n\geq 0} \big(\mathsf{H}(t)\Theta\big)^{\star n},
\end{equation}
it is profitable to interpret $\mathsf{H}(t)\Theta$ as the adjacency matrix of a graph $\mathcal{G}$. Indeed, \cite[p. 337]{Flajolet2009} asserts that the powers of a graph's adjacency matrix enumerate all possible walks on the graph (see also \cite[Lemma~2.5]{Biggs1993}). This observation indicates that by associating  $\mathsf{H}(t)\Theta$ with the graph $\mathcal{G}$ whose vertices are quantum states and edges are permissible transitions between those states,
then any entry $i,j$ of the matricial $\star$-resolvent of $\mathsf{H}(t)\Theta$ is the sum of the weights of all walks from vertex $|j\rangle$ to vertex $|i\rangle$ on $\mathcal{G}$ \cite[Proposition V.6]{Flajolet2009}. By construction, the dynamical  weight assigned to a transition (i.e., an edge) from state $|j\rangle$ to state $|i\rangle$ is given by $\mathsf{H}_{ij}(t)\Theta$ and $\mathcal{G}$ encodes the discrete structure of the quantum state space.
In this context, the weight of an individual walk is defined as the ordered $\star$-product of the weights of the edges it traverses. 

Since any $\left(\mathsf{Id}_{\star}-\mathsf{H}(t)\Theta\right)^{\star-1}_{ij}$ is, formally, a sum of walks, this interpretation motivates a resummation technique based on the combinatorial structure of walk sets, the path-sum method \cite{Giscard2012,Giscard2015}.
In its most general form, this method originates from a fundamental property of walks on graphs: the existence and uniqueness of their factorization into simple paths and simple cycles, walks which do not visit any vertex more than once. This property entails that the series of all walks between any two vertices of a graph can be reduced to a continued fraction that terminates at finite depth and only involves its simple cycles. This, in turn, implies that all entries of the $\star$-resolvent can be represented as such fractions over the simple cycles of $\mathcal{G}$. Since $\mathcal{G}$ represents the quantum state space, these cycles are the fundamental, irreducible physical processes generating the system's evolution. We present, in Appendix~\ref{PS2x2}, the method and its output in the case of interest here, namely Hamiltonians of the form Eq.~(\ref{HamForm}). In practice, the method yields any entry of a $\star$-resolvent--and thence, by Eq.~(\ref{SolForm}), of the evolution operator--exactly in terms of a few $\star$-products and scalar $\star$-resolvents, a major improvement over the infinite, matrix-valued, Dyson series \footnote{While we work here in a quantum physics context, both the $\star$-algebra and the path-sum approach are more broadly applicable outside of this context. This means that their applicability does not depend on any physics-based assumptions.}.

\subsection{Omega calculus}\label{SecOC}
The problem of evaluating $\star$-products and $\star$-resolvents using only ordinary operations finds an unexpected solution in MacMahon's Partition Analysis, also known as Omega calculus. This was originally introduced in the early twentieth century as a combinatorial analysis tool to describe the solution of linear Diophantine systems composed of equalities and inequalities \cite{macmahon2001combinatory, Andrews2001}. At the heart of MacMahon's strategy is the $\Omega$-operator, a linear operator that extracts certain coefficients of a convergent Laurent series \cite{Han2003, Neto2020}. Using this operator, \cite{Neto2023, Neto2025} showed recently that it is possible to obtain an integral-free representation of quantities involving iterated integrals, as is, for instance, the case for $\star$-powers of a function. 
 
 In spite of its combinatorial origins and implementation, Omega calculus can be understood as playing a role analogous to that of  classical integral transforms (Laplace, Fourier), albeit with two major differences: no integral is ever performed; and it operates well with $\star$-products, even when they differ from convolutions. 
  In practice, calculations involve three steps: i) passing from time-dependent functions to so-called `crude generating functions' depending on Omega variables, i.e. effectively going to an \textit{Omega domain}; ii) doing the required operations (such as taking $\star$-powers) in the Omega domain; and iii) eliminating the Omega variables, i.e. effectively going back to the \emph{time domain}. Remarkably, none of these steps involve anything but basic calculus. Time variables are retained in the Omega domain, only they play a much simplified role there. Step i) is achieved by a standard universal formula; step ii) comprises only ordinary products and sums, while in the present context step iii) boils down to little more than an identification of variables to specific values. Of particular importance is the result that rational functions in the Omega domain become divided-difference exponential functions in the time domain. 
  
Formally, the center piece of this mathematical machinery is the Omega operator which is defined as follows:
\begin{equation}
\overset{\lambda}{\underset{=
}{\Omega}}
\sum_{n\in\mathbb{Z}}\,a_n\, \lambda^n=a_0.
\end{equation} 
Here, the Omega operator acts on a Laurent series simply by extracting the coefficient of its constant term $\lambda^0$. This process is called the elimination of the Omega variable $\lambda$ and the Laurent series is called a crude generating function. We have, for instance:
\begin{equation}
\overset{\lambda}{\underset{=
}{\Omega}}\,e^{\lambda}\left(1+\lambda^{-2}\right)=
\overset{\lambda}{\underset{=
}{\Omega}}\,\Big(\underline{1}+\lambda+\underline{\underline{\lambda^2/2!}}+\cdots\Big)\!\!\left(\underline{1}+\underline{\underline{\lambda^{-2}}}\right)=1+\frac{1}{2!}=\frac{3}{2},
\end{equation} 
where, to obtain the second equality, the definition of the Omega operator was used, extracting the coefficient of $\lambda^0$ in the product $e^{\lambda}(1+\lambda^{-2})$ (contributions to $\lambda^0$ are the terms underlined above).
This definition immediately generalizes to more Omega variables $\lambda_1,\ldots ,\lambda_N$, $N\geq 1$, with $\overset{\lambda_1,\ldots,\lambda_N}{\underset{=
}{\Omega}}$ extracting the coefficient of $\lambda_1^0\cdots \lambda_N^0$ from any Laurent series in these variables. 

Therefore, a crude generating function belongs to the Omega domain as long as it has Omega variables, while a time domain expression does not have such variables. Therefore and by eliminating these, the Omega operator makes the transition from the Omega domain back to the time domain. The process of going to the Omega domain is achieved through an elementary universal formula. Consider $f(t)=\sum_n f_n t^n/n!$ an analytic function and the purely formal power series $f_\star(\lambda)=\sum_n f_n \lambda^n$. Then, this definition reveals the following: 
\begin{equation}
\overset{\lambda}{\underset{=
}{\Omega}} \,e^{\lambda t} f_\star(\lambda)=f(t).
\end{equation}
We say that $e^{\lambda t} f_\star(\lambda)$ is the Omega transform of $f(t)$ and, conversely, $f(t)$ is the Borel transform of $f_\star(\lambda)$.
Crucially, this construction and ensuing results extend to functions of two variables: 
\begin{equation}
\overset{\lambda_1,\lambda_2}{\underset{=
}{\Omega}} \,e^{\lambda_1 t} f_\star(\lambda_1,\lambda_2)e^{\lambda_2 s}=f(t,s).
\end{equation}
 One can observe that the Omega transform $e^{\lambda_1 t} f_\star(\lambda_1,\lambda_2)e^{\lambda_2 s}$ still depends on both $t$ and $s$ but in a much simpler way than $f(t,s)$ since it only involves a product of two ordinary exponentials of $t$ and of $s$. This is what makes Omega calculus so well suited to $\star$-products, because it turns complicated integrals into integrals of ordinary exponential functions:
\begin{equation}
f(t,s)\Theta\star g(t,s)\Theta=
\overset{\pmb{\lambda}}{\underset{=
}{\Omega}} \,e^{\lambda_1 t} f_\star(\lambda_1,\lambda_2)\left(\int_s^te^{\lambda_2 \tau}e^{\lambda_3 \tau} d\tau\right)\, g_\star(\lambda_3,\lambda_4) e^{\lambda_4 s}\,\Theta,
\end{equation}
 $\pmb{\lambda}=(\lambda_1,\lambda_2,\lambda_3,\lambda_4)$. In practice, evaluating $\star$-products and $\star$-resolvents in the Omega domain give rise, through repeated integration of ordinary exponentials, to divided-difference functions, which we now briefly present.

\subsection{Divided-Differences}
The divided-difference exponential is a generalization of the ordinary exponential function defined by: 
\begin{align}\label{expdivideddifference}
e^{[a_0,\ldots,a_n]t}
=\sum_{m\geq 0}
\frac{t^m}{m!}[a_0,\ldots, a_n]^m,
\end{align} with $a_0,\ldots,a_n$ meaning complex variables and
$[a_0,\ldots, a_n]^{m+n}=h_m(a_0,\ldots,a_n)$ for $m>0$ (otherwise $0$ for $m<0$ and 1 for $m=0$). In this expression, $h_m$ is the complete homogeneous symmetric polynomial of degree $m$ \cite[Chapter~1]{macdonald1998symmetric}. 
This implies that $e^{[a_0,\ldots,a_n]t}$ is invariant under any permutation of $(a_0,\ldots,a_n)$. Despite providing divided-difference exponentials in full explicit form, these functions also satisfy a recurrence relation that makes for easier computations:
 \cite{milne2000calculus}, 
\begin{align}\label{rrdd}
e^{[a_0,\ldots,a_n]t}=\frac{e^{[a_0,\ldots,a_{n-1}]t}-e^{[a_1,\ldots, a_n]t}}{a_0-a_n}.
\end{align}
This definition entails $e^{[a_0]t}=e^{a_0 t}$ while in general divided-difference exponentials are a compact formulation of intricate linear combinations of ordinary exponentials \cite{baxter2011functionals,kalev2021integral-free,lesch2017divided},
e.g.,
\begin{align}\label{e[a,b]}
&e^{[a,b]t}=\frac{e^{bt}-e^{at}}{b-a},\quad e^{[a,b,c]t}=\Scale[1.21]{\frac{\frac{e^{ct}-e^{bt}}{c-b}-\frac{e^{bt}-e^{at}}{b-a}}{c-a}},\nonumber\\
&e^{[a,b,c,d]t}=\Scale[1.22]{\frac{\frac{\frac{e^{dt}-e^{ct}}{d-c}-\frac{e^{ct}-e^{bt}}{c-b}}{d-b}-\frac{\frac{e^{ct}-e^{bt}}{c-b}-\frac{e^{bt}-e^{at}}{b-a}}{c-a}}{d-a}}.
\end{align} We also have
\begin{equation}\label{ddrepeat}
e^{[a,\ldots,a]}=t^ke^{at},
\end{equation} where $a$ appears $(k+1)$-times in the LHS of Eq.~(\ref{ddrepeat}) which follows from the pattern given in Eq.~(\ref{e[a,b]}) taking a limit procedure.
Because $\star$-products and $\star$-powers are reduced to iterated integrals of ordinary exponentials in the Omega domain, by the Hermite-Genocchi formula, divided-difference exponentials naturally appear in the exact solutions of all quantum problems considered here. Mathematically this is encapsulated by the following relation (see also Appendix~\ref{OC}),
\begin{align}\label{Id4}
\overset{\lambda}{\underset{=
}{\Omega}}\,\, e^{\lambda t}\,\frac{1}{\lambda^{n}\prod_{k=0}^{n}(1-a_k/\lambda)}=e^{[a_0,a_1,\ldots,a_{n}]t}.
\end{align}
That is, divided-difference exponentials are the time domain equivalent of rational functions with trivial numerators in the Omega domain. 
This result extends to other rational functions via divided-differences of polynomials times exponentials. An additional advantage of Omega calculus is that advanced relations between such divided-difference functions all reduce to elementary partial fraction decompositions in the Omega domain, a powerful tool to obtain compact time domain expressions for scalar $\star$-resolvent and perform exact resummations of series of divided-difference functions.

\subsection{Overview}
To summarize, in the $\star$-algebra, the Green's function of \textit{any} linear differential system is a matrix resolvent with respect to the $\star$-product and the evolution operator is the $\star$-product of this with a Heaviside Theta function. By the method of path-sum, any entry of the matricial $\star$-resolvent is a continued fraction of $\star$-products and scalar $\star$-resolvents that \textit{terminates at finite depth}. These products and resolvents are evaluated in the Omega domain where the time-dependence of all functions is carried by ordinary exponentials while Omega variables enter rational functions. Eliminating these variables yields divided-difference exponentials and the path-sum fraction evaluates to an unconditionally convergent series of those.

\section{\Large First case \texorpdfstring{$\epsilon(t)=e_0
\cos(\omega_0 t)$, $f(t)=1$}{cosine modulation of ground state}}\label{FirstCase}
\noindent We first consider the time dependent Hamiltonian:
\begin{equation}\label{Hfirstcase}
\mathsf{H}(t) = 
\begin{pmatrix}
S_0 + e_0
\cos(\omega_0 t) & g \\
\bar{g} & S_1
\end{pmatrix},
\end{equation}
corresponding to the very simple case where the static coupling $g$ between the ground and the excited states of the system is supposed to bring a part of the small $e_0$ energy periodic oscillations of the $S_0$ ground to the $S_1$ excited states. A direct path-sum treatment of the above is possible (see Appendix~\ref{PS2x2}). But we first perform a standard frame change to be on line with existing work, e.g. \cite{Warnock2025}. We emphasize that this preliminary step is not required by the theoretical machinery presented above. Let $\mathsf{H_{frame}}(t)=\text{diag}\big(
S_0 + e_0
\cos(\omega_0 t) , S_1
\big)$,
the Hamiltonian in the new frame is given by:
\begin{align}
\mathsf{H}_{\text{rot}}(t)&= \mathsf{U}_0^{\dagger}(t)\mathsf{H}(t)\mathsf{U}_0(t)-i\mathsf{U}_0^{\dagger}(t)(d/dt)\mathsf{U}_0(t),\nonumber\\
&=\begin{pmatrix}
0 & g e^{i S_0 t + \frac{i e_0}{\omega_0}\sin{\omega_0 t}} e^{-i S_1 t}\\ 
\overline{g} e^{-i S_0 t - \frac{i e_0}{\omega_0}\sin{\omega_0 t}} e^{i S_1  t} & 0  
\end{pmatrix},
\end{align}
using $\mathsf{U}=\mathsf{U}_0\mathsf{U}_{\text{rot}}$ and $(d/dt)\left(\mathsf{U},\mathsf{U}_{\text{rot}}\right)=-i\left(\mathsf{H}\mathsf{U},\mathsf{H}_{\text{rot}}\mathsf{U}_{\text{rot}}\right)$, where
\begin{equation}
\mathsf{U}_0(t)=e^{-i\int\mathsf{H}_{\text{frame}}(t)dt}
=\begin{pmatrix}
e^{-iS_0t -i\frac{e_0}{\omega_0}\sin\omega_0 t}  & 0\\ 
0 & e^{-iS_1t}
\end{pmatrix}.
\end{equation}

\subsection{Exact analytical solution}
Using the Jacobi-Anger relation (see Appendix~\ref{JacobiAnger}, in particular Eq.~(\ref{J-A}) to expand the exponentials of trigonometric functions, it comes:
\begin{align}
\mathsf{H}_{\text{rot}}(t) &= \begin{pmatrix}
0 & g \sum_n J_n(\frac{e_0}{\omega_0}) e^{i (S_0 -S_1 + n \omega_0) t }\\ 
\overline{g} \sum_n J_n(\frac{e_0}{\omega_0}) e^{-i (S_0 -S_1 + n \omega_0) t }& 0  
\end{pmatrix},\nonumber\\
&=\begin{pmatrix}
0 & H_{12}(t)\\
H_{21}(t)&0
\end{pmatrix}.
\end{align}
The goal of using the Jacobi-Anger relation in Eq.~(\ref{J-A}) is to obtain kernels that involve sums of exponentials to recover series of divided-difference exponentials in the solution. This step is not a necessity: using the Omega calculus, one may obtain the solution regardless of the availability (or existence) of the Fourier series expansion of the Hamiltonian. Thus, we warn that while the method appears close to the Floquet formalism results, it is only superficially so in that the proximity stems only from us adopting a Fourier series decomposition to recover divided-difference exponentials. 
In this frame we have, per path-sums,
\begin{subequations}\label{PSUequations}
\begin{align}
U_{ii}&=\Theta\star (1_\star -(-i)^2 H_{ij}\Theta\star H_{ji}\Theta)^{\star -1},\nonumber\\
&=\Theta\star \sum_{k\geq 0} (-i)^{2k}\big(H_{ij}\Theta\star H_{ji}\Theta\big)^{\star k},\label{U_ii}\\
U_{ij}&=\Theta\star(-i)H_{ij}U_{jj}\Theta,\label{U_ij}
\end{align} 
\end{subequations}
both for $i,j\in\{1,2\}$ and $i\neq j$. Note that we only need $U_{11}$ and $U_{12}$ since
\begin{align}\label{Symcc}
U_{11}|_{\varepsilon_0\rightarrow \bar{\varepsilon}_0}=\bar{U}_{22}\quad {\rm and}
\quad \bar{g}U_{12}|_{\varepsilon_0\rightarrow \bar{\varepsilon}_0}=-g\bar{U}_{21}.
\end{align} We can now evaluate these $\star$-products and $
\star$-resolvents exactly with Omega calculus using divided-difference exponentials. 
Let $\pmb{m}_k:=(m_1,\ldots,m_k)\in\mathbb{N}^k$ and similarly for $\pmb{n}_k$. Let $J_n$ be the $n$th Bessel function of Eq.~(\ref{Bessel}) and define
\begin{equation}\label{BesselSymbolsDef}
J_{\pmb{n}_k,\pmb{m}_k}:=\prod_{i=1}^k J_{n_i}(e_0/\omega_0)\, J_{m_i}(e_0/\omega_0)
\;\; {\rm and} \;\; J_{n,\pmb{n}_k,\pmb{m}_k}:=J_n(e_0/\omega_0)J_{\pmb{n}_k,\pmb{m}_k}(e_0/\omega_0).
\end{equation} 
Then we have
\begin{equation}
\mathsf{U}_{\text{lab}}(t)=\underbrace{\begin{pmatrix}
e^{-iS_0t -i\frac{e_0}{\omega_0}\sin\omega_0 t}  & 0\\ 
0 & e^{-iS_1t}  
\end{pmatrix}}_{=\mathsf{U}_0(t)}\underbrace{\begin{pmatrix}
U_{11}(t) & U_{12}(t)\\ 
U_{21}(t) & U_{22}(t)  
\end{pmatrix}}_{=\mathsf{U}_{\text{rot}}(t)},
\end{equation}
where, using Eq.~(\ref{Id1}),
\begin{subequations}
\begin{align}
U_{11}(t) &= 
\sum_{k\geq 0}|g|^{2k}\sum_{\pmb{m}_k,\pmb{n}_k} J_{\pmb{n}_k,\pmb{m}_k}\,\, 
e^{i[A_1,\varepsilon_0+B_1,A_2,\varepsilon_0+B_2,\ldots,A_k,\varepsilon_0+B_k,0]t},\label{U11Case1}\\
U_{12}(t) &= 
 \sum_{k\geq 0}g|g|^{2k} \sum_{n,\pmb{m}_k,\pmb{n}_k}J_{n,\pmb{n}_k,\pmb{m}_k}\,e^{i(\varepsilon_0+n\omega_0)t}\nonumber\\ 
 &\hspace{15mm} \times e^{-i[A_1,\varepsilon_0+B_1,A_2,\varepsilon_0+B_2,\ldots, A_k,\varepsilon_0+B_k, \varepsilon_0+n\omega_0,0]t},\label{U12Case1}
\end{align}
\end{subequations}
with $\varepsilon_0 = S_0 - S_1$ (this is different from $e_0$), $A_i:=N_i-M_i$, and $B_i:=N_i-M_{i-1}$ such that for $j\geq 1$, $M_j = \Sigma_{i=1}^{j} m_i \omega_0$ otherwise $M_0=0$, and for $j\geq 1$, $N_j = \Sigma_{i=1}^{j} n_i \omega_0$. 
\begin{remark*}
In the above and from now on we use the convention 
\begin{equation}
[X_1,Y_1,\ldots,X_k,Y_k,\pmb{a}]|_{k=0}=[\pmb{a}].
\end{equation} So, e.g., if $X_i=A_i$ and $Y_i=\varepsilon_0+B_i$ we get $[X_1,Y_1,\ldots,X_k,Y_k,\pmb{a}]|_{k=0}=[0]=0$ if $\pmb{a}=0$ and $[X_1,Y_1,\ldots,X_k,Y_k,\pmb{a}]|_{k=0}=[\varepsilon_0+n\omega_0,0]$ if $\pmb{a}=(\varepsilon_0+n\omega_0,0)$.
\end{remark*}
 The term of order $k$ in $U_{11}$ involves a product of $2k$ Bessel functions and divided-difference exponentials with $2k+1$ arguments. For $U_{12}$, the term of order $k$ comprises a product of $2k+1$ Bessel functions and divided-difference exponentials with $2k+2$ arguments. Given that $|J_m(x)|\leq1$ for all $x$ and $m$, the amplitude of these terms decay very fast, enabling an excellent approximation through truncation of the above series. This decay is further accelerated in the situations where $|g|/\omega_0 \ll 1$. The series given here are nonetheless \emph{unconditionally convergent}, a property they inherit from the $\star$-Neumann series. This is manifested by the following fact about divided-difference exponentials $|e^{[a_1,\ldots,a_n]t}|\leq (M t)^n/n!$ for some finite $M$ as long as $a_i$ are finite \cite{Zeng2025}. 
 As a further final check, we prove in Appendix~\ref{FirstCaseTimeIndep} that in the case $e_0=0$--for which the Hamiltonian becomes time-independent--the solution above correctly reduces to the expected results.\\ 

\begin{figure}[t!]
 \centering
 \includegraphics[scale=0.5]{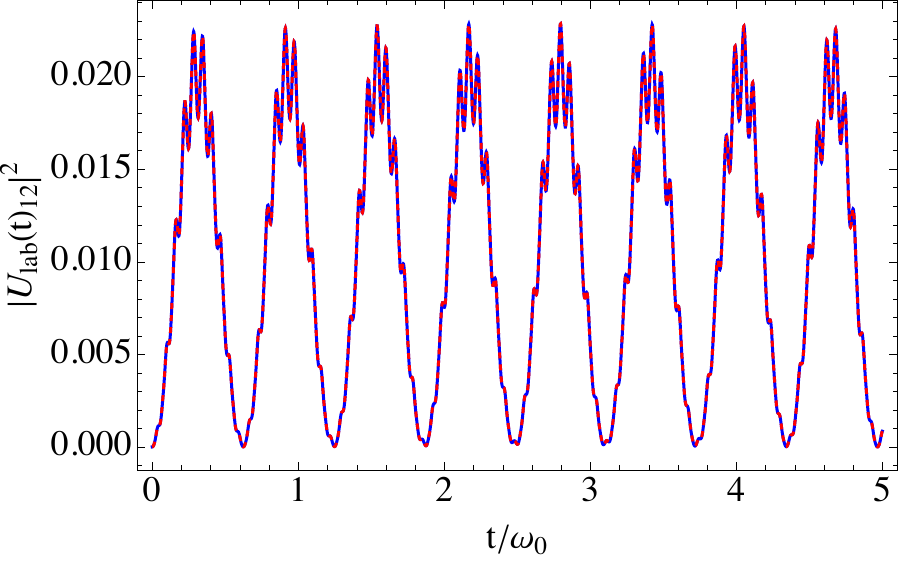} \hspace{5mm}\includegraphics[scale=0.48]{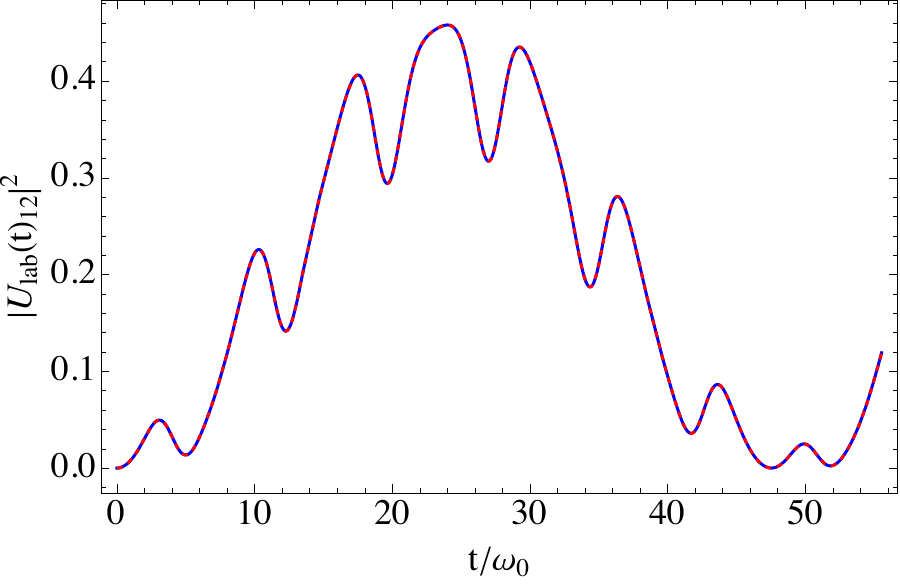} 
 \caption{Evolution of the transition probability $P_{|0\rangle\to |1\rangle}(t):=|U_\text{lab}(t)_{12}|^2=1-|U_\text{lab}(t)_{11}|^2$ as a function of time as determined by a fully numerical solver (solid blue line) and exact analytical solution  Eq.~(\ref{U12Case1}) plotted from its  truncation $|U_{12}^{(3,2)}|^2$ (dashed red line). Parameters for the \textbf{top} figure: $\omega_0/|S_1-S_0|=10$, $g=0.1$, $e_0=\omega_0$. This is a high-frequency case since $\omega_0\gg S_1-S_0$. Parameters for the \textbf{bottom} figure: $\omega_0/|S_1-S_0|=0.9$, $g=0.1$, $e_0=\omega_0$. 
 Remark that the parameters of the up figure are not suited to the high-frequency approximation, as $\omega_0\sim |S_1-S_0|$, and also lie beyond the RWA approximation. In both cases, the analytical results flawlessly match the fully numerical simulations.}
 \label{fig:Case1}
\end{figure}

Various approximation strategies can be devised from the exact solution given above in order to plot it, help interpret it, or relate it to existing perturbative results. Plotting is best achieved by naively truncating Eqs.~(\ref{U11Case1}, \ref{U12Case1}) at any finite order $K$ as well as truncating all terms with Bessel amplitude $J_{m}$ whose parameter $m$ runs over some threshold $|m|>L$. We let $U^{(K,L)}$ be the resulting approximation for the evolution operator. Rigorously, we let
\begin{equation}\label{Amn}
A_{n,m}(t,s)\,\Theta=
i|g|^2  J_n(e_0/\omega_0)
J_m(e_0/\omega_0)e^{i(n -m)\omega_0t}e^{i[\varepsilon_0+m\omega_0,0](t-s)}~\Theta,
\end{equation}
so that the exact solution is, per Eq.~(\ref{U_ii}),
$
U_{11}=\Theta\star \big(1_\star - \sum_{m,n\in\mathbb{Z}}A_{m,n}\Theta\big)^{\star-1}
$ and ultimately yields Eq.~(\ref{U11Case1}).
Then $U^{(K,L)}$ is given by
\begin{equation}
U^{(K,L)}_{11}=\Theta\star \sum_{k=0}^K\Big(\sum_{m,n\in\mathbb{Z}\atop |m|,|n|\leq L}A_{m,n}\Theta\Big)^{\star k}.
\end{equation}
This strategy is particularly successful in deciphering the solution's behavior at finite times and over all parameter values; see Fig.~\ref{fig:Case1}. 
If the need arises--for example to help physical interpretations--alternative analytical expressions for any of the $U^{(K,L)}$ or the general solution can be obtained via partial fraction expansions in the Omega domain. 
 This way we can, for instance, obtain a representation of 
 $U^{(3,2)}_{11}$ involving products of divided-differences of cardinal sine functions. 
 
Alternative approximation strategies reveal the link with existing approaches, such as perturbative expansions. For example, 
supposing that $e_0\ll \omega_0$, we observe that $J_n(e_0/\omega_0)\simeq 0$ for all $n\neq 0$. So we may retain only terms whose coefficients involve solely powers of $J_0$ in the exact solution. This yields
$
U_{11}
\simeq \Theta \star (1_\star - A_{0,0}\Theta)^{\star-1}=:U^{(\infty,0)}_{11}.
$
Using Omega calculus, this is immediately found to be (see Appendix~\ref{Approx}), 
\begin{subequations}
\begin{align}
U^{(\infty,0)}_{11}&=1+J_0^2|g|^2e^{i[r_{0+},r_{0-},0]t}
= e^{\frac{i \varepsilon_0 t}{2}} \left(\cos( \Omega t/2 )-\frac{i \varepsilon_0}{\Omega} \sin (\Omega t/2)\right),\\
U^{(\infty,0)}_{12}&=-\frac{2iJ_0g}{\Omega}e^{\frac{i \varepsilon_0 t}{2}}\sin(\Omega t/2),
\label{eq:Case1_approx}
\end{align}
\end{subequations}
where $r_{0\pm}=\frac{1}{2}\Big(\varepsilon_0\pm \sqrt{\varepsilon_0^2+4J_0^2|g|^2}\Big)$ and $\Omega =r_{0+}-r_{0-}=\sqrt{\varepsilon_0^2+4 |g|^2 J_0^2}$. 
This is also precisely the output of the lowest-order of the standard high-frequency approximation \cite{Eckardt2015} for the Hamiltonian considered here, valid for $\omega_0$ much larger than other system parameters (causing again $e_0/\omega_0$ to be small). Higher orders are produced similarly, by retaining more families of terms. We emphasize that contrary to the high-frequency expansion under its various guises \cite{Eckardt2015,Mikami2016} (Brillouin-Wigner theory, van-Vleck perturbation theory, Floquet-Magnus approximation) or under other approximation schemes e.g. \cite{Vogl2019}, we access here approximated results from the exact one. In particular, the goal in doing this is not to determine the time-evolution, which is now exactly and explicitly known, but rather to connect with past successful approaches or to help interpretations from simpler expressions in the quest for physical meaning. 

One interesting case is to be positioned down in frequency below the resonances. While we can obtain the stored energy exactly using Eq.~(\ref{eq:Energy_Stored_Large}) by substituting in Eqs.~(\ref{U11Case1}) and (\ref{U12Case1}), a simpler expression can be found as discussed above. Starting in the ground state and using Eq.~(\ref{eq:Case1_approx}), the very low amplitude energy oscillation of the ground state is now showing up in the excited state since in this case:
$\Delta E(t)/S_1 = (|g|^2/\Omega^2)\,\sin^2(\Omega t/2)$.

\subsection{Population transfer at resonance}
\begin{figure}[t!]
 \centering
 \includegraphics[width = \textwidth]{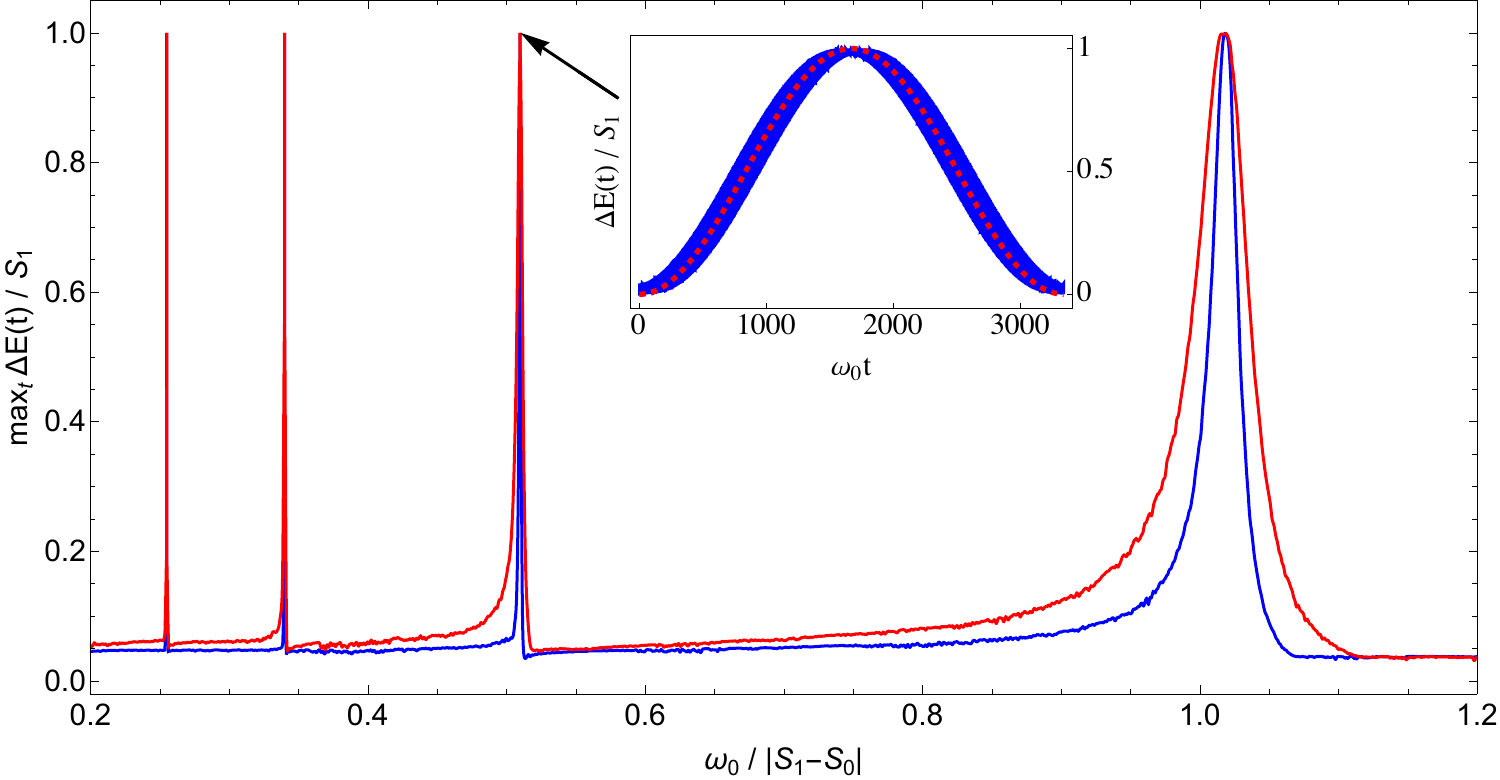} 
 \caption{Numerical simulations of the quantity $\max_t \Delta E(t)/S_1=\max_t P_{|0\rangle\to|1\rangle}(t)$ as a function of $\omega_0$ for $S_0=\Gamma=0$, $S_1=1$, $\varepsilon_0=S_1-S_0=1$, $g=|S_1-S_0|/10$ and $e_0=|S_1-S_0|/10$ (blue line) or $e_0=|S_1-S_0|/5$ (red line). 
 Parametric resonances occur very close to the low fractional frequencies $\omega_0\simeq|S_1-S_0|$, $\omega_0\simeq|S_1-S_0|$/2, $\omega_0\simeq|S_1-S_0|$/3 etc. They are shifted from those exact values by Bloch-Siegert-like shifts. Further resonant peaks for $m\geq 5$ are not shown here. Numerical simulations for those get harder as $m$ increases: not only are those resonances narrower but the simulation times required to reliably evaluate $\max_t P_{|0\rangle\to|1\rangle}(t)$ grow extremely fast with $m$. Indeed resonances at $\omega_0\simeq |S_1-S_0|/m$ manifest themselves as ultra-slow oscillations occurring at an effective frequency of $|g J_m(m\,e_0/\varepsilon_0)|\sim |g| (m\,e_0/\varepsilon_0)^m/(2^m m!)$ and so require a simulation time of at least $2^m m!/(|g| (m\,e_0/\varepsilon_0)^m)$ to be reliably picked up which quickly becomes untractable. In inset, plot of $\Delta E(t)/S_1$ as a function of $t$ for $m=2$ and $e_0=|S_1-S_0|/10$ (solid blue curve), together with the theoretically-predicted main population $\sin(|g J_2(e_0/\omega_0)|t)^2$ (dashed red-line) for the resonant $\omega_0=0.50988 |S_1-S_0|$. The thick appearance of the blue curve is due to ultra-high frequency oscillations in the exact solution atop of the main population contributor. 
 }
 \label{fig:DeltaECase1}
\end{figure}
Interestingly and as presented in Fig.~\ref{fig:DeltaECase1}, parametric resonances are showing up when plotting the $\max_t \Delta E(t)/S_1$ normalized maximum stored energy as a function of the $S_0$ perturbation modulation frequency $\omega_0$. No need here to use a large $e_0$ to bring a large population on the excited state. At integer fractions of the energy difference between states $\omega_0\simeq |S_1-S_0|/m$, $m\in\mathbb{N}\backslash\{0\}$, those dynamical resonances are called ``parametric'' because it is the $S_0$ structural characteristic of the quantum system which is here modulated. 

Analytically, this effect is correctly produced by the solution Eq.~(\ref{U12Case1}). Indeed, for any positive integer $m$, if $\omega_0=|S_1-S_0|/m$ then $n=-m$ implies $\varepsilon_0+n\omega_0=0$ which creates a resonance of families of divided-difference exponentials. Mathematically, resonant divided-difference exponentials produce terms that are polynomial in time, which when summed produce a low-frequency oscillation allowing $P_{|0\rangle\to|1\rangle}(t)$ to reach $1$ over very long times. For instance, take the leading term of Eq.~(\ref{U12Case1}), which is $-g\sum_{n\in\mathbb{Z}}J_n e^{i[\varepsilon_0+n\omega_0,0]t}$. If $n=-m$ then this reduces to $-g J_{-m} e^{i[0,0]t}= g J_{m} i t $, here producing the first-order term of the Taylor expansion of the sine oscillation with an effective frequency of $|gJ_m(m\,e_0/\varepsilon_0)|$. More rigorously, for $\omega_0=\varepsilon_0/m$, the $\star$-kernel $(-i)^2 H_{12}\Theta\star H_{21}\Theta$
is dominated by the single resonant term $A_{m,m}$. This implies that $U_{11}(t)\simeq\Theta\star (1_\star - A_{m,m}\Theta)^{\star -1}=\cos\big(|g J_m(m\,e_0/\varepsilon_0)| t\big)$ and $U_{12}(t)=\Theta\star(-i)A_{m,m}U_{11}=\sin(|g J_m(m\,e_0/\varepsilon_0)| t)$ as per Eqs.~(\ref{U_ii}) and (\ref{U_ij}), respectively. Thus, at resonance, the main contribution to the population in $\Delta E(t)/S_1=P_{|0\rangle\to|1\rangle}(t)$ is given by 
\begin{equation}\label{PopResCase1}
 P_{|0\rangle\to|1\rangle}(t)\sim \sin^2\!\big(|g|J_m(e_0/\omega_0) t\big)\sim\sin^2\!\big(|g|J_m(m\,e_0/\varepsilon_0) t\big),
\end{equation}
as also confirmed through numerical simulations; see the inset plot in Fig.~\ref{fig:DeltaECase1}. We discuss in more detail in \S\ref{BSCaseSection} how to systematically determine the resonances and related physical quantities (width, shifts, effective models) from the exact solutions.

\section{\Large Second case \texorpdfstring{$\epsilon(t)=e_0
\cos(\omega_0 t)$, $f(t)=\cos(\omega t)$}{}}\label{SecondCase}
\begin{figure}[t!]
 \centering
 \includegraphics[scale=0.5]{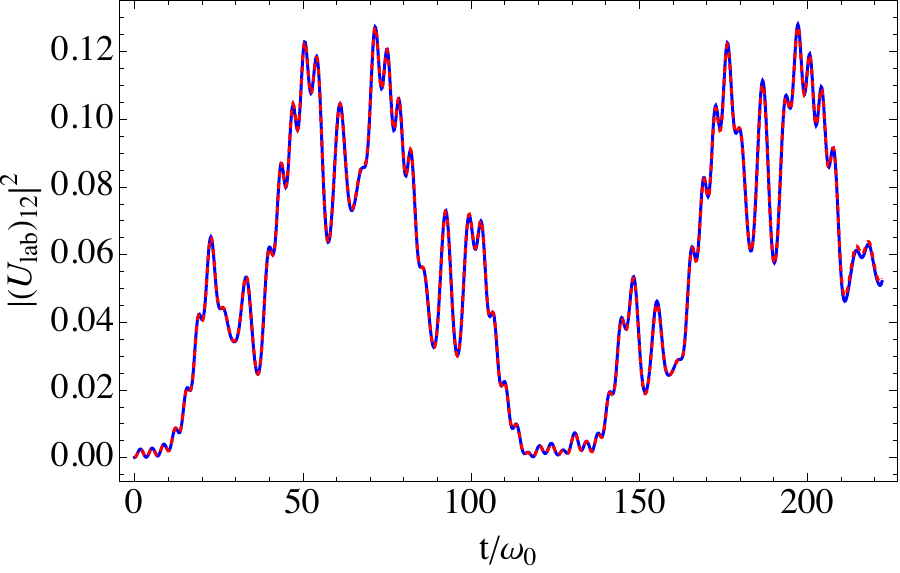} 
 \hspace{5mm}
 \includegraphics[scale=0.475]{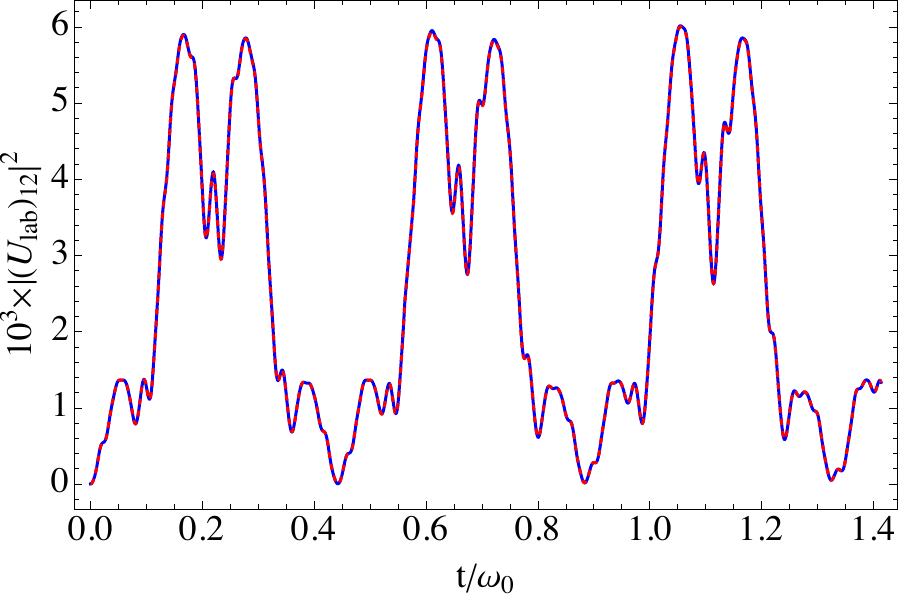} 
 \caption{Evolution of the transition probability $P_{|0\rangle\to |1\rangle}(t):=|U_\text{lab}(t)_{12}|^2=1-|U_\text{lab}(t)_{11}|^2$ as a function of time as determined by a fully numerical solver (solid blue line) and analytical truncation $|U_{12}^{(3,4)}|^2$  (\textbf{top} figure, dashed red line) or $|U_{12}^{(2,2)}|^2$  (\textbf{bottom} figure, dashed red line) of the exact solution Eqs.~(\ref{U11Case2}, \ref{U12Case2}). Parameters for the \textbf{top} figure: $\omega_0/|S_1-S_0|=0.45$, $\omega/|S_0-S_1|=2$, $g=0.1$, $e_0=2\omega_0$. In this situation, all three frequencies in the system are comparable, a regime unsuitable to both the RWA and the high-frequency expansion. The ``erratic'' behavior of the solution due to the mixing comparable system frequencies is perfectly captured by the analytical expression, which is indistinguishable from the numerical solution. Parameters for the \textbf{bottom} figure: $\omega_0/|S_1-S_0|=10\sqrt{2}$, $\omega/|S_0-S_1|=2$, $g=0.1$, $e_0=\omega_0$. In this situation $\omega_0$ and $\omega$ are not commensurate and the Hamiltonian is not periodic, yet Eqs.~(\ref{U11Case2}, \ref{U12Case2}) continue to hold.}
 \label{fig:Case2}
\end{figure}
Adding an oscillatory part in the $g$ coupling of Eq.~(\ref{Hfirstcase}) reflects the standard way in physics to bring more population from the ground to the excited states. This is well known for $e_0 =0$, because of the absence of $\omega_0$ in this case in Eq.~(\ref{Hfirstcase}). This give rise to the famous I.I. Rabi resonance when $\omega =|S_1-S_0|$. Considering also a small ground state energy oscillation, i.e., $e_0\ne 0$ adds up two more resonances possibilities while comparing again $\omega_0$ with $|S_1-S_0|$ and now $\omega_0$ with $\omega$. Our new Hamiltonian reads:
\begin{equation}
\mathsf{H}(t) = 
\begin{pmatrix}
S_0 + e_0
\cos(\omega_0 t) & g \cos(\omega t) \\
\bar{g}\cos(\omega t) & S_1
\end{pmatrix}.
\end{equation}
We shall make no assumption on the two frequencies $\omega$ and $\omega_0$, in particular, they need not be commensurate. 
We proceed as in Case 1, first performing a standard frame change, then using the Jacobi-Anger relation, so we work with $H_{11}=0=H_{22}$ and
\begin{subequations}
\begin{align}
&H_{12}(t) =  \frac{g}{2} \sum_n \sum_{\alpha=\pm 1} J_n(e_0/\omega_0)e^{i(S_0-S_1-i\Gamma+ n \omega_0+\alpha\omega) t},\\
&H_{21}(t)=\frac{\bar{g}}{2} \sum_n\sum_{\alpha=\pm1} J_n(e_0/\omega_0)e^{-i(S_0-S_1-i\Gamma+ n \omega_0+\alpha\omega) t}.
\end{align}
\end{subequations}
The path-sum results of Eqs.~(\ref{U_ii}, \ref{U_ij}) remain valid in this situation and yield the path-sum kernel
\begin{align}
&(-i)^2\,H_{12}\Theta\star H_{21}\Theta
:=\\
&i|g/2|^2 \sum_{n,m\in\mathbb{Z}\atop\beta,\alpha= \pm1} J_n(e_0/\omega_0)
J_m(e_0/\omega_0)e^{i\{(n -m)\omega_0+(\beta-\alpha)\omega\}t}e^{i[\varepsilon_0+m\omega_0+\alpha\omega,0](t-s)}\,\Theta.\nonumber
\end{align} 
Just as in the first case, the evolution operator is now given by the formalism of Appendix~\ref{GenProof}.

\subsection{Exact analytical solution}
The evolution operator in the laboratory frame is given by
\begin{equation}
\mathsf{U}_{\text{lab}}(t)=\begin{pmatrix}
e^{-iS_0t -i\frac{e_0}{\omega_0}\sin\omega_0 t}  & 0\\ 
0 & e^{-iS_1 t}  
\end{pmatrix}\begin{pmatrix}
U_{11}(t) & U_{12}(t)\\ 
U_{21}(t) & U_{22}(t)  
\end{pmatrix}.
\end{equation}
In this situation as in the previous one, Eq.~(\ref{Symcc}) holds and, therefore, we need only to determine $U_{11}$ and $U_{12}$ exactly.
Let $\pmb{m}_k:=(m_1,\ldots,m_k)\in\mathbb{N}^k$, similarly for $\pmb{n}_k$ and $\pmb{\alpha}_k:=(\alpha_1,\ldots,\alpha_k)\in\{-1,1\}^k$, similarly for $\pmb{\beta}_k$. Evaluating the general path-sum solution (Appendix~\ref{PS2x2}) using Omega calculus, we find 
\begin{subequations}\label{Case2GenSol}
\begin{align}
U_{11}(t)&=\sum_{k\geq 0}\Big|\frac{g}{2}\Big|^{2k}\sum_{\pmb{m}_k,\pmb{n}_k\atop\pmb{\alpha}_k,\pmb{\beta}_k}\!\!\!J_{\pmb{n}_k,\pmb{m}_k}\,e^{i[A_1, \varepsilon_0+B_1,A_2,\varepsilon_0B_2,\dots,A_k,\varepsilon_0 +B_k,0]t},\label{U11Case2}
\end{align} and
\begin{align}
U_{12}(t)&=
\sum_{k\geq 0}\frac{g}{2}\Big|\frac{g}{2}\Big|^{2k}\!\!\!\sum_{n,\pmb{m}_k,\pmb{n}_k\atop \alpha,\pmb{\alpha}_k,\pmb{\beta}_k}\!\!\!J_{n,\pmb{n}_k,\pmb{m}_k}\,e^{i(\varepsilon_0+n\omega_0+\alpha \omega)t}\nonumber\\
&\hspace{15mm} \times e^{-i[A_1,\varepsilon_0+B_1,
A_2,\varepsilon_0+B_2,\dots,A_k,\varepsilon_0+B_k,\varepsilon_0+n\omega_0+\alpha \omega,0]t}.\label{U12Case2}
\end{align}
\end{subequations}
In these expressions, we have again
$\varepsilon_0 = S_0 - S_1 $, $A_i:=N_i-M_i$, and  $B_i:=N_i-M_{i-1}$ with $M_j=\sum_{i=1}^j(m_i\omega_0+\alpha_i\omega)$ and
$N_j=\sum_{i=1}^j(n_i\omega_0+\beta_i\omega)$. The Bessel function symbols are defined as in Eq.~(\ref{BesselSymbolsDef}).
At an integer resonance, i.e., when $\omega$ and $\omega_0$ are commensurate, we obtain a shift in $m_i\rightarrow m_i+\alpha_i$ and $n_i\rightarrow n_i+\beta_i$. An alternative representation of the solution without the sums over $\pmb{\beta}_k$ and $\pmb{\alpha}_k$ is accessible using Omega calculus but involves divided-difference functions beyond the exponential and heavier notation.

An example of the time evolution of the transition probability $P_{|0\rangle\to |1\rangle}(t):=|U_\text{lab}(t)_{12}|^2$ is shown in Fig.~\ref{fig:Case2} which confirms the agreement between fully numerical solutions and the analytical results above. It is noteworthy that this agreement is maintained in the parameter range where all frequencies present in the system are comparable but not commensurate. The solution is almost periodic and exhibits erratic behavior. It is also the case when the Hamiltonian is not periodic due to the large difference between the frequencies involved.

\subsection{Population transfer: resonances \& double resonances}

\subsubsection{Bloch-Siegert case \texorpdfstring{$e_0=0$}{}.}\label{BSCaseSection}
\begin{figure}[t!]
 \centering
 \includegraphics[width = \textwidth]{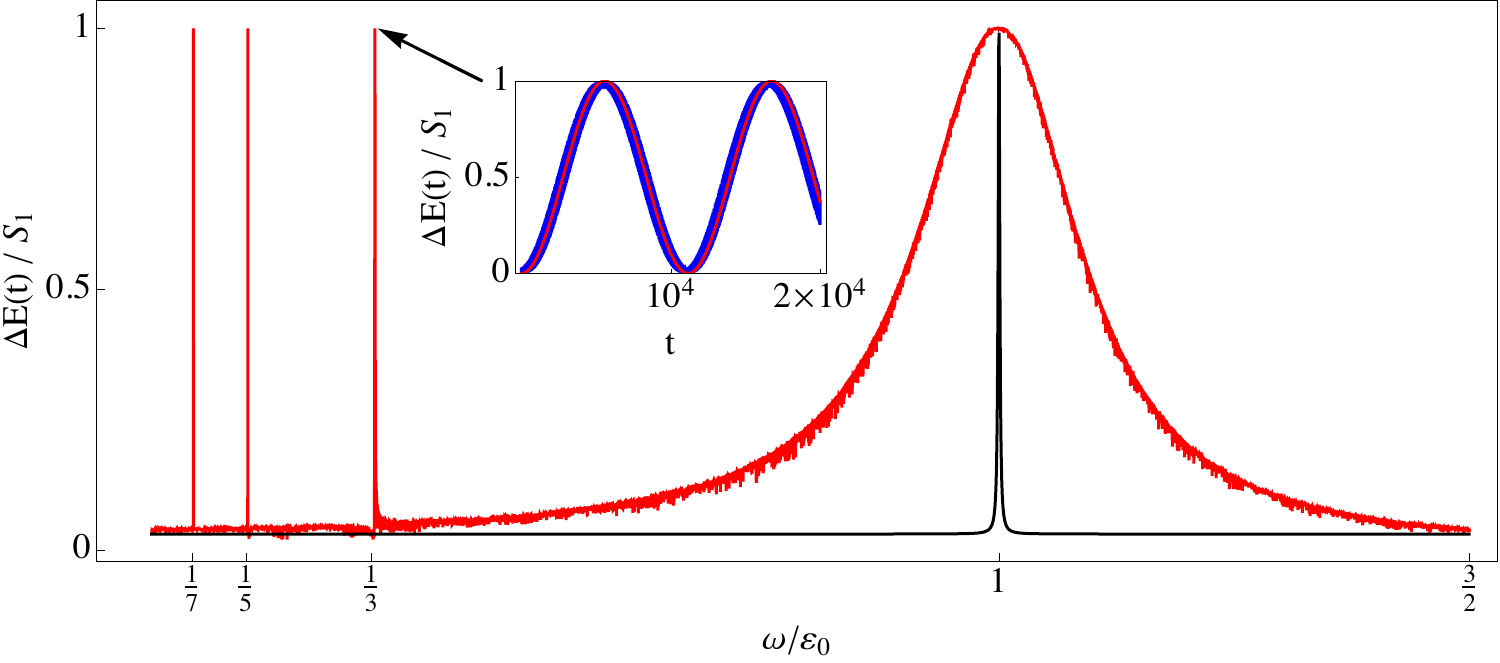} 
 \caption{Profile of $\max_t \Delta E(t)/S_1$ as a function of $\omega$ in the Bloch-Siegert Hamiltonian for $g=|S_1-S_0|/10$ (red curve) and $g=|S_1-S_0|/1000$ (black curve). Resonances occur whenever $\varepsilon_0$ is an odd multiple of $\omega$.
 In \textbf{inset}, numerical simulation of $\Delta E(t)/S_1=P_{|0\rangle\to|1\rangle}(t)$ as a function of $t$ for $\varepsilon_0=\omega/0.3371$ (blue line) together with the prediction of Eq.~(\ref{Carrier3Res}) (solid red line) for the long-time carrier. 
 Parameters: $S_0=\Gamma=0$, $S_1=1$, $\varepsilon_0=S_0-S_1-i\Gamma=-1$ and $e_0=0$. Numerically, the resonances are located at $\omega/\varepsilon_0\simeq0.3371,\, 0.202068$ and $0.1443048$. These values are larger than $1/3$, $1/5$ and $1/7$ due to $\varepsilon_0<0$.}
  \label{fig:BSResonances}
\end{figure}

The situation where $e_0=0$ corresponds to the well-known I.I. Rabi physical case where at least for $|g|\ll|S_0-S_1|$, the out-of-diagonal $g\cos(\omega t)$ term is there for the $S_0$ excited state to be populated by a resonance between this external $g\cos(\omega t)$ classical field and the $\hbar |S_0-S_1|$ internal structural frequency. More generally and when $g$ is not small as compared to $|S_0-S_1|$, it leads to the standard Bloch-Siegert Hamiltonian where the path-sum kernel generating the solution simplifies to: 
\begin{align}\label{KernelBSORIGINAL}
(-i)^2\,H_{12}\Theta\star H_{21}\Theta |_{e_0=0}&
=i|g/2|^2 \sum_{\alpha,\beta=\pm1} e^{i(\beta-\alpha)\omega t}e^{i[\varepsilon_0+\alpha\omega,0](t-s)}\,\Theta,
\end{align} using $J_n(e_0/\omega)=J_n(0)=\delta_{n,0}$ which follows from Eq.~(\ref{Bessel}) and the solution now only involves sums over the $\pm1$ indices,
\begin{subequations}\label{BSSol}
\begin{align}
U_{11}(t)|_{e_0=0}&=\sum_{k\geq 0}\Big|\frac{g}{2}\Big|^{2k}\sum_{\pmb{\alpha}_k,\pmb{\beta}_k} e^{i[A_1, \varepsilon_0+B_1,A_2,\varepsilon_0+B_2,\dots,A_k,\varepsilon_0 +B_k,0]t},\label{U11Case2BS}\\
U_{12}(t)|_{e_0=0}&= 
\sum_{k\geq 0}\frac{g}{2}\Big|\frac{g}{2 }\Big|^{2k}\!\!\!\sum_{\alpha,\pmb{\alpha}_k,\pmb{\beta}_k} e^{i(\varepsilon_0+\alpha \omega) t}\nonumber\\
&\hspace{15mm}\times e^{-i[A_1,\varepsilon_0+B_1,
A_2,\varepsilon_0+B_2,\dots,A_k,\varepsilon_0+B_k,\varepsilon_0+\alpha \omega,0]t},\label{U12Case2BS}
\end{align}
\end{subequations}
with $A_i:=N_i-M_i$, and $B_i:=N_i-M_{i-1}$ with $M_j=\sum_{i=1}^j\alpha_i\omega$ and
$N_j=\sum_{i=1}^j\beta_i\omega$. Assuming further that $g\ll 1$, the above can be approximated via perturbative formulas. 
Just as in Case 1, the solution of Eqs.~(\ref{BSSol}) undergoes resonances \cite{Shirley1965}, although by a different mechanism. As previously, the resonances occur when divided-difference functions present repeated arguments. In order to systematically track those, we begin by reformulating the mathematical kernel generating the solution, Eq.~(\ref{KernelBSORIGINAL}), into the divided-difference of a single function namely (see Appendix~\ref{AltKBS}),
\begin{align}\label{KernelBS}
&(-i)^2\,H_{12}\Theta\star H_{21}\Theta|_{e_0=0}\nonumber\\
&\hspace{15mm}=i|g/2|^2
\Big(A(\bullet,e^{i\omega s})e^{i\bullet(t-s)}\Big)[\varepsilon_0+\omega,\varepsilon_0-\omega,2\omega,-2\omega,0],
\end{align} 
which should be understood as the divided-difference in $[\varepsilon_0+\omega,\varepsilon_0-\omega,2\omega,-2\omega,0]$ of the function $F(X):=A(X,Y)e^{iX(t-s)}$ with 
\begin{align}\label{A}
A(X,Y)&:=2(X-\varepsilon_0)(X^2-4\omega^2)\nonumber\\
&\hspace{15mm}+X(X-\varepsilon_0-\omega)(X-2\omega)Y^{-2}\nonumber\\
&\hspace{25mm}+X(X-\varepsilon_0+\omega)(X+2\omega)Y^2.
\end{align} 
Now we observe that repeated arguments in $[\varepsilon_0+\omega,\varepsilon_0-\omega,2\omega,-2\omega,0]$ occur if and only if $\varepsilon_0=\omega$ or $3\omega$. These two kernel resonances are responsible for generating infinitely many secondary ``replica'' ones through the $\star$-powers of the kernel. Given that the kernel dependency in the $s$ time variable is of the form $e^{\pm 2i\omega s}$, each $\star$-power of the kernel shifts the two resonant conditions $\varepsilon_0=\omega,3\omega$ by $\pm2\omega$. This produces resonances for all $\varepsilon_0$ that are odd-integer multiples of $\omega$, $\varepsilon_0=(2n+1)\omega$, $n\in\mathbb{N}$ as first discovered by Winter \cite{Winter1959} and Shirley \cite{Shirley1965}, see Fig.~\ref{fig:BSResonances} and Appendix \ref{AltKBS} for a mathematical proof. Since the resonance at $\varepsilon_0=(2n+1)\omega$ requires $n$ shifts by $2\omega$, it stems from  the $n$th $\star$-power of the kernel  
and every quantity related to the resonance appears in the exact solution in this order with a $g^{2n+1}$ prefactor. For instance, on resonance we get $P_{|0\rangle\to|1\rangle}(t)\simeq \sin^2\!\big(\Omega_{\text{eff}}t\big)$ and the effective frequency of evolution  $\Omega_{\text{eff}}\propto g^{2n+1}$ is the coefficient of the time variable $t$ in the resonant terms. This is because, on resonance, divided-differences with repeated arguments produce polynomials in time, which add up to the long-time resonant behavior of the solution. This immediately leads to
\begin{subequations}
\begin{align}\label{Carrier3Res}
\Omega_{\text{eff}}\big|_{\varepsilon_0=\omega}&=\frac{g}{2}-\frac{g^3}{64 \omega ^2}-\frac{13 g^5}{4096 \omega ^4}-\frac{81 g^7}{131072 \omega ^6}-\frac{1677
   g^9}{16777216 \omega ^8}+\cdots,\\
\Omega_{\text{eff}}\big|_{\varepsilon_0=3\omega}&=\frac{9 g^3}{32
   \varepsilon_0^2}-\frac{81 g^5}{256 \varepsilon_0^4}+\frac{2187 g^7}{8192 \varepsilon_0^6}-\frac{6561 g^9}{32768 \varepsilon_0^8}+\frac{295245 g^{11}}{2097152 \varepsilon_0^{10}}+\cdots,\\
\Omega_{\text{eff}}\big|_{\varepsilon_0=5\omega}&=\frac{625 g^5}{2048 \varepsilon_0^4}-\frac{15625 g^7}{32768 \varepsilon_0^6}+\frac{2734375 g^9}{2097152 \varepsilon_0^8}+\frac{68359375 g^{11}}{16777216 \varepsilon_0^{10}}+\cdots,\\  \Omega_{\text{eff}}\big|_{\varepsilon_0=7\omega}&=\frac{117649 g^7}{294912 \varepsilon_0^6}-\frac{5764801 g^9}{18874368 \varepsilon_0^8}+\frac{21750594173 g^{11}}{10871635968 \varepsilon_0^{10}}+\cdots,\\
\Omega_{\text{eff}}\big|_{\varepsilon_0=9\omega}&=-\frac{4782969 g^9}{8388608 \varepsilon_0^8}+\frac{387420489 g^{11}}{536870912 \varepsilon_0^{10}}+\cdots,\\
\Omega_{\text{eff}}\big|_{\varepsilon_0=11\omega}&=\frac{25937424601 g^{11}}{30198988800 \varepsilon_0^{10}}+\cdots,\\
\vdots\nonumber&
\end{align}
\end{subequations}
This recovers and goes beyond known results, e.g., at the $3\omega$ resonance, \cite[Eqs.~(21)~and~(23)]{Shirley1965} give $q=b^3/4\omega^2=(g/2)^3/\big(4(\varepsilon_0/3)^2\big)=9g^3/32 \varepsilon_0^2$. Alternatively $\Omega_{\text{eff}}$ can also be found from the effective Hamiltonian, see below.

The type of analysis performed here from the exact solution closely resembles Floquet perturbation theory in its outcomes, but this is fortuitous. Firstly, by using the Jacobi-Anger expansion, we obtained a representation of the solution into series of (divided-difference) complex exponential, effectively mimicking Fourier analysis. This is because the models considered here are very simple,  larger systems typically yield series of divided-differences of other functions (e.g., exponential polynomials) which do not parallel Fourier analysis. Second, because of the frame change the kernel carries a $g$ prefactor and so the $\star$-Neumann series \textit{looks} like ca perturbative expansion in $g$. Although this eases the analysis of the solution by further accelerating its convergence for small $g$, the series solution we obtain is, in fact, \textit{convergent for any $|g|$}. This may  not be the case for perturbative quantities which are not intrinsic to the true solution but theoretical devices that help interpret it. For this reason, these quantities may not exist, e.g., for $|g|\gg1 $. This indicates that the picture they provide is not always a valid description of Eqs.~(\ref{BSSol}), which provide the unchanged and unconditionally exact solution. In contrast, intrinsic quantities such as repeated arguments in divided-differences are universal mathematical signatures of resonances, throughout all parameter regimes and all models. In general, known perturbative or approximation-based approaches can be derived from the exact solution Eqs.~(\ref{BSSol}). For instance, the monodromy matrix $\mathsf{M}:=\mathsf{U}_{\text{eff}}(\mathrm{T})$ from Floquet theory and related effective Hamiltonian $\mathsf{H}_{\text{eff}}$ both follow from those equations. In particular,  
the latter can be given an exact and explicit formula based on Eqs.~(\ref{BSSol}) by setting $t=\mathrm{T}$, $\mathrm{T}:=2\pi/\omega$ and letting $\varepsilon_0\rightarrow \omega$, that is,
\(
\mathsf{U}_{\text{eff}}:=e^{-i\mathsf{H}_{\text{eff}}\mathrm{T}}=\lim_{\varepsilon_0\rightarrow \omega}\mathsf{U}_{\text{rot}}(\mathrm{T})
\). The exact solution gives $\mathsf{U}_{\text{eff}}(\mathrm{T})$ (see also Appendix~\ref{FMFromStar} for a proof of the Floquet-Magnus expansion in the $\star$-algebra),
and so the effective Hamiltonian becomes explicitly known (proof in Appendix \ref{Monodromy}),
\begin{align}\label{ExactHeff}
    \mathsf{H}_{\text{eff}}&=-\left(\frac{1}{i\mathrm{T}}\sum_{k\geq 1}\frac{g^{2k}}{4^{k}\omega^{2k}}\sum_{\pmb{\alpha}_k,\pmb{\beta}_k}e^{i\omega \mathrm{T}[a_1, 1+b_1,a_2,1+b_2,\dots,a_k,1+b_k,0]}\right)\!\Bigg|_{\mathrm{T}=0}\sigma_z\\
    &-\left(\frac{1}{i\mathrm{T}}\sum_{k\geq 0}\frac{g^{2k+1}}{2^{2k+1}\omega^{2k+1}}\!\!\!\sum_{\alpha,\pmb{\alpha}_k,\pmb{\beta}_k} e^{-i\omega \mathrm{T}[a_1,1+b_1,
a_2,1+b_2,\dots,a_k,1+b_k,1+\alpha,0]}\right)\!\Bigg|_{\mathrm{T}=0}\sigma_x,
\nonumber\end{align} 
with $a_i:=n_i-m_i$, $b_i:=n_i-m_{i-1}$ with $m_j=\sum_{i=1}^j\alpha_i$ and
$n_j=\sum_{i=1}^j\beta_i$, and we assumed $g\in\mathbb{R}$ to alleviate an already cumbersome result. A general expression of the effective Hamiltonian in terms of matrix-valued interated integrals is known to exist since the 1960s \cite{Bialynicki1969}; by contrast Eq.~(\ref{ExactHeff}) is explicit, completely integral-free and only involves scalar functions.
By Eq.~(\ref{Symcc}), the effective Hamiltonian is guaranteed to be Hermitian, see Appendix~\ref{Monodromy}. 
The formula recovers and go beyond the latest Floquet results \cite{Dey2025,Zeuch2018}: 
\begin{align}
\mathsf{H}_{\text{eff}}&=\left(\frac{g}{2}-\frac{g^3}{32 \omega ^2}-\frac{3 g^5}{256
   \omega ^4}-\frac{341 g^7}{98304 \omega ^6}-\frac{21745 g^9}{28311552 \omega ^8}+\cdots\right)\sigma_x\\
   &-\left(\frac{g^2}{8 \omega }+\frac{g^4}{32 \omega
   ^3}+\frac{61 g^6}{12288 \omega ^5}+\frac{937
   g^8}{7077888 \omega ^7}-\frac{5033593 g^{10}}{20384317440 \omega ^9}+\cdots\right)\sigma_z,\nonumber
\end{align}
with $\sigma_{x,y,z}$ the Pauli matrices and  $g\in\mathbb{R}$. The standard Rabi (RWA) and Bloch-Siegert Hamiltonians correspond to the first and second orders, respectively. The explicit formula of Eq.~(\ref{ExactHeff}) avoids the proliferation of nested commutators typical of high-frequency expansions in the Floquet-Magnus and van Vleck methods as well as in some forms of Brillouin-Wigner perturbation theory \cite{Mikami2016}. A formula for $\mathsf{H}_{\text{eff}}$ is also available for arbitrary $\varepsilon_0\neq \omega$ on setting $t=\mathrm{T}$ in Eqs.~(\ref{BSSol}) then following the steps of Appendix \ref{Monodromy}. This notably facilitates the investigation of the system dynamics near any chosen resonance, by endowing the full solution with a simpler, effective representation. Note also that the present approach lies beyond the usual perturbation theory and preserves structural features of the evolution in another remarkable aspect. Indeed, by computing $\mathsf{H}_{\text{eff}}$ within the present framework we preserve unitary evolution irrespective of the order of $\omega^{-1}$ as described in \cite{Zeuch2018}. More generally, this strategy will yield exact and explicit formulas for the effective Hamiltonian associated to any time-dependent periodic Hamiltonians. Finally, effective Hamiltonians themselves are a convenient tool for interpreting the system's behavior, but not an essential one anymore since the true evolution operator is always exactly and explicitly knowable using the present method.

Beyond their interpretive value, these effective Hamiltonians also carry
an operational meaning. In the frame rotating with the drive,
$\mathsf{H}_{\text{eff}}$ generates the single-qubit gate produced over one
drive period. Writing its two coefficients as $c_x(g,\omega)$ and
$c_z(g,\omega)$, we have
$\mathsf{H}_{\text{eff}}=c_x\sigma_x+c_z\sigma_z
=\Omega_{\text{eff}}\,\hat{n}\!\cdot\!\vec{\sigma}$, with
$\vec{\sigma}=(\sigma_x,\sigma_y,\sigma_z)$,
$\Omega_{\text{eff}}=\sqrt{c_x^2+c_z^2}$ and
$\hat{n}=(c_x,0,c_z)/\Omega_{\text{eff}}$. The propagator over one period
is then the $\mathrm{SU}(2)$ rotation
\begin{equation}\label{HeffGate}
\mathsf{U}_{\text{eff}}(\mathrm{T})=e^{-i\mathsf{H}_{\text{eff}}\mathrm{T}}
=\cos\!\big(\Omega_{\text{eff}}\mathrm{T}\big)\,\mathsf{Id}
-i\sin\!\big(\Omega_{\text{eff}}\mathrm{T}\big)\,\hat{n}\!\cdot\!\vec{\sigma},
\end{equation}
whose axis is $\hat{n}$ and whose rotation angle on the Bloch sphere is
$\Phi=2\,\Omega_{\text{eff}}\mathrm{T}$. This is an exact statement, not an
approximate one. Here $\Omega_{\text{eff}}=\sqrt{c_x^2+c_z^2}$ is the
magnitude of the quasi-energy, and as such it coincides with the resonant
carrier $\Omega_{\text{eff}}|_{\varepsilon_0=\omega}$ found above; the two
expansions agree term by term, up to the order retained here. The same
$\Omega_{\text{eff}}$ thus governs both the gate and the long-time
population $P_{|0\rangle\to|1\rangle}(t)= \frac{c_x^2}{c_x^2 +c_z ^2}\sin^2(\Omega_{\text{eff}}t)$,
and what the gate picture adds is the rotation axis. The orientation of
this axis is itself a beyond-rotating-wave effect. The rotating-wave
generator $\mathsf{H}_{\text{RWA}}=(g/2)\sigma_x$ is a rotation about
$\hat{x}$, and it is the generated $\sigma_z$ term, the Bloch--Siegert
coefficient
$c_z=-(g^2/8\omega+\cdots)$~\cite{Bloch1940,Yan2015}, that
tilts $\hat{n}$ off the equator. The departure from the rotating-wave gate
is therefore a coherent error, unitary and not dissipative. It is
generated by the Hermitian difference
$\mathsf{H}_{\text{eff}}-\mathsf{H}_{\text{RWA}}$, of norm
$\varepsilon(g,\omega)=\lVert\mathsf{H}_{\text{eff}}-\mathsf{H}_{\text{RWA}}\rVert
=g^2/8\omega+\mathcal{O}(g^4/\omega^3)$. This quantity is known in closed
form; it fixes the size of the error per period, and hence the scale of the
gate infidelity. Being unitary, the error is absorbed by recalibration and
does not degrade coherence. The relation can also be read backwards: since
$c_x$ and $c_z$ are explicit functions of $(g,\omega)$, solving
$\mathsf{H}_{\text{eff}}(g,\omega)=c_x\sigma_x+c_z\sigma_z$ returns the
drive parameters that realize a target rotation in the $x$--$z$ plane,
exactly and beyond the rotating-wave approximation; the off-resonance
formula ($\varepsilon_0\neq\omega$) quoted above enlarges the set of
reachable targets. 

Such a regime is met in superconducting qubits, where the drive can be
made a sizable fraction of the level spacing. There the Floquet
quasi-energies of the strongly driven two-level system are read out by
cavity absorption, and corrections reaching well beyond the lowest-order
Bloch--Siegert shift are required to reproduce them~\cite{Tuorila2010};
this is the all-orders, beyond-rotating-wave situation considered here. All
of the above concerns the stroboscopic map at the period $\mathrm{T}$ and
its quasi-energies $\pm\Omega_{\text{eff}}$. The single-generator picture
holds only as long as these remain inside the first Floquet--Brillouin
zone, $\Omega_{\text{eff}}<\omega/2$. At the boundary the two quasi-energies
meet, their gap reaching one drive quantum, and the principal branch of the
logarithm that defines $\mathsf{H}_{\text{eff}}$ becomes singular; this
degeneracy is of the same nature as the conical points reported in such
driven-qubit spectra.

When now turning the Hamiltonian into a self-commuting one or, equivalently, assuming parameters values so that all self-commutators of the Hamiltonian are zero (equivalent to a 0th order Magnus expansion), we can reach another approximation going over the standard I.I. Rabi (RWA) population approximation of the $S_0$ excited state where now the $P_{|0\rangle\to|1\rangle}(t)$ prefactor becomes time-dependent (See Appendix~\ref{ChapterToulouse}):
\begin{align}
&P_{|0\rangle\to|1\rangle}(t)
=
\frac{4|g|^2 \cos^2(\omega t)}
     {(\Delta - \hbar \omega)^2 + 4|g|^2 \cos^2(\omega t)}\\
&\hspace{35mm}\sin^2\!\left(
\int_0^t
\frac{\sqrt{(\Delta-\hbar \omega)^2 + 4|g|^2 \cos^2(\omega \tau)}}
     {2\hbar}
\,d\tau
\right),\nonumber
\end{align}
which is also 
\begin{align}
    P_{|0\rangle\to|1\rangle}(t)=
\frac{k^2\cos^2\omega t}{1-k^2\sin^2\omega t}\,
\sin^2\!\left(\frac{\Omega_0}{\omega}\,E(\omega t,k^2)\right),
\end{align}
where $\delta=\Delta-\hbar\omega$, $ \Omega_R=\sqrt{\delta^2+4|g|^2}$, $
\Omega_0=\frac{\Omega_R}{2\hbar}$, $k=\frac{2|g|}{\Omega_R}$ and $E(\omega t,k^2)$ is the incomplete elliptic integral of the second kind.

\subsubsection{General case \texorpdfstring{$e_0\neq0$}{}.}
\begin{figure}[t!]
 \centering
 \includegraphics[width = \textwidth]{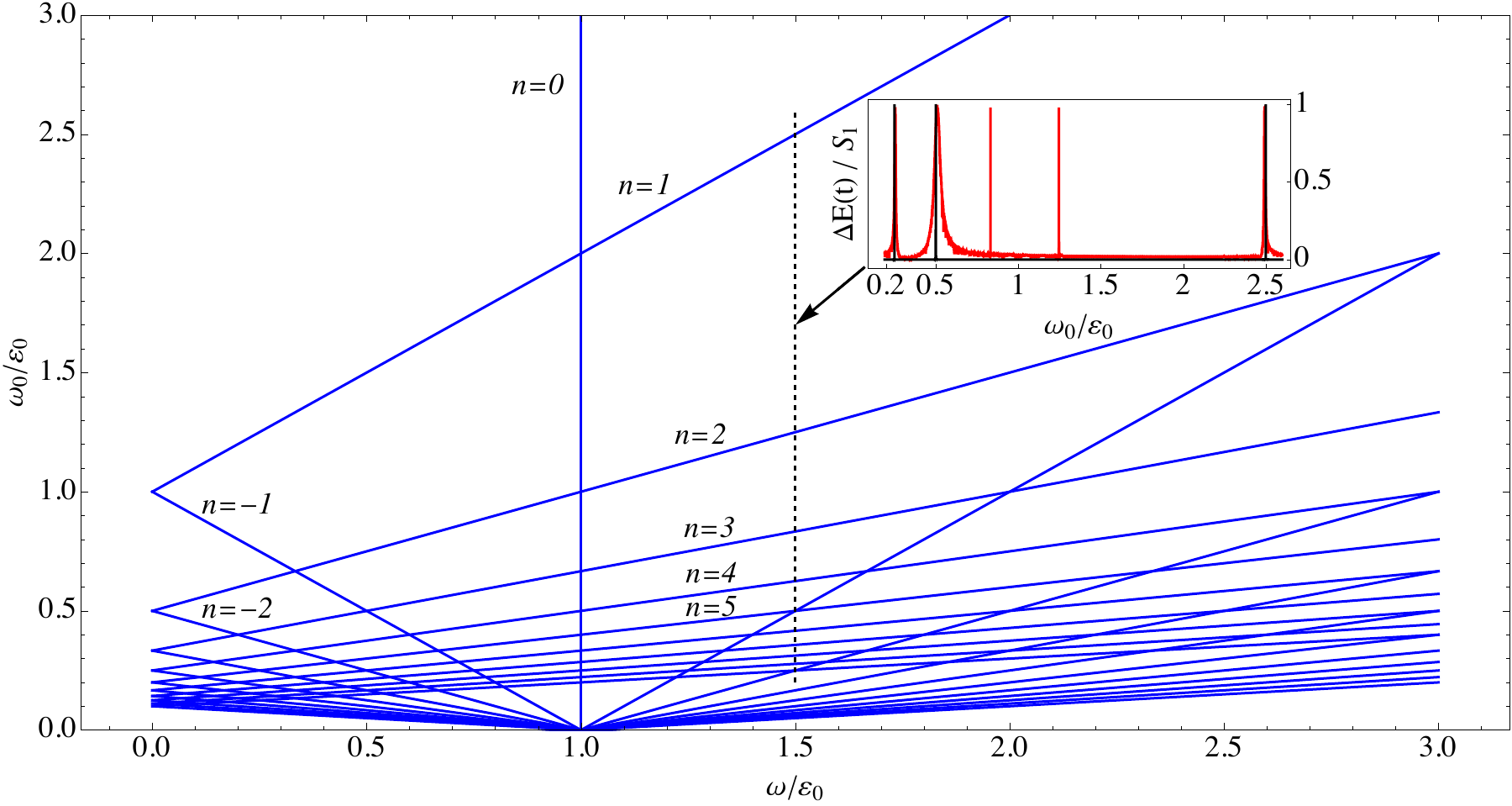} 
 \caption{Location of the resonances (solid blue lines) in the $(\omega,\omega_0)$ parameter space. As indicated by Eq.~(\ref{ResLoc}) the resonances are indexed by integer $n\in\mathbb{Z}$, here showing only $-10\leq n\leq 10$ for the sake of clarity.
 In \textbf{inset}, numerical simulations of $\max_t \Delta E(t)/S_1$ at fixed $\omega/\varepsilon_0=3/2$ as a function of $\omega_0$ for $g=|S_1-S_0|/10$ (red line) and $g=|S_1-S_0|/1000$ (black line). Note that resonances with $n=4,\,6,\,7$ and $8$ are not visible owing to numerical issues (too narrow, required simulation times too long) and, as $g$ diminishes, each resonance becomes narrower. Resonances at $\omega_0/\varepsilon_0\simeq 0.5$ and $0.25$ are strongly enhanced double resonances. For $\omega/\varepsilon_0=0.5$, the $n=5$ line resonates with the $+\omega$ frequency component at the same time as the $n=-1$ line resonates with the $-\omega$ frequency component. For $\omega_0/\varepsilon_0\simeq 0.25$, the resonances involved are the $n=10$ line with the $+\omega$ frequency component and the $n=-2$ line with the $-\omega$ frequency component. 
 Parameters: $S_0=\Gamma=0$, $S_1=1$, $\varepsilon_0=S_1-S_0-i\Gamma=1$ and $e_0=|S_1-S_0|/5$.}
  \label{fig:DeltaECase2}
\end{figure}
Just as in the previous case and for the same reasons, the quantity $\max_t \Delta E(t)$ undergoes resonances as both $\omega_0$ and $\omega$ are tuned. 
As indicated by the exact solution in Eq.~(\ref{U12Case2}), the resonances occur in the presence of repeated arguments of the divided-difference exponentials, i.e. when
$\varepsilon_0+n\omega_0+\alpha \omega=0$. Since $\alpha=\pm 1$ and $n$ must be an integer, this implies that 
\begin{equation}\label{ResLoc}
\frac{\varepsilon_0\pm \omega}{\omega_0}=n\in\mathbb{Z}\backslash\{0\},\quad\text{or}\quad n=0 \,\text{ that is }\, \omega=\varepsilon_0,
\end{equation}
while the resonance widths are proportional to $|g|^n$. We show in Fig.~\ref{fig:DeltaECase2} the prediction of Eq.~(\ref{ResLoc}) as a function of both $\omega$ and $\omega_0$ together with the result of purely numerical simulations of $\max_t \Delta E(t)$ confirming some of the theoretical predictions. By the same reasoning as in the previous case we find that, on the $n$th-resonance, the long-time population dynamics is given by
\begin{align}
P_{|0\rangle\to|1\rangle}(t)\simeq \sin^2\!\big(|g/2| \,J_n(e_0/\omega_0) t\big).
\end{align}
Here, the additional factor $1/2$ compared to Eq.~(\ref{PopResCase1}) comes from the fact that only one of the two frequency components of $\cos(\omega t)$ participates in the resonance. Condition (\ref{ResLoc}) allows double resonances to occur for this reason. These are values of $\omega$ and $\omega_0$ for which both $(\varepsilon_0+ \omega)/\omega_0=:n_+$ and $(\varepsilon_0- \omega)/\omega_0=:n_-$ are integers, corresponding to different kernel terms that resonate with each of the frequency components $\pm \omega$ of the $\cos(\omega t)$ driving. For such double resonances, the long-time population dynamics follows
\begin{align}
P_{|0\rangle\to|1\rangle}(t)\simeq \sin^2\!\left(|g/2| \,\sqrt{J_{n_+}^2(e_0/\omega_0)+J_{n_-}^2(e_0/\omega_0)}\, t\right),
\end{align}
where $n_{\pm}$ are the two resonant integers. The resonance width is proportional to $|g|^{\min{(n_-,n_+)}}$. These findings are in agreement with \cite[Ref.~15]{Shirley1965} where it was pointed out that the addition of $\cos(\omega t)$ on the main diagonal is responsible for resonances at even multiples of $\omega$.

Under the assumption that $e_0$ is very small or $\omega_0$ is very large, more rigorously $e_0\ll \omega_0$, the general solution Eqs.~(\ref{Case2GenSol}) is well approximated by the Bloch-Siegert one provided one rescales $g\to gJ_0$. Indeed, as in the previous case we can approximate $J_n\to 0$ for $n\neq 0$. Consequently, the sums over $n$, $\pmb{n}_k$ and $\pmb{m}_k$ in Eqs.~(\ref{Case2GenSol}) reduce to the single term where all $n_i,m_j$ are 0. This yields $J_0^{2k}$ and $J_0^{2k+1}$ prefactors in $U_{11}$ and $U_{12}$, respectively, corresponding to a simple rescaling $g\to gJ_0$ of the Bloch-Siegert solution Eqs.~(\ref{BSSol}).

\section{\Large Third case\texorpdfstring{ $\epsilon(t)$ random and $f(t)=1$ or $f(t) = \cos(\omega t)$}{: noise and constant or modulated coupling }}\label{ThirdCase}

Next, instead of the small $e_0\cos(\omega_0t)$  perturbation of the $S_0$ ground state, we consider the effect of a white noise perturbing this $S_0$ ground state. It corresponds to a two-level quantum system coupled to a supporting surface with a thermal noise perturbing this $S_0$ ground state. For this case, Eq.~(\ref{HamForm}) is modified such that $\epsilon (t) = e_0 \sqrt{\alpha k_B T} \,\eta(t)$ is the energy of the thermal noise where $e_0$ is a dimensionless parameter to be able to change the amplitude of the thermal noise fluctuations, $\alpha$ is the friction of the quantum system to the surface to ensure thermalization, $k_B$ is Boltzmann's constant, $T$ is the temperature, and $\eta(t)$ is a normalized Gaussian white noise process representing thermal fluctuations. In the following, $g$ is the coupling between the $S_0$ and the $S_1$ states modulated or not by a $f(t)$ contribution. Here again the problem is to calculate how the $S_1$ state population can benefit from the surface thermal source first with a constant $f(t)$ and then with the contribution (resonant or not) of a given $f(t)$. Mathematically, the Schr\"odinger-Langevin equation with noisy perturbation corresponds to the stochastic Schr\"odinger equation
\begin{equation}
d\mathsf{U}(t)=-i\mathsf{A}(t)\mathsf{U}(t)dt - i\mathsf{B}(t)\mathsf{U}(t) \circ dW_t,
\end{equation}
where $W_t$ is the Brownian motion process with its formal derivative, $dW_t/dt$, corresponding to white noise with $\mathbb{E}[W_tW_s]=\delta(t-s)$ and $\mathbb{E}[W_t]=0$. This stochastic equation is of the Stratonovich type. If we were to continue with the above, we first require It\^o calculus to introduce a drift term as has been performed previously \cite{barchielli2010stochastic,semina2014stochastic}, where both the above and the drift corrected It\^o stochastic equation have formal solution
\begin{equation}
\mathsf{U}(t) = \mathscr{T}e^{-i\int_{0}^t \mathsf{A}(s)ds - i\int_{0}^t \mathsf{B}(s) dW_s } \mathsf{U}(0),
\end{equation}
where $\mathscr{T}$ is the time-ordering operator. To solve the stochastic equation, we use the Wong-Zakai theorem (see, e.g., \cite[Theorem~2.1]{twardowska1996wong} and references therein)  with a particular regular approximation of the Wiener process given by the Karhunen-Lo\`eve theorem. This allows us to rewrite the stochastic equation as a random ordinary differential equation using a finite truncation of the Karhunen-Lo\`eve expansion for the Brownian motion, similar to \cite{williams2006polynomial,hodgkinson2020stochastic}. Once the solution of this is reached, extending the truncation rank to infinity produces a solution of the original stochastic equation.  

The Karhunen-Lo\`eve theorem represents a stochastic process as a linear combination of orthogonal functions within a finite time interval given the covariance function $K(t,s)$ of the stochastic process. For a Wiener process representing Brownian motion, the covariance function, $K(t,s) = \text{min}(s,t)$, results in an expansion in trigonometric functions. Explicitly, the expansion in a finite time interval $[0,\mathcal{T}]$, is
\begin{equation}
W_t =  \sum_{k=0}^\infty Z_k \frac{2\sqrt{2\mathcal{T}}}{(2k+1)\pi}\sin\left(\frac{(2k+1)\pi t}{2\mathcal{T}}\right),
\end{equation}
where $Z_i$ is a sequence of independent Gaussian random variables with zero mean and variance $1$. Approximating the noisy process by truncating the expansion and substituting in the derivative for white noise as $\eta(t) = dW_t/dt$, we find the ``noisy'' Hamiltonian   

\begin{align}
\mathsf{H}^{(K)}(t) = 
\begin{pmatrix}
S_0 + e_0 \sqrt{\alpha k_B T} \,
(d/dt)W_t^{(K)} & gf(t)  \\
\bar{g}\bar{f}(t) & S_1
\end{pmatrix},
\end{align} 
where the coefficient $\gamma = e_0 \sqrt{\alpha k_B T}$ is the amplitude of the white noise. It can be tuned either via the surface friction $\alpha$, by the parameter $e_0$, or through variations of the surface temperature $T$. Substituting into the Schr\"odinger equation results in a non-autonomous random differential equation. As before, we move to the interaction frame, except here we find a complex exponential term with argument given by the Brownian motion. We decompose the complex exponential of Brownian motion into a complex Fourier basis with coefficients given by the Generalized Bessel Functions (GBF) \cite{dattoli1996theory}, see also Appendix~\ref{JacobiAnger}. Then we have: 
\begin{subequations}\label{H12H21Case3}
\begin{align}
 &H_{12}(t) =  g f(t)\sum_n \ \mathcal{J}_n e^{i(S_0-S_1 + n \omega_{\eta}) t},\\
&H_{21}(t)=\bar{g} f(t) \sum_n \mathcal{J}_ne^{-i(S_0-S_1+ n \omega_{\eta}) t}.
\end{align}
\end{subequations}
Here, the GBF, $\mathcal{J}_n$, has arguments given by the sequence $\left\lbrace Z_k \frac{2\gamma\sqrt{2\mathcal{T}}}{(2k+1)\pi}\right\rbrace$ and the fundamental frequency $\omega_{\eta} = \pi/(2\mathcal{T})$. 

\subsection{Constant \texorpdfstring{$f(t) = 1$:}{coupling:} the exact solution}
We consider $f(t) = 1$, reducing the Hamiltonian to 
\begin{align}
\mathsf{H}^{(K)}(t) = 
\begin{pmatrix}
S_0 + \gamma
(d/dt)W_t^{(K)} & g \\
\bar{g} & S_1
\end{pmatrix}.
\end{align}
After moving to the interaction frame and using Eq.~(\ref{H12H21Case3}), the path-sum kernel is
\begin{equation}
K(t,s)\,\Theta:=
i|g|^2  \sum_{n,m}\mathcal{J}_n
\mathcal{J}_me^{i(n -m)\omega_0t}e^{i[\varepsilon_0+m\omega_0,0](t-s)}~\Theta.
\end{equation}
Finally, Let $\pmb{m}_k:=(m_1,\ldots,m_k)\in\mathbb{N}^k$ and similarly for $\pmb{n}_k$. Defining
\begin{equation}
\mathcal{J}_{\pmb{n}_k,\pmb{m}_k}:=\prod_{i=1}^k \mathcal{J}_{n_i}\, \mathcal{J}_{m_i}
\quad {\rm and} \quad \mathcal{J}_{n,\pmb{n}_k,\pmb{m}_k}:=\mathcal{J}_n \mathcal{J}_{\pmb{n}_k,\pmb{m}_k},
\end{equation}
\begin{figure}[t!]
 \centering
 \includegraphics[scale=0.35]{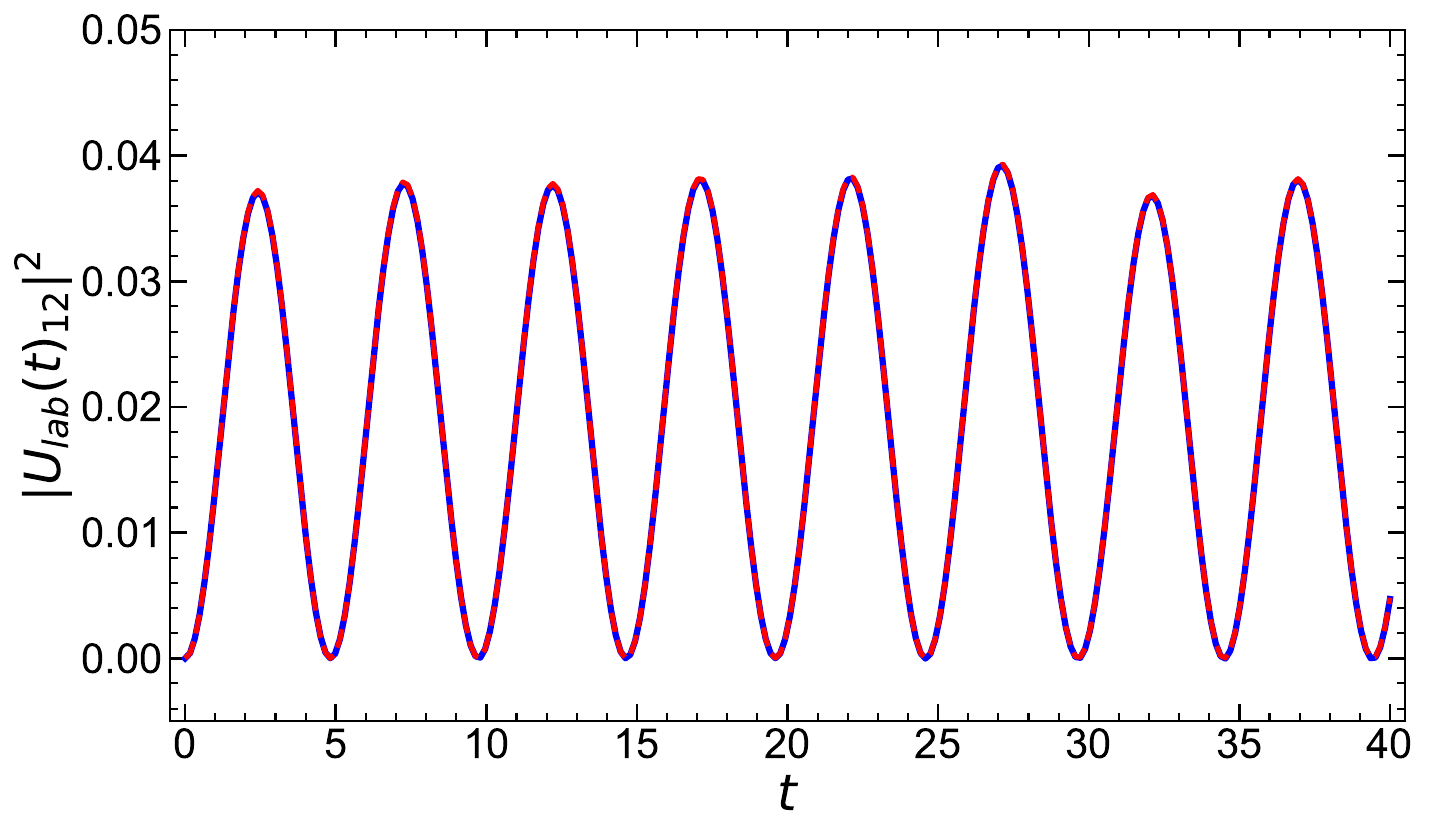} 
 \hspace{5mm}
 \includegraphics[scale=0.35]{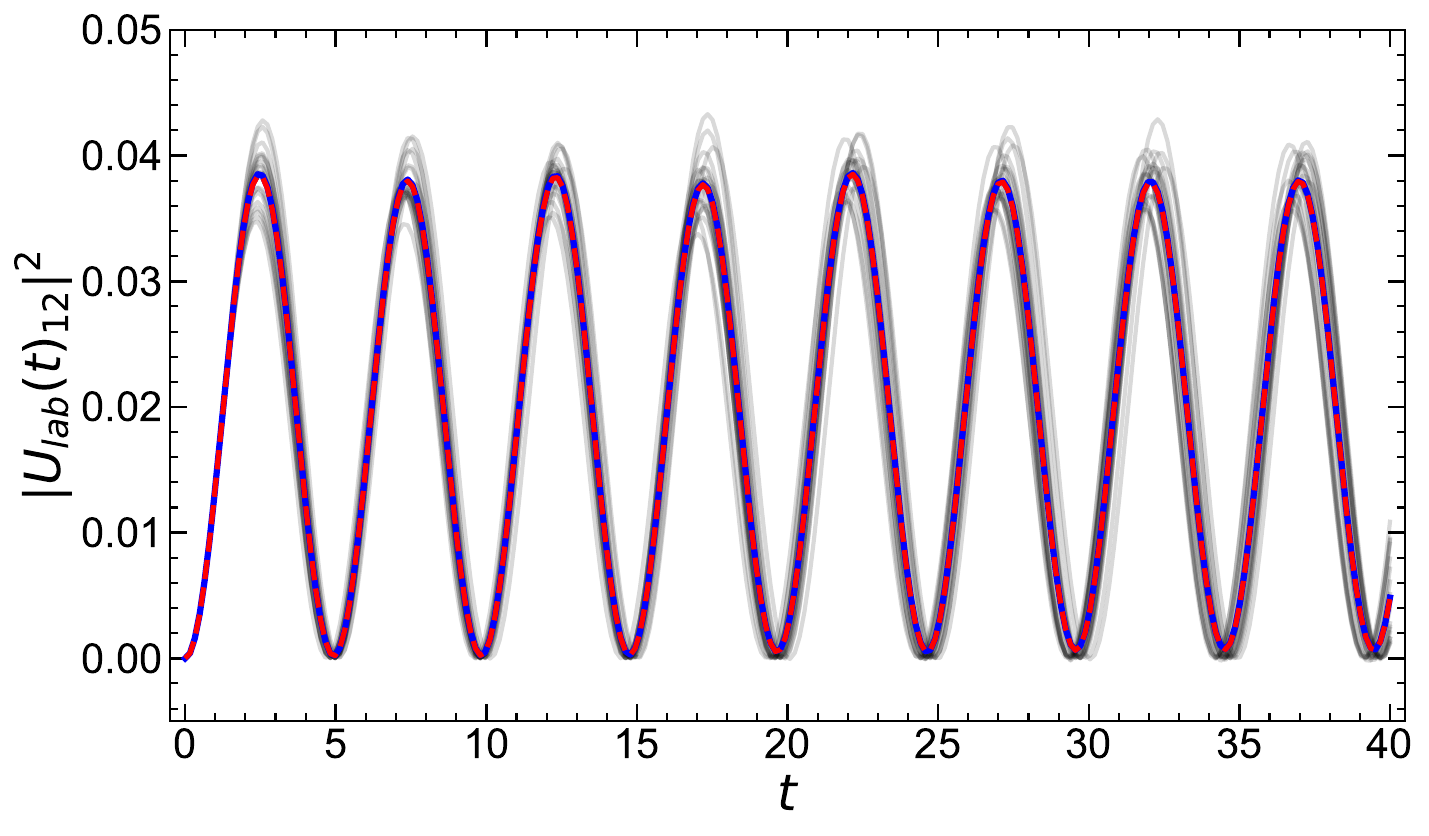} 
 \caption{Evolution of the stochastic transition probability $P_{|0\rangle\to |1\rangle}(t):=|U_\text{lab}(t)_{12}|^2=1-|U_\text{lab}(t)_{11}|^2$ as a function of time as determined by a fully numerical solver (solid blue line) and analytical truncation $|U_{12}^{(3,6)}|^2$  (\textbf{top} figure, dashed red line) for one realization of noisy variables. Stochastic transition probability of $P_{|0\rangle\to |1\rangle}(t):=|U_\text{lab}(t)_{12}|^2=1-|U_\text{lab}(t)_{11}|^2$(\textbf{bottom} figure) over 10 realizations (red solid lines), the analytical mean (blue line), and the mean of the numerical solver (black line) of the exact solution. Parameters for both figures:  $|S_1-S_0|=1$, $\mathcal{T} = 40$, $\gamma = 0.1$, $K=10$, $g=0.1$, $\Gamma=0$.  In the above situations, the frequency of the driving depends on the length of the simulation $\omega_\eta = 2\pi/\mathcal{T}$, leading to low driving frequencies, a case that is beyond approximating regimes.  The resulting oscillations remain low amplitude due to the amplitudes of the resonance requiring time greater than $\mathcal{T}$ to cause a transition.  For lower values of $g$, the amplitudes of the oscillations would be lower.  We note that the single realization appears as a smooth trajectory, despite the stochastic forcing that would typically produce a highly irregular evolution. This behavior arises from the use of a truncated Karhunen–Lo\`eve expansion, in which the stochastic process is represented by a finite sum of smooth basis functions. Consequently, the resulting noise realization is itself smooth, and the corresponding solution inherits this regularity. In contrast, a simulation based on the Euler–Maruyama method, which directly approximates the Brownian increments, would exhibit a noticeably more jagged trajectory that more clearly reflects the rough path properties of the underlying stochastic process.}
 \label{fig:Case3}
\end{figure}
we have:
\begin{equation}
\mathsf{U}_{\text{lab}}(t)=\begin{pmatrix}
e^{-iS_0t -i \gamma \sum_{k=0}^K Z_k \frac{2\sqrt{2\mathcal{T}}}{(2k+1)\pi}\sin\left(\frac{(2k+1)\pi t}{2T}\right)}  & 0\\ 
0 & e^{-iS_1t}  
\end{pmatrix}\begin{pmatrix}
U_{11}(t) & U_{12}(t)\\ 
U_{21}(t) & U_{22}(t)  
\end{pmatrix},
\end{equation}
where $U_{ij}$ are given by Eqs.~(\ref{PSUequations}), 
\begin{subequations}
\begin{align}
U_{11}(t) &= 
\sum_{k\geq 0} |g|^{2k}\sum_{\pmb{m}_k,\pmb{n}_k} \mathcal{J}_{\pmb{n}_k,\pmb{m}_k}\,\, 
e^{i[A_1,\varepsilon_0+B_1,A_2,\varepsilon_0+B_2,\ldots,A_k,\varepsilon_0+B_k,0]t},\\
U_{12}(t)&= 
 \sum_{k\geq 0}g|g|^{2k}\sum_{n,\pmb{m}_k,\pmb{n}_k}
e^{i(\varepsilon_0+n\omega_\eta)t}\mathcal{J}_{n,\pmb{n}_k,\pmb{m}_k}\nonumber\\
&\hspace{15mm} \times e^{-i[A_1,\varepsilon_0+B_1,A_2,\varepsilon_0+B_2,\ldots,A_k,\varepsilon_0+B_k,\varepsilon_0+n\omega_\eta,0]t},
\end{align}
\end{subequations}
with $\varepsilon_0 = S_0 - S_1$, $A_i:=N_i-M_i$, and  $B_i:=N_i-M_{i-1}$ such that $M_j = \Sigma_{i=1}^{j} m_i \omega_{\eta}$ and $N_j = \Sigma_{i=1}^{j} n_i \omega_{\eta}$. Finally, by the Wong-Zakai theorem, the above result almost surely converges to the solution of the Stratonovich stochastic differential equation as $K\rightarrow \infty$.

\subsubsection{Physical Interpretation}

 Similarly to Case 1, the term of order $k$ in $U_{11}$ involves a product of $2k$ Generalized Bessel functions and divided-difference exponentials with $2k+1$ arguments. For $U_{12}$, the term of order $k$ comprises a product of $2k+1$ Generalized Bessel Functions (GBF) and divided-difference exponentials with $2k+2$ arguments. The series given here are  \emph{unconditionally convergent}, a property they inherit from the $\star$-Neumann series.  In the case $e_0=0$--for which the Hamiltonian becomes time-independent--the solution above correctly reduces to the expected results, similar to Case 1. With the above expressions, we are able to flawlessly reproduce the numerical solution for $P_{|0\rangle\to|1\rangle}(t)$ for one specific realization of noisy variables, as well as the mean averaged over many realizations, see Fig.~\ref{fig:Case3}.  

Mathematically, this case is very similar to Case 1. In particular, due to the Jacobi-Anger relation and the GBF, the analytical signature of anomalous resonances is the same: these arise from the exact solution for any choice of parameters such that they are equalities in the arguments of the divided-difference exponentials. However, the physics of the noisy Hamiltonian discussed here is profoundly altered by the fact that the driving frequency from the Karhunen-Lo\`eve expansion is extremely low. This is because the fundamental frequency, $\omega_\eta = \pi/(2T)$, is inversely proportional to the overall time evolution of the system. But in the dynamics, anomalous resonances manifest themselves as low-frequency oscillations of $P_{|0\rangle\to|1\rangle}(t)$ to reach $1$ over very long times, hence must here correspond to very low $\omega_\eta$.
In Case 1, the resonances obeyed the criterion $\varepsilon_0 + n \omega_0 = 0$ and here they follow a similar one, namely $\varepsilon_0 + n \omega_\eta = 0$. For low frequency driving as is the case here, integer $n$ needs to be extremely large to satisfy the above equation unless $\varepsilon_0$ is itself vanishingly small. If $N\gg 1$ satisfies $\varepsilon_0 + N \omega_\eta = 0$ the present Hamiltonian drives a long time resonance with effective frequency $|g \mathcal{J}_N|$, in principle allowing for a complete population transfer with  $P_{|0\rangle\to|1\rangle}(t)$ reaching 1.  In the physical situation of interest here, the two-level quantum system is only weakly coupled to the thermal fluctuations of the surface: the noise amplitude is thus very small ($\gamma\ll 1$) so that, effectively, $|\mathcal{J}_n|\ll 1$ whenever $n \ne 0$.  This indicates that for small noise and resonant integer $N$ the effective frequency of transition from state $|0\rangle$ to $|1\rangle$ on resonance is $|g \mathcal{J}_N|\ll 1$, i.e. almost zero. This suggests that while the noisy Hamiltonian considered here can in principle drive anomalous resonances, in practice these will not yield sizable population transfers within the time frame of evolution $\mathcal{T}$. 

Approximation strategies may be derived from the exact expression, as in Case 1. Due to the low noise amplitude and with the majority of the spectral content of the Brownian motion concentrated in $\mathcal{J}_0$, a natural approximation strategy consists in retaining only those terms whose coefficients solely involve powers of $\mathcal{J}_0$ in the exact solution. This yields
$
U_{11}
\simeq \Theta \star (1_\star - A_{0,0}\Theta)^{\star-1}=:U^{(\infty,0)}_{11}.
$
Using Omega calculus, this is immediately found to be (see Appendix~\ref{Approx}): 
\begin{subequations}
\begin{align}
U^{(\infty,0)}_{11}&=1+\mathcal{J}_0^2|g|^2e^{i[r_{0+},r_{0-},0]t}
= e^{\frac{i \varepsilon_0 t}{2}} \left(\cos( \Omega t/2 )-\frac{i \varepsilon_0}{\Omega} \sin (\Omega t/2)\right),\\
U^{(\infty,0)}_{12}&=-\frac{2i\mathcal{J}_0g}{\Omega}e^{\frac{i \varepsilon_0 t}{2}}\sin(\Omega t/2),
\end{align}\label{eq:Case3_approx}
\end{subequations}
where $r_{0\pm}:=\frac{1}{2}\Big(\varepsilon_0\pm \sqrt{\varepsilon_0^2+4\mathcal{J}_0^2|g|^2}\Big)$ and $\Omega:=r_{0+}-r_{0-}=\sqrt{4 |g|^2 \mathcal{J}_0^2+\varepsilon_0^2}$. The difference with Case 1 here is that the derived amplitude and frequencies are noise-dependent and now vary slightly over different realizations of the noisy parameters, $Z_i$. Even for moderate increases of the noise, more GBF terms must be considered, invalidating the above approximation and requiring novel re-summation strategies or a full retreat to the exact solution.

\subsection{Noisy Bloch-Siegert Hamiltonian \texorpdfstring{$f(t) = \cos(\omega t)$:}{:} exact  solution}
By the virtue of the out-of-diagonal time-dependent term in Eqs.~(\ref{eq:Case3_approx}) and to appreciate how part of the noise can be transferred to the $S_1$ excited state, we now turn to the Hamiltonian:
\begin{align}
\mathsf{H}^{(K)}(t) = 
\begin{pmatrix}
S_0 + \gamma
(d/dt)W_t^{(K)} & g\cos(\omega t)  \\
\bar{g}\cos(\omega t) & S_1
\end{pmatrix}.
\end{align}
Using again the rotating reference frame and also the generalized Jacobi-Anger relation, we obtain the kernel:
\begin{align}
K(t,s)\Theta&
:=i|g/2|^2 \sum_{n,m\in\mathbb{Z}\atop\beta,\alpha= \pm1} \mathcal{J}_n
\mathcal{J}_me^{i\{(n -m)\omega_\eta+(\beta-\alpha)\omega\}t}e^{i[\varepsilon_0+m\omega_\eta+\alpha\omega,0](t-s)}\,\Theta.
\end{align}
Let $\pmb{m}_k:=(m_1,\ldots,m_k)\in\mathbb{N}^k$ and similarly for $\pmb{n}_k$. 
The evolution operator in the laboratory frame is now given by:
\begin{equation}
\mathsf{U}_{\text{lab}}(t)=\begin{pmatrix}
e^{-iS_0t -i \gamma \sum_{k=0}^K Z_k \frac{2\sqrt{2\mathcal{T}}}{(2k+1)\pi}\sin\left(\frac{(2k+1)\pi t}{2T}\right)}  & 0\\ 
0 & e^{-iS_1t}  
\end{pmatrix}\begin{pmatrix}
U_{11}(t) & U_{12}(t)\\ 
U_{21}(t) & U_{22}(t)  
\end{pmatrix}.
\end{equation}
In this situation as in the previous one, Eq.~(\ref{Symcc}) holds and, therefore, we need only to determine $U_{11}$ and $U_{12}$ exactly.

Let $\pmb{m}_k:=(m_1,\ldots,m_k)\in\mathbb{N}^k$, similarly for $\pmb{n}_k$ and $\pmb{\alpha}_k:=(\alpha_1,\ldots,\alpha_k)\in\{-1,1\}^k$, similarly for $\pmb{\beta}_k$. We find: 
\begin{subequations}\label{SolutionCase3}
\begin{align}
U_{11}(t)&=\sum_{k\geq 0}\Big|\frac{g}{2}\Big|^{2k}\sum_{\pmb{m}_k,\pmb{n}_k\atop\pmb{\alpha}_k,\pmb{\beta}_k}\!\!\!\mathcal{J}_{\pmb{n}_k,\pmb{m}_k}\,e^{i[A_1,\varepsilon_0+B_1,A_2,\varepsilon_0+B_2,\dots,A_k,\varepsilon_0 +B_k,0]t},\label{U11Case3}\\
U_{12}(t)&= 
\sum_{k\geq 0}\frac{g}{2}\Big|\frac{g}{2}\Big|^{2k}\!\!\!\sum_{n,\pmb{m}_k,\pmb{n}_k\atop \alpha,\pmb{\alpha}_k,\pmb{\beta}_k}\!\!\!\mathcal{J}_{n,\pmb{n}_k,\pmb{m}_k}\,e^{i(\varepsilon_0+n\omega_\eta+\alpha \omega)t}\nonumber\\
&\hspace{15mm} \times e^{-i[A_1,\varepsilon_0+B_1,
A_2,\varepsilon_0+B_2,\dots,A_k,\varepsilon_0+B_k,\varepsilon_0+n\omega_\eta+\alpha \omega,0]t}.\label{U12Case3}
\end{align}
\end{subequations}
In these expressions, we have again:
$\varepsilon_0 = S_0 - S_1$, $A_i:=N_i-M_i$, and  $B_i:=N_i-M_{i-1}$ with $M_j=\sum_{i=1}^j(m_i\omega_\eta+\alpha_i\omega)$ and
$N_j=\sum_{i=1}^j(n_i\omega_\eta+\beta_i\omega)$. 
An example of the time evolution of the transition probability $P_{|0\rangle\to |1\rangle}(t):=|U_\text{lab}(t)_{12}|^2$ is presented in Fig.~\ref{fig:Case3_BS}. Here and at resonance, the noise is clearly perturbing the transition toward the excited state. Finally, by the Wong-Zakai theorem, the above result almost surely converges to the solution of the Stratonovich stochastic differential equation as $K\rightarrow \infty$.

\subsubsection{Physical Interpretations: resonances and the role of the noise}\label{PhysIntNoise}
For zero noise ($\gamma=0$), we recover the Bloch-Siegert Hamiltonian and its unitary evolution, reproducing for small $g$ the Rabi resonance frequency $\Omega_{\text{eff}} = |g/2|$. In other regimes---specifically when analyzing the population transfer from the ground to the excited state and the resonances of the Bloch-Siegert Hamiltonian---we find that many large-amplitude, low-frequency oscillations interact, depending on the coupling strength to the thermal noise.
As indicated by the exact solution, and similarly to Case 2, the resonances occur in the presence of repeated arguments of the divided-difference exponentials and since $\alpha=\pm 1$ and $n$ must be an integer, this implies that: 
\begin{equation}\label{ResLoc_noise}
\frac{\varepsilon_0\pm \omega}{\omega_\eta}=n\in\mathbb{Z}\backslash\{0\},\quad\text{or}\quad n=0 \,\text{ that is }\, \omega=\varepsilon_0.
\end{equation}
However, as mentioned previously and for the low-frequency oscillations associated with $\omega_\eta$, satisfying the resonance condition requires a large integer $N$. This leads to extremely small-amplitude oscillations because $\mathcal{J}_N \ll 1$. Since $\mathcal{J}_N\propto 1/N!$ is extremely small, the time required to observe a population transfer would be much larger than the time parameter $\mathcal{T}$ used for the Karhunen-Lo\`eve expansion.  Even if the noise strength is increased and for a non-negligible $\mathcal{J}_N$, all terms $n<N$ would contribute to low frequency, high amplitude oscillations, causing destructive interferences.

\begin{figure}[t!]
 \centering
\includegraphics[width=.85\textwidth]{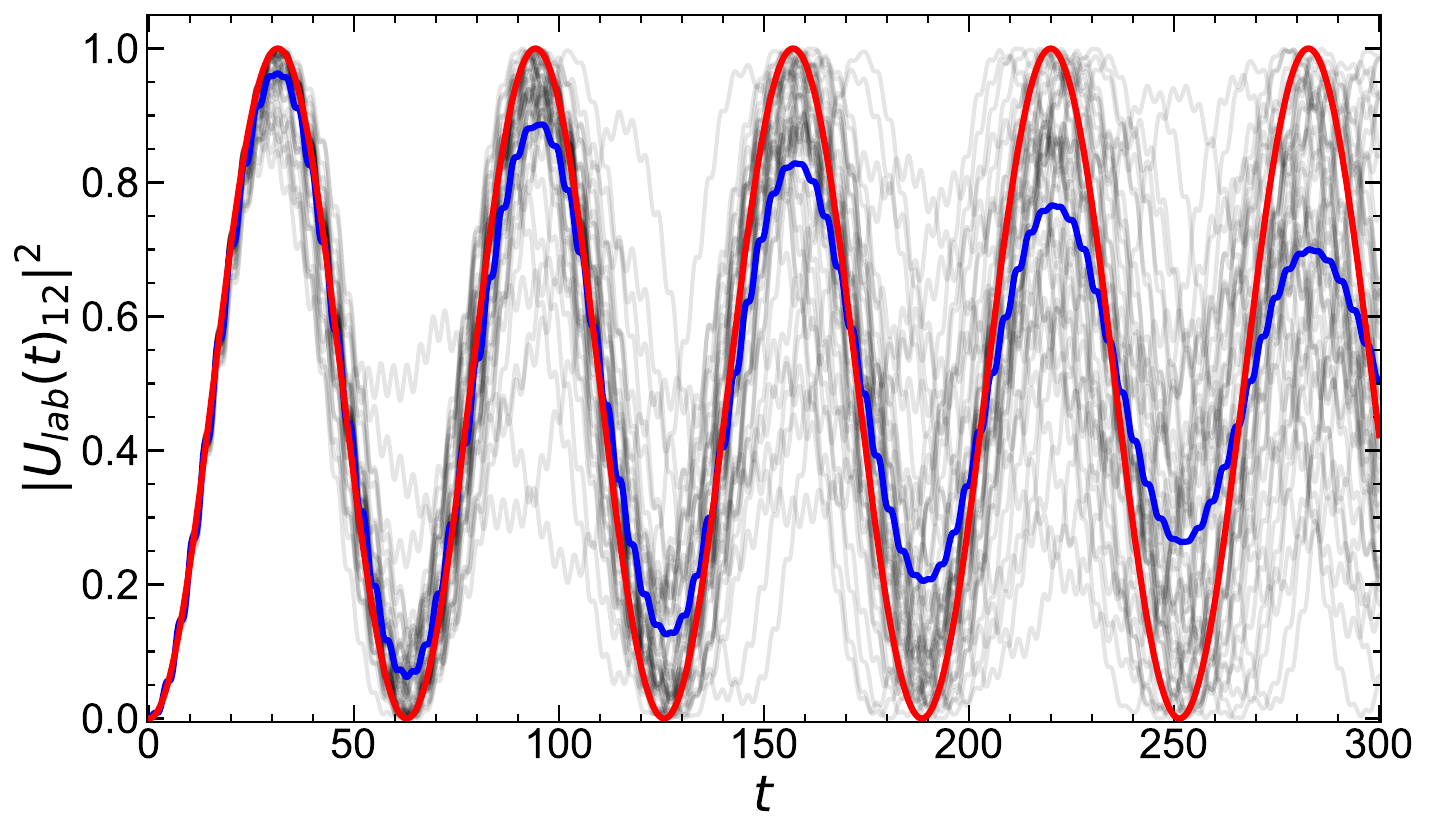}  
 \caption{Evolution of the transition probability $P_{|0\rangle\to |1\rangle}(t):=|U_\text{lab}(t)_{12}|^2=1-|U_\text{lab}(t)_{11}|^2$ of the noisy Bloch-Siegert Hamiltonian as a function of time as determined by a fully numerical solver for 50 realizations (solid gray lines) with the mean (solid blue line) and the analytical evolution with no noise (solid red line). Parameters for the figure: $\omega/|S_1-S_0|=1$, $g=0.1$, $\gamma=0.1$ and $K=50$. In this situation, the cosine driving is on resonance causing oscillations transitioning between the ground and excited state with frequencies dependent upon the noise amplitude or temperature $T$.  With extremely low noise ($<0.01$), the frequency of oscillations are re-normalized by the zeroth order Generalized Bessel function. By increasing the noise, the spectral content of the noise is spread out over many GBFs, generating additional long lived oscillations associated with $|g\mathcal{J}_n/2|$ for $n\ne0$, allowing for interferences to significantly alter the shape in comparison to the Bloch-Siegert evolution with no noise. When averaged over many realizations the overall frequency matches the frequency obtained with noise. However, the amplitude of the oscillations decrease.  While the oscillations are chaotic, the analytical expressions continue to hold.}
 \label{fig:Case3_BS}
\end{figure}

For small noise amplitude, GBFs with non-zero indices have very small amplitude. For $\varepsilon_0=\omega$, terms of Eq.~(\ref{U12Case3}) with $n=0$ and index $\alpha = -1$ are resonant and the behavior of the exact solution may be approximated from terms involving $\mathcal{J}_0$. Doing so yields a modified $U_{12}$, very similar to that obtained from the Bloch-Siegert Hamiltonian, except with the Rabi frequency rescaled to $\Omega_{\text{eff}}=g\mathcal{J}_0/2$. As a consequence, the system is capable of making the transition from the ground to the excited state within the finite time interval governed by the Karhunen-Lo\`eve expansion. Another surprising byproduct of the rescaling $g/2\to g\mathcal{J}_0/2$ is that the Bloch-Siegert shift and effective Hamiltonians listed in Case 2 are now noise-dependent at all resonances. More precisely, we get the following resonant effective frequencies:  
 \begin{subequations}
\begin{align}\label{Carrier3ResJ0}
\Omega_{\text{eff}}\big|_{\varepsilon_0=\omega}&=\frac{g\mathcal{J}_0}{2}-\frac{g^3\mathcal{J}_0^3}{64 \omega ^2}-\frac{13 g^5\mathcal{J}_0^5}{4096 \omega ^4}-\frac{81 g^7\mathcal{J}_0^7}{131072 \omega ^6\mathcal{J}_0^3}-\frac{1677
   g^9\mathcal{J}_0^9}{16777216 \omega ^8}+\cdots,\\
\Omega_{\text{eff}}\big|_{\varepsilon_0=3\omega}&=\frac{9 g^3 \mathcal{J}_0^3}{32
   \varepsilon_0^2}-\frac{81 g^5 \mathcal{J}_0^5}{256 \varepsilon_0^4}+\frac{2187 g^7 \mathcal{J}_0^7}{8192 \varepsilon_0^6}-\frac{6561 g^9 \mathcal{J}_0^9}{32768 \varepsilon_0^8}+\cdots,\\
\Omega_{\text{eff}}\big|_{\varepsilon_0=5\omega}&=\frac{625 g^5 \mathcal{J}_0^5}{2048 \varepsilon_0^4}-\frac{15625 g^7 \mathcal{J}_0^7}{32768 \varepsilon_0^6}+\frac{2734375 g^9 \mathcal{J}_0^9}{2097152 \varepsilon_0^8}+\cdots,\\   \Omega_{\text{eff}}\big|_{\varepsilon_0=7\omega}&=\frac{117649 g^7 \mathcal{J}_0^7}{294912 \varepsilon_0^6}-\frac{5764801 g^9 \mathcal{J}_0^9}{18874368 \varepsilon_0^8}+\cdots,\\
\vdots\nonumber&
\end{align}
\end{subequations}
and for the effective Hamiltonian:
\begin{align}
&\mathsf{H}_{\text{eff}}=\frac{g \mathcal{J}_0}{2}\sigma_x-\frac{g^2 \mathcal{J}_0^2}{8\omega}\sigma_z-\frac{g^3 \mathcal{J}_0^3}{32\omega^2}\sigma_x+\cdots .
\end{align}
For this class of approximations, assuming a low noise in the ground state is critical. While increasing the noise amplitude or $g$, additional low-frequency oscillations are picked up and superimpose themselves onto the resonant  modes in a way that cannot be neglected anymore. For instance, while the resonance equation is solved exactly for the above, it can be satisfied approximately for small $n$. Specifically, setting $n = \pm 1$ with a sufficiently small $\omega_{\eta}$ yields $\varepsilon_0 + n \omega_{\eta} + \alpha \omega = \pm \omega_{\eta} \approx 0$, where $g/2$ is now rescaled by $|\mathcal{J}_1|$. The resulting large-amplitude oscillations have a frequency of $|g \mathcal{J}_1|/2$, which satisfies $|g \mathcal{J}_1|/2 < |g \mathcal{J}_0|/2 < g/2$. Consequently, multiple large-amplitude, low-frequency oscillations combine to drive chaotic dynamics within individual realizations. This causes the average transition probability to decrease when averaged over many realizations, as illustrated in Fig.~\ref{fig:Case3_BS}. 

For increasing noise amplitude, an alternative approximation strategy consists in including an entire family of terms in the dominant sinusoidal function for approximating $U_{12}(t)$. On resonance $\varepsilon_0 = \omega$ and for not too large values of $g$ and $\gamma$, one may approximate the excited state population $U_{12}$ by:
\begin{equation}
    U_{12} = \sin{\left( \frac{g}{2}\theta(t)\right)},
\end{equation}
where $\theta(t) = \int_0^t e^{i\gamma W(\tau)}d\tau$.  This is best seen from a direct evaluation of the exact solution from which the dominant terms form the Taylor series expansion of the sine function; $\left( g\theta(t)/2 \right)^k$ appearing at order $k$ with additional terms that are safely averaged out in this context. However, with an even moderate further increase in the noise, these additional terms must be retained. Constructing accurate approximation strategies becomes increasingly more complex as the noise amplitude grows.

\section{\Large Fourth case \texorpdfstring{$\epsilon(t)$ random, $f(t)$ Gaussian or modulated Gaussian}{with white noise and Gaussian or modulated Gaussian coupling}}\label{FourthCase}

Finally, we consider a noisy Hamiltonian Eq.~(\ref{HamForm}) with white noise and Gaussian or modulated Gaussian coupling function $f(t)$. By adding a supposed external Gaussian control in the out-of-diagonal term of the Hamiltonian, it was shown in \cite{Gemme2024} that the population of the excited state can be stabilized in time when $e_0=0$ i.e. turning off the noise. 

\subsection{Gaussian \texorpdfstring{$f(t)$:}{coupling:} Exact analytical solution}
We begin with a Gaussian control function but first with no resonant pumping: 
\begin{equation}
f(t) = \sqrt{\frac{\pi}{2}}\frac{1}{g\sigma} e^{-\frac{(t-t_m)^2}{2\sigma^2}}, 
\end{equation}
 after substituting for the truncated Karhunen-Lo\`eve expansion, we have the following Hamiltonian: 
\begin{align}
\mathsf{H}^{(K)}(t) &= \begin{pmatrix}
S_0 + \gamma (d/dt)W_t^{(K)} & g\sqrt{\frac{\pi}{2}}\frac{1}{g\sigma}e^{-\frac{(t-t_m)^2}{2\sigma^2}}  \\
\bar{g}\sqrt{\frac{\pi}{2}}\frac{1}{\bar{g}\sigma}e^{-\frac{(t-t_m)^2}{2\sigma^2}} & S_1
\end{pmatrix}, \nonumber\\
&=
\begin{pmatrix}
S_0 + \gamma \sqrt{2/\mathcal{T}}
\sum_{k=0}^K Z_k \cos\left(\frac{(2k+1)\pi t}{2\mathcal{T}}\right) & g\sqrt{\frac{\pi}{2}}\frac{1}{g\sigma}e^{-\frac{(t-t_m)^2}{2\sigma^2}}  \\
\bar{g}\sqrt{\frac{\pi}{2}}\frac{1}{\bar{g}\sigma}e^{-\frac{(t-t_m)^2}{2\sigma^2}} & S_1
\end{pmatrix},
\end{align}
expecting that the Gaussian can filter the noise and stabilize some population in the excited state. We perform the standard frame change and decompose the Gaussian function into a complex Fourier series with coefficients given in terms of erf functions. We find:
\begin{subequations}
\begin{align}
&H_{12}(t) = \sqrt{\frac{\pi}{2}}\frac{1}{\sigma} \sum_n \sum_{\ell} \mathcal{J}_n c_\ell e^{i(S_0-S_1+ n \omega_{\eta}+\ell \omega_{f}) t},\\
&H_{21}(t)=\sqrt{\frac{\pi}{2}}\frac{1}{\sigma} \sum_n\sum_{\ell} \mathcal{J}_n \bar{c}_\ell e^{-i(S_0-S_1+ n \omega_{\eta}+\ell \omega_{f}) t}.
\end{align}
\end{subequations}
In these expressions, the arguments of the GBF are given by the sequence $\left\lbrace Z_k \frac{2 \gamma \sqrt{2\mathcal{T}}}{(2k+1)\pi}\right\rbrace$ and $c_\ell$ are the coefficients of the Fourier series of the Gaussian window function with a base frequency of $\omega_f$. Due to the base frequencies for both the decomposition of the Gaussian window and the Jacobi-Anger relation depending on the length of the finite time interval, a convolution is possible.  That is, as stated previously, the fundamental frequency of the Brownian motion is $\omega_\eta = \pi/(2 \mathcal{T})$. However, the period of evolution for the Gaussian function is $\mathcal{T}$, resulting in a fundamental frequency of $\omega_f = 2\pi/\mathcal{T} = 4\omega_\eta$. This results in
\begin{subequations}
\begin{align}
&H_{12}(t) = \sqrt{\frac{\pi}{2}}\frac{1}{\sigma} \sum_n  \mathcal{J}^c_n e^{i(S_0-S_1+ n \omega_{\eta}) t},\\
&H_{21}(t)=\sqrt{\frac{\pi}{2}}\frac{1}{\sigma} \sum_n \bar{\mathcal{J}}^c_n e^{-i(S_0-S_1+ n \omega_{\eta}) t},
\end{align}
\end{subequations}
where
$
\mathcal{J}^c_\ell = \sum_n \mathcal{J}_{\ell - 4n} c_{n}
$. Using the fact that $\omega_f = 4 \omega_\eta$, the convolution sum includes a GBF shifted by $4n$, as stated previously.  This results in the kernel
\begin{equation}
K(t,s)\,\Theta:=
i|g|^2  \sum_{n,m}\mathcal{J}^c_n
\bar{\mathcal{J}}^c_me^{i(n -m)\omega_0t}e^{i[\varepsilon_0+m\omega_0,0](t-s)}~\Theta.
\end{equation}
Let $\pmb{m}_k:=(m_1,\ldots,m_k)\in\mathbb{N}^k$ and similarly for $\pmb{n}_k$. It follows that
\begin{figure}[t!]
 \centering
 \includegraphics[scale=0.35]{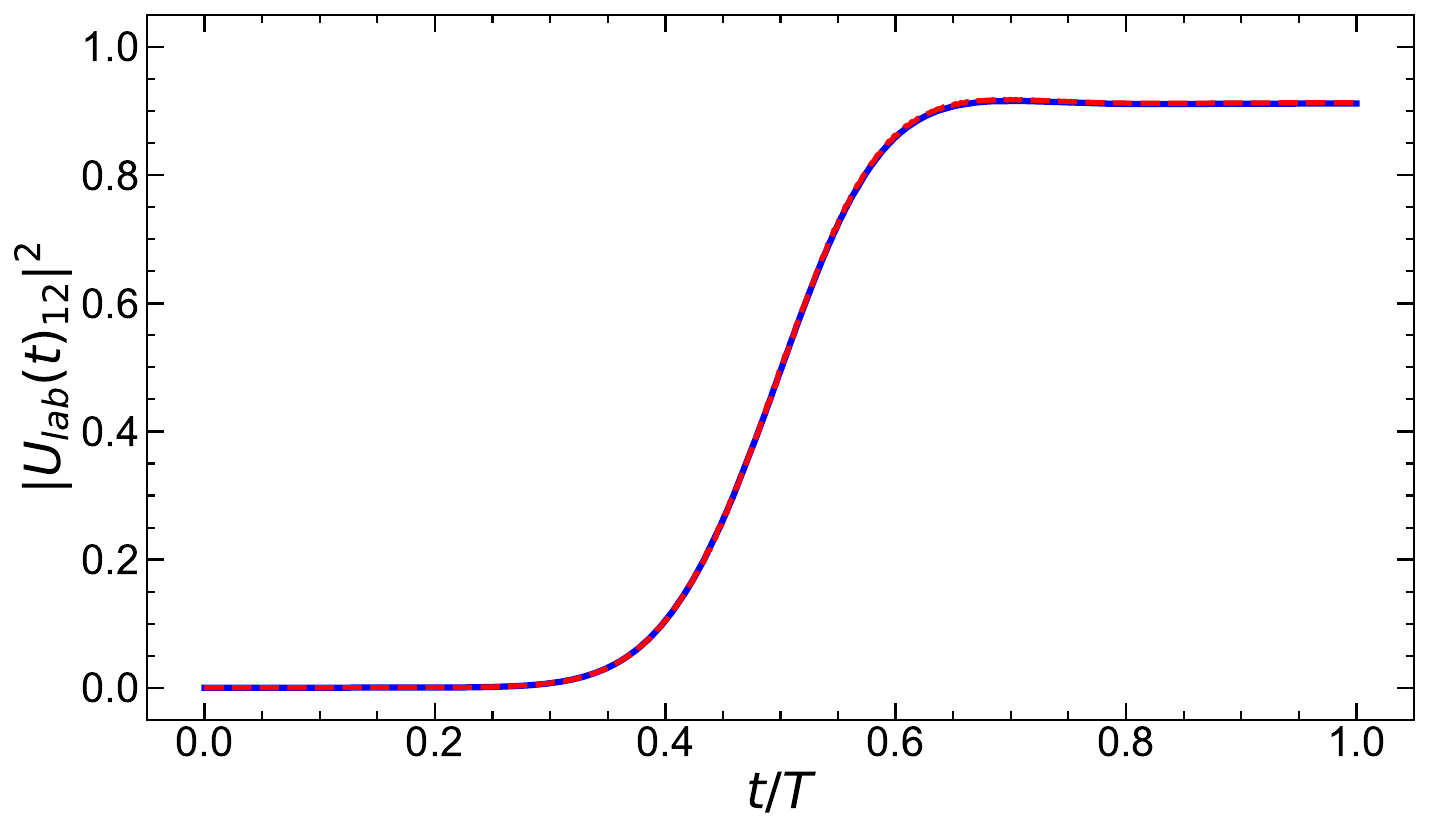} 
 \hspace{5mm}
 \includegraphics[scale=0.35]{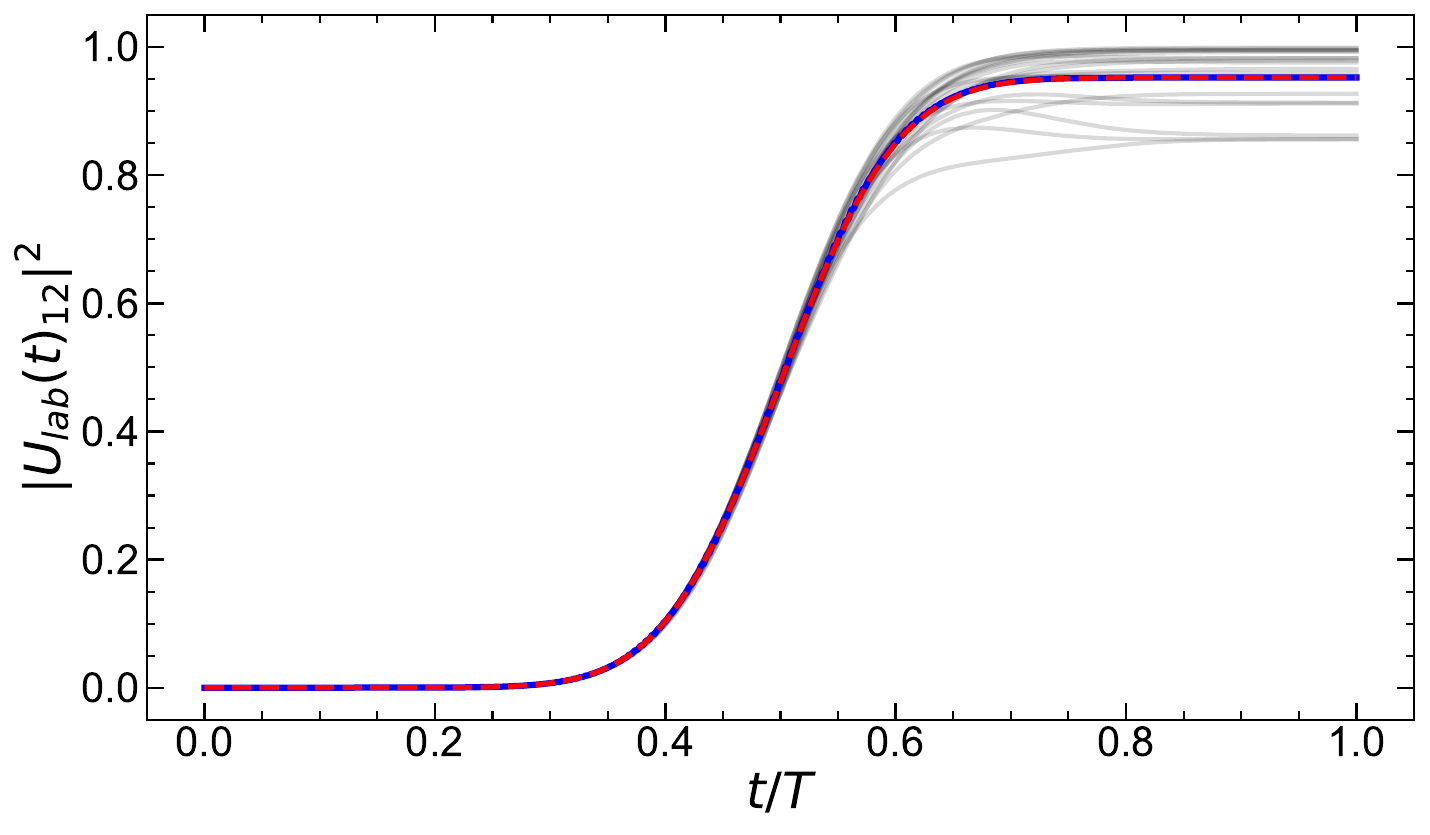} 
 \caption{Evolution of the stochastic transition probability $P_{|0\rangle\to |1\rangle}(t):=|U_\text{lab}(t)_{12}|^2=1-|U_\text{lab}(t)_{11}|^2$ as a function of time as determined by a fully numerical solver (solid blue line) and analytical formula (\textbf{top} figure, dashed red line) for one realization of noisy variables.  Stochastic transition probability of $P_{|0\rangle\to |1\rangle}(t):=|U_\text{lab}(t)_{12}|^2=1-|U_\text{lab}(t)_{11}|^2$ (\textbf{bottom} figure) over 10 realizations (solid gray lines), the analytical mean (dashed red line), and the numerically calculated mean (solid blue line) of the exact solution. Parameters for the figures: $\omega/|S_1-S_0|=1$, $\mathcal{T} = 600$, $t_m = \mathcal{T}/2$, $\sigma = \mathcal{T}/8$, $\gamma=0.1$ and $K=10$. In the above situations, the frequency of the driving depends on the length of the simulation $\omega_\eta = \pi/(2\mathcal{T})$, leading to low driving frequencies that are enveloped by a Gaussian, non-periodic waveform, a case that is beyond standard approximating regimes.  When there is no noise, the Gaussian is constructed as to allow the transitions to the excited state when $\omega=\varepsilon_0$. However, with the introduction of a noise, the Gaussian is altered, therefore, in some cases, the quantum system is not able to populate fully the excited state. For lower noise amplitude, the transition probability is closed to the zero noise case.}
 \label{fig:Case4}
\end{figure}
\begin{equation}
\mathsf{U_{lab}}(t)=\begin{pmatrix}
e^{-iS_0t -i\gamma W_t}  & 0\\ 
0 & e^{-iS_1t}  
\end{pmatrix}\begin{pmatrix}
U_{11}(t) & U_{12}(t)\\ 
U_{21}(t) & U_{22}(t)  
\end{pmatrix},
\end{equation}
where
\begin{subequations}\label{ExactCase4Sol}
\begin{align}
U_{11}(t)&=\sum_{k\geq 0} \left(\sqrt{\frac{\pi}{2}}\frac{1}{\sigma}\right)^{\!2k}\sum_{\pmb{m}_k,\pmb{n}_k}\mathcal{J}^c_{\pmb{n}_k,\pmb{m}_k} 
e^{i[A_1,\varepsilon_0+B_1,A_2,\varepsilon_0+B_2,\dots,A_k,\varepsilon_0+B_k,0]t},\\
U_{12}(t)&=\sum_{k\geq 0}\left(\sqrt{\frac{\pi}{2}}\frac{1}{\sigma}\right)^{2k+1}\sum_{n,\pmb{m}_k,\pmb{n}_k}e^{i(\varepsilon_0+n\omega_{\eta})t}\mathcal{J}^c_{n,\pmb{n}_k,\pmb{m}_k}\nonumber\\
&\hspace{15mm} \times e^{-i[A_1, \varepsilon_0+B_1,A_2,\varepsilon_0+B_2,\dots,A_k,\varepsilon_0+B_k,\varepsilon_0+n\omega_{\eta},0]t}.
\end{align} 
\end{subequations}
In these expressions, we again used
$\varepsilon_0 = S_0 - S_1$, $A_i:=N_i-M_i$, and  $B_i:=N_i-M_{i-1}$ with $M_j=\sum_{i=1}^jm_i\omega_\eta$ and
$N_j=\sum_{i=1}^jn_i\omega_\eta$. As before, in the limit $K\rightarrow \infty$, the above gives the exact solution of the Stratonovich stochastic differential equation by the Wong-Zakai theorem.

\subsubsection{Physical interpretation}
For $\gamma=0$, the exact solutions as described by Eqs.~(\ref{ExactCase4Sol}) simplifies to
\begin{subequations}\label{ExactGamma0}
\begin{align}
U_{11}(t)&=\sum_{k\geq 0} \left(\sqrt{\frac{\pi}{2}}\frac{1}{\sigma}\right)^{\!2k}\sum_{\pmb{m}_k,\pmb{n}_k}c_{\pmb{n}_k,\pmb{m}_k} 
e^{i[A_1,\varepsilon_0+B_1,A_2,\varepsilon_0+B_2,\dots,A_k,\varepsilon_0+B_k,0]t},\\
U_{12}(t)&=\sum_{k\geq 0}\left(\sqrt{\frac{\pi}{2}}\frac{1}{\sigma}\right)^{2k+1}\sum_{n,\pmb{m}_k,\pmb{n}_k}e^{i(\varepsilon_0+n\omega_{\eta})t}\nonumber\\
&\hspace{15mm} \times c_{n,\pmb{n}_k,\pmb{m}_k}e^{-i[A_1, \varepsilon_0+B_1,A_2,\varepsilon_0+B_2,\dots,A_k,\varepsilon_0+B_k,\varepsilon_0+n\omega_{\eta},0]t}.
\end{align} 
\end{subequations}
In these equations, we define
\begin{equation}
c_{\pmb{n}_k,\pmb{m}_k}:=\prod_{i=1}^k c_{n_i}\, \bar{c}_{m_i}
\quad {\rm and} \quad c_{n,\pmb{n}_k,\pmb{m}_k}:=c_n c_{\pmb{n}_k,\pmb{m}_k},
\end{equation}
using the Fourier series coefficients of the Gaussian functions. In this case, we can apply similar observations as used in the first subcase of Case 3: the resonances occur at integer values $n$ that satisfy the condition $\varepsilon_0+n\omega_\eta=0$. But again, the integer $n$ must be very large if the equation above is to be satisfied. However, the Fourier coefficient of order $n$ of the Gaussian function decays as $e^{-n^2}$. Therefore, the divided-difference exponentials significantly contributing to any resonance have low frequencies. Due to this and when $g$ is small, a longer time than $\mathcal{T}$ is required to obtain a transition toward the excited state. Alternatively, the decay of the Fourier coefficients is also proportional to $e^{-\sigma^2}$. Therefore and for sufficiently small Gaussian variance, the excited state can be reached in a very short time. Nonetheless, the amplitude of the Gaussian will no longer be small as it scales as $1/\sigma$ resulting in fast oscillations from a non-harmonic function represented by the exact solution Eqs.~(\ref{ExactGamma0}). When $\gamma \ne0$ similar arguments hold, except now the coefficients are convolved with the GBFs. At integer $N$, the coefficient corresponding to a resonance would be $\mathcal{J}_N^c = \sum_\ell \mathcal{J}_{N-4\ell} c_{\ell}$. The Gaussian decay of the $c_{\ell}$ coefficients  would be mitigated at $\ell=0$. However, the magnitude of $\mathcal{J}_N \ll 1$ for $\gamma \ll 1$ requires here a time longer than $\mathcal{T}$ to trigger a transition towards the excited state.    

\subsection{Modulated Gaussian \texorpdfstring{$f(t)$:}{coupling:} Exact analytical solution}
We now consider the situation described in \cite{Gemme2024} where the off-diagonal elements are given by $g f(t)$ where $f(t)$ is a modulated Gaussian of the $\cos(\omega t)$ Rabi like pumping term:
\begin{equation}
f(t) = \sqrt{\frac{\pi}{2}}\frac{1}{g\sigma} e^{-\frac{(t-t_m)^2}{2\sigma^2}} \cos(\omega t).
\end{equation}
The Hamiltonian we consider is now 
\begin{align}
\mathsf{H}^{(K)}(t) = \begin{pmatrix}
S_0 + \gamma (d/dt)W_t^{(K)} & g\sqrt{\frac{\pi}{2}}\frac{1}{g\sigma}e^{-\frac{(t-t_m)^2}{2\sigma^2}}\cos(\omega t)  \\
\bar{g}\sqrt{\frac{\pi}{2}}\frac{1}{\bar{g}\sigma}e^{-\frac{(t-t_m)^2}{2\sigma^2}}\cos(\omega t) & S_1
\end{pmatrix}. 
\end{align}
To treat this situation, after substituting for the truncated Karhunen-Lo\`eve expansion, we proceed as before to obtain
\begin{subequations}
\begin{align}
&H_{12}(t)=\frac{1}{2}\sqrt{\frac{\pi}{2}}\frac{1}{\sigma}\sum_{n,\ell,\alpha} \mathcal{J}_n c_\ell e^{i(S_0-S_1+ n \omega_{\eta}+\ell \omega_{f}+\alpha\omega) t},\\
&H_{21}(t)=\frac{1}{2}\sqrt{\frac{\pi}{2}}\frac{1}{\sigma}\sum_{n,\ell,\alpha} \mathcal{J}_n \bar{c}_\ell e^{-i(S_0-S_1+ n \omega_{\eta}+\ell \omega_{f}+\alpha\omega) t}.
\end{align}
\end{subequations}
We can again convolve the coefficients $c_\ell$ with the GBFs to obtain $
\mathcal{J}^c_n
$, so that the path-sum kernel is
\begin{align}\label{Case4Kernel}
&K(t,s)\Theta:=\\
&i|g/2|^2 \sum_{n,m\in\mathbb{Z}\atop\beta,\alpha= \pm1} \mathcal{J}^c_n
\bar{\mathcal{J}}^c_me^{i\{(n -m)\omega_\eta+(\beta-\alpha)\omega\}t}e^{i[\varepsilon_0+m\omega_\eta+\alpha\omega,0](t-s)}\,\Theta.\nonumber
\end{align}
Consequently, by Eqs.~(\ref{PSUequations}), the entries of the evolution operator are 
\begin{subequations}\label{Case4Sols}
\begin{align}
U_{11}(t)&=\sum_{k\geq 0} \left(\frac{1}{2}\sqrt{\frac{\pi}{2}}\frac{1}{\sigma}\right)^{\!2k}\sum_{\pmb{m}_k,\pmb{n}_k\atop\pmb{\alpha}_k,\pmb{\beta}_k}\!\!\!\mathcal{J}^c_{\pmb{n}_k,\pmb{m}_k} 
e^{i[A_1,\varepsilon_0+B_1,A_2,\varepsilon_0+B_2,\dots,A_k,\varepsilon_0+B_k,0]t},\\
U_{12}(t)&=\sum_{k\geq 0}\left(\frac{1}{2}\sqrt{\frac{\pi}{2}}\frac{1}{\sigma}\right)^{\!2k+1}\sum_{n,\pmb{m}_k,\pmb{n}_k\atop \alpha,\pmb{\alpha}_k,\pmb{\beta}_k}\!\!\!e^{i(\varepsilon_0+n\omega_{\eta}+\alpha \omega)t}\mathcal{J}^c_{n,\pmb{n}_k,\pmb{m}_k}\\
&\hspace{15mm} \times e^{-i[A_1, \varepsilon_0+B_1,A_2,\varepsilon_0+B_2,\dots,A_k,\varepsilon_0+B_k,\varepsilon_0+n\omega_{\eta}+\alpha \omega,0]t},\nonumber
\end{align} 
\end{subequations} where $A_i:=N_i-M_i$ and  $B_i:=N_i-M_{i-1}$ with $M_j=\sum_{i=1}^j(m_i\omega_\eta+\alpha_i\omega)$ and
$N_j=\sum_{i=1}^j(n_i\omega_\eta+\beta_i\omega)$.  In the limit $K\rightarrow \infty$, the above gives the exact solution of the Stratonovich stochastic differential equation by the Wong-Zakai theorem.

\subsubsection{Physical Interpretation}

As considered previously, this model undergoes numerous resonances whose mathematical signature is the presence of repeated arguments in the divided-difference exponentials of the path-sum kernel Eqs.~(\ref{Case4Kernel}). Given that $\alpha=\pm 1$ and $n\in \mathbb{Z}$ we obtain 
\begin{equation}\label{ResLoc_noise2}
\frac{\varepsilon_0\pm \omega}{\omega_\eta}=n\in\mathbb{Z}\backslash\{0\},\quad\text{or}\quad n=0 \,\text{ that is }\, \omega=\varepsilon_0.
\end{equation}
The mathematical analysis of the resonance is similar to the previous cases. When resonating, the long time behavior of the population transfer $P_{|0\rangle\to|1\rangle}$ follows a $\sin^2(\Omega_{\text{eff}} t)$ time dependence with an effective Rabi frequency given directly by the coefficient of the $t$ polynomial in the exact solution described by Eqs.~(\ref{Case4Sols}). When $\gamma = 0$, we recover the results obtained in \cite{Gemme2024}. Starting from the ground state and as demonstrated in \cite{Gemme2024} simply using the RWA approximation, the energy stored reduces to: 
\begin{equation}\label{sin_envelope}
\Delta E(t) = \varepsilon_0 \sin^2\big(\theta(t)/2\big),
\end{equation}
where $\theta(t) = g \int_0^t f(\tau)d\tau = \left(\frac{1}{2}\sqrt{\frac{\pi}{2}}\frac{1}{\sigma}\right)\sum_n c_n e^{i[n \omega_\eta,0]t} $ when $\varepsilon_0 = \omega$. Starting with the exact expression for $U_{12}$ with zero noise, $\varepsilon_0 = \omega$, and $g\ll\varepsilon_0$, the divided-difference exponentials exhibit a resonance, as discussed previously, allowing for the resummation of the exact solution. We have:
\begin{equation}
U_{12}(t)\approx\sum_{k\geq 0}\left(\frac{1}{2}\sqrt{\frac{\pi}{2}}\frac{1}{\sigma}\right)^{\!2k+1}\!\!\!\sum_{n,\pmb{m}_k,\pmb{n}_k}\!\!\!e^{in\omega_{\eta}t}c_{n,\pmb{n}_k,\pmb{m}_k}e^{-i[A_1, B_1,A_2,B_2,\dots,A_k,B_k,n\omega_{\eta},0]t},
\end{equation}
where $A_i:=N_i-M_i$ and  $B_i:=N_i-M_{i-1}$ with $M_j=\sum_{i=1}^jm_i\omega_\eta$ and
$N_j=\sum_{i=1}^jn_i\omega_\eta$. As noted above, the leading-order contribution to $U_{12}$ reproduces the Taylor expansion of the sine function, with the $k=1$ term leading to the $\theta(t)/2$ term in Eq.~(\ref{ResLoc_noise2}). More generally, upon expanding the sine series and expressing the powers of $\theta(t)$ in terms of divided-difference exponentials, one finds that a distinguished subset of the terms appearing in the $\star$-power expansion coincides exactly with the powers of $\theta(t)$. The remaining contributions consist of rapidly oscillating secular terms of the form $t^k e^{i\Omega t}$, which average to zero in a regime where the amplitude of the driving is low. A systematic treatment of a generalization of this resummation procedure is presented in Appendix~\ref{AppTDHeff}.

For the symmetric Gaussian envelope considered here, the erf coefficients $c_n$ are exactly real valued and Eq.~(\ref{sin_envelope}) follows directly. Consequently, the Gaussian envelope enters as an argument in the sine function of Eq.~(\ref{ResLoc_noise2}) through its integral. As already observed in \cite{Gemme2024}, it results in a smooth control of the complete population inversion: $P_{|0\rangle\to|1\rangle}(t)\simeq 1$ when the Gaussian amplitude has become negligible.

When $\gamma\neq0$, an analogous expression may be derived in the small amplitude noise regime, with now $\theta(t)$ including the Gaussian envelope modulated by the complex Brownian motion. In this case, the smooth transition $P_{|0\rangle\to|1\rangle}(t)=1$ towards the excited state is largely preserved. However, as the noise amplitude is increasing, the spectral content becomes distributed across an increasingly broad set of GBFs, resulting in destructive interference of the smooth population transfer. As a result, the smooth population transfer characteristic of the noiseless case no longer approaches the complete inversion, as illustrated in Fig.~\ref{fig:Case4}.

\section{\Large Conclusion}\label{Conclusion}

This work has established exact representations of the evolution operators for four families of time-dependent $2\times2$ Hamiltonians arising in thermal energy pumping, including two models subject to stochastic driving. A central feature of the results is that the solutions are exact, analytical, and explicit simultaneously. By these we mean that they are expressed as unconditionally convergent series involving only elementary operations on standard functions with fully specified coefficients. Furthermore, no perturbative assumptions or parameter restrictions are required, allowing the resulting formulas to remain valid throughout the entire parameter space, including strongly driven, non-periodic, and noise-dominated regimes.

The derivations rely for the first time on a unified and fruitful combination of three complementary mathematical developments: the $\star$-algebra, path-sums, and Omega calculus which are briefly reviewed in Subsections \ref{SecStarProd}, \ref{SecPS}, and \ref{SecOC}, respectively. The $\star$-algebra provides the mathematical framework within which non-autonomous differential systems are amenable to ordinary linear algebraic tools. The path-sum theorem then formulates the solutions into continued fractions that terminate at finite depth and optimally exploit the system's sparsity pattern. These fractions involve $\star$-products and inverses, which are finally evaluated into divided differences of standard functions thanks to Omega calculus.  Omega calculus relies only on basic manipulation of crude rational functions and the simplest elimination rules from combinatorial analysis, typically producing sums of divided-difference exponential polynomials. The combination of these techniques yields exact evolution operators in a form that is both analytically practical and, using numerical implementations of $\star$-algebras, can be translated into computationally efficient codes \cite{Pozza2023a,Pozza2023Legendre}. 

The exact solutions also provide a natural starting point for constructing effective descriptions from first principles; that is, directly from the evolution operator describing the dynamics without recourse to uncontrolled approximations or ad hoc procedures. In two cases, explicit effective Hamiltonians were obtained directly from the exact evolution operators, leading in particular to an all-orders commutator-free representation that extends conventional rotating-wave and Rabi-type approximations. See, e.g., the exact expression in Eq.~(\ref{ExactHeff}) for $\mathsf{H}_{\text{eff}}$ valid to all orders in $\omega^{-1}$. Rather than relying on asymptotic arguments, these effective generators emerge from the exact dynamics themselves, thereby retaining validity beyond the regimes where standard approximations are typically justified.

The stochastic models also illustrate the versatility of the formalism. In the weak-noise limit, the exact solution reduces to a deterministic evolution dressed by a noise-dependent contribution, making it possible to analytically identify noise-induced modifications of familiar phenomena such as the Bloch--Siegert shift or thermal energy pumping in a Gaussian pulsed system. See, e.g., Subsection \ref{PhysIntNoise}. More broadly, the exact expressions obtained  expose resonance structures in every model considered irrespective of being deterministic or stochastic, including parameter regimes that are inaccessible or difficult to characterize using conventional perturbative techniques.

Finally, the exact analytical treatment as shown here need not be confined to isolated $2\times 2$ deterministic or stochastic quantum solvable models. 
The breadth of applicability of the underlying methodology suggests that its impact extends well beyond the specific quantum systems studied here, offering a powerful and systematic approach to a wide range of non-autonomous problems including higher dimensional ones encountered across physics, engineering, and applied mathematics.

\section*{Acknowledgements}
M. Warnock acknowledges funds from the Naval Undersea Warfare Center In-House Laboratory Independent Research (ILIR) program from the Office of Naval Research (ONR) under N0001425GI00297.

\section*{Appendices}
\appendix
\section{Path-sums}\label{PS2x2}
We present the general path-sum treatment of a $2\times 2$ non-autonomous differential system, then apply it to the present cases of interest. In general, we consider solving
\begin{equation}\mathsf{U}(t)=\mathsf{A}(t)\mathsf{U}(t,s),\end{equation} again with $\mathsf{U}(s,s)=\mathsf{Id}$, $t\geq s$ and
\begin{equation}
\mathsf{A}(t) = \begin{pmatrix}a(t)&b(t)\\c(t)& d(t)\end{pmatrix}.
\end{equation}
The theory of the $\star$-algebra indicates that $\mathsf{U}=\Theta\mathsf{Id}\star \mathsf{G}$, with $\mathsf{G}:=(\mathsf{Id}_\star-\mathsf{A}\Theta)^{\star -1}$ being the system Green function. In particular,  
\begin{equation}
U_{11}=\Theta \star G_{11} = \int_s^t G_{11}(\tau,s)d\tau,
\end{equation}
where $G_{11}(t,s)=(1_\star - k_{11}\Theta )^{\star -1}$ with $1_\star=\delta(t-s)$ and, per path-sums, 
\begin{align}
k_{11}(t,s)\Theta &= \overbrace{A_{11}\Theta}^{\text{Loop }1\,\CircleArrow}+\underbrace{A_{12}\Theta\star (1_\star-\overbrace{A_{22}\Theta}^{\text{Loop }2\,\CircleArrow})^{\star -1}\star A_{21}\Theta}_{\text{Backtrack }1\leftarrow 2\leftarrow 1},\nonumber\\
&=a(t)\,\Theta + b(t) \int_s^t e^{\int_{\sigma}^t d(\tau) d\tau}c(\sigma)d\sigma\,\Theta.
\end{align}
The reader may have seen that this result can be recast into the Schur complement formula for the inverse of a $2\times 2$ matrix but in the $\star$-algebra (i.e., replacing matrix products and inverses by $\star$-products and inverses). This is because the graph has only two vertices. In general path-sums differ from a recursive implementation of Schur complements but can still be understood as a kind of generalization of it that maximally exploits the graph sparsity patterns. A further property of path-sums, scale invariance, manifests itself only for $3\times 3$ systems upwards and so is left out of the discussion here. See, e.g. \cite{GisBon} for further details. Similarly, we have
$
U_{22}=\Theta \star G_{22}=\int_s^t G_{22}(\tau,s)d\tau,
$
where $G_{22}(t,s)=(1_\star - k_{22}\Theta )^{\star -1}$ with
\begin{align}
k_{22}(t,s)\Theta&=\overbrace{A_{22}\Theta}^{\text{Loop }2\,\CircleArrow}+\underbrace{A_{21}\Theta\star (1_\star-\overbrace{A_{11}\Theta}^{\text{Loop }1\,\CircleArrow})^{\star -1}\star A_{12}\Theta}_{\text{Backtrack }2\leftarrow 1\leftarrow 2},\nonumber\\
&= d(t)\,\Theta + c(t) \int_s^t e^{\int_{\sigma}^t a(\tau) d\tau}b(\sigma)d\sigma\,\Theta.
\end{align}
The off-diagonal entries are
\begin{align}\label{U12GenPS}
U_{12}(t,s)&=\Theta\star G_{12},\nonumber\\
&=\Theta\star\underbrace{\overbrace{(1_\star - A_{11}\Theta)^{\star-1}}^{\text{Cycles }1\leftarrow1\text{ on }\mathcal{G}\backslash\{2\}}\star\, \overbrace{A_{12}\Theta}^{\text{Edge }1\leftarrow2}\,\star\!\!\!\! \overbrace{G_{22}}^{\text{All cycles }2\leftarrow 2}}_{\text{Path }1\leftarrow 2},\nonumber\\
&=\int_s^te^{\int^t_{\sigma} a(\tau)d\tau}b(\sigma) U_{22}(\sigma,s)d\sigma\,\Theta.
\end{align}
Reading the matrix indices from right to left reveals the simple path undertaken on the graph $\mathcal{G}$. Here $\mathcal{G}\backslash\{2\}$ denotes the graph on which vertex 2 and all edges attached to it have been removed. Given the indices ordering, time can be seen to run right to left in all algebraic expressions coming out of path-sums. One can equally reverse the arrow of time though, such time reversal in path-sums implies that the above is, in fact, equal to
\begin{align}\label{U12GenPS2}
U_{12}(t,s)&=\Theta\star\!\!\!\! \overbrace{G_{11}}^{\text{All cycles }1\rightarrow 1}\!\!\!\!\star\, \underbrace{\overbrace{A_{12}\Theta}^{\text{Edge }1\to 2}\star \overbrace{(1_\star-A_{22}\Theta)^{\star-1}}^{\text{Cycles }2\to 2\text{ on }\mathcal{G}\backslash\{1\}}}_{\text{Path }1\to 2},\nonumber\\
&=\int_s^tU_{11}(t,\sigma)b(\sigma)e^{\int_s^{\sigma}d(\tau)d\tau}d\sigma\,\Theta,
\end{align}
which is not obvious from the integral expressions in Eqs.~(\ref{U12GenPS}, \ref{U12GenPS2}).
For the other off-diagonal entry we have
\begin{align}
U_{21}(t,s)&=\Theta\star G_{21},\nonumber\\
&=\Theta\star(1_\star - A_{22}\Theta)^{\star-1}\star A_{21}\Theta\star G_{11},\nonumber\\
&=\int_s^te^{\int^t_{\sigma} d(\tau)d\tau}c(\sigma) U_{11}(\sigma,s)d\sigma\,\Theta,
\end{align}
and, by time reversal, we obtain
\begin{equation}
U_{21}(t,s)=\Theta\star G_{22}\star A_{21}\Theta\star (1_\star-A_{11}\Theta)^{\star-1}=\int_s^t U_{22}(t,\sigma)c(\sigma)e^{\int_s^{\sigma}a(\tau)d\tau}d\sigma\,\Theta.
\end{equation}
Combining a result and its time reverse version allows for further simplifications. For example, one can obtain all entries of $\mathsf{U}$ from sole $U_{11}$. Indeed, since
\begin{equation}\label{ReversePSG12}
G_{12}=(1_\star-A_{11}\Theta)^{\star-1}\star A_{12}\Theta\star G_{22}=G_{11}\star A_{12}\Theta\star (1_\star-A_{22}\Theta)^{\star-1},
\end{equation}
it follows that, assuming $A_{12}(t)$ is not identically 0 for all $t$ (in which case the problem is, in fact, trivial),
\begin{equation}
G_{22}=A_{12}^{-1}\delta'\star (1_\star - A_{11}\Theta)\star G_{11}\star A_{12}\Theta\star (1_\star-A_{22}\Theta)^{\star-1},
\end{equation}
which remains well defined even if $A_{12}$ intermittently passes through $0$ \cite{Giscard2020}. In this expression $\delta'(t-s)$ is the derivative of the Dirac delta distribution. These observations generalize to higher dimensions so that in general all entries of an evolution operator can be found from some of its diagonal entries. 
In the particular situation of interest here, $\mathsf{A}=-i \mathsf{H}$ with
\begin{equation}
\mathsf{H}(t) = 
\begin{pmatrix}
S_0 + \epsilon(t) & g f(t)\\
\bar{g} f(t)& S_1+ i \Gamma
\end{pmatrix},
\end{equation}
and so
\begin{align}
U_{11}(t) = &\Theta\star\Big(1_\star-(-i/\hbar)(S_0+\epsilon(t)\big)\Theta\\&\hspace{15mm}-(-i/\hbar)^2|g|^2 f(t)e^{-i\frac{S_1+i \Gamma}{\hbar}(t-s)}\Theta\star f(t)\Theta \Big)^{\star -1},\nonumber
\end{align}
then
\begin{align}
U_{22}(t) = &\Theta\star\Big(1_\star-(-i/\hbar)(S_1+i \Gamma\big)\Theta\\
&\hspace{15mm}-(-i/\hbar)^2|g|^2 f(t)e^{(-i/\hbar)S_0(t-s)+(-i/\hbar)\int_s^t\epsilon(\tau)d\tau} \Theta\star f(t)\Theta \Big)^{\star -1},\nonumber
\end{align}
and 
\begin{subequations}
\begin{align}
&U_{12}(t)=(-i/\hbar)g\,e^{-\frac{iS_0}{\hbar}(t-s)-\frac{i}{\hbar}\int_s^t\epsilon(\tau)d\tau}\Theta\star f(t)U_{22}\Theta,\\
&U_{21}(t)=(-i/\hbar)\bar{g}\,e^{-i\frac{S_1+i \Gamma}{\hbar}(t-s)}\Theta\star f(t)U_{11}\Theta.
\end{align}
\end{subequations}
In all of this, the main difficulty is the calculation of the scalar $\star$-resolvents. This is achieved by Omega calculus. We remark that performing a frame change before path-sum only serves to simplify somewhat the equations above, as a preconditioner stage. Nevertheless, the formalism is obviously valid in all frames.

\section{Jacobi-Anger and Generalized  Jacobi-Anger Expansions}\label{JacobiAnger}

The Jacobi-Anger expansion \cite{korenev2019bessel} and \cite[entry 9.1.41]{abramowitz1965handbook} is given by
\begin{align}\label{J-A}
e^{iz\sin\theta}=\sum_{n\in \mathbb{Z}}J_n(z)e^{in\theta},
\end{align} 
where $J_n(z)$ is the Bessel function of integer order $n\in\mathbb{Z}$, defined through the series 
\begin{align}\label{Bessel}
J_n(z)=\sum_{k=0}^\infty (-1)^k \frac{(z/2)^{2k+n}}{(k+n)!k!}.
\end{align}
For $n$ an integer, we have
$J_{-n}(z)=(-1)^nJ_n(z)$. 
This expansion is a special case of the generalized Jacobi-Anger expansion
\begin{equation}\label{GeneralizedJacoAng}
e^{i \sum_{k=1}^K z_k \sin(k\theta)} = \sum_{n\in \mathbb{Z}} \mathcal{J}_n(z_1,\dots, z_k)e^{in\theta},
\end{equation}
where $\mathcal{J}$ is a Generalized Bessel function (GBF) defined by \cite{dattoli1996theory} as a convolutional sum of two or more Bessel functions. For example, the two-dimensional GBF is 
\begin{equation}
\mathcal{J}_n(x,y) = \sum_{k=-\infty}^{\infty} J_{n-\ell k}(x) J_k(y),
\end{equation}
where $\ell$ may be any integer and we used the notation $\mathcal{J}_n(x,y)\equiv \mathcal{J}_n(z_1,\dots, z_k)$ with $z_i=0$ for all $i\neq 1,\ell$ such that $z_1\equiv x$ and $z_{\ell}\equiv y$. See, e.g., the generalizations considered in \cite{Korsch2006}. Higher-dimensional versions are obtained similarly, $\mathcal{J}_n(x,y,z)$ being the convolution of three Bessel functions, etc.
The generalized Jacobi-Anger relation of Eq.~(\ref{GeneralizedJacoAng}) appears naturally in Fourier series decompositions of many of the functions appearing in analytical spin dynamics.

\section{Proof of the equations for \texorpdfstring{$U_{ij}$}{the evolution operator}}\label{GenProof}

\subsection{Preliminary results concerning Omega calculus}\label{OC}

We here collect four identities pertaining to Omega calculus that are useful in the proofs of the results of the main text.\\

\textbullet~Identity 1 (Multiplication invariance under the Omega operator) Multiplying Omega variables by auxiliary variables does not change the result after the elimination of those variables,
\begin{align}\label{Id0}
\overset{\lambda}{\underset{=
}{\Omega}} \,F(\lambda)=\overset{\lambda}{\underset{=
}{\Omega}}\, F(z\lambda).
\end{align} 
Indeed, only $\lambda^0=(z\lambda)^0$ survives the action of the Omega operator so that one can freely replace $\lambda\rightarrow z\lambda$. This is convenient to ensure, via $F(z\lambda)$, that we work with convergent series in the Omega domain.

\textbullet~Identity 2 (Omega domain representation of the divided-difference exponential). We have
\begin{align}\label{hsp}
\overset{\lambda}{\underset{=
}{\Omega}} \frac{e^{\lambda t}}{\lambda^{n}\prod_{k=0}^{n}(1-a_k/\lambda)}
=e^{[a_0,a_1,\ldots,a_{n}]t}.
\end{align} 
This important result originates from the ordinary generating function of the complete homogeneous symmetric polynomials which gives, for any $n\in\mathbb{Z}$ and $m\in\mathbb{N}$ \cite[Chapter I, Sec. 2]{macdonald1998symmetric},
\begin{align}
\overset{\lambda}{\underset{=
}{\Omega}}\frac{\lambda^n}{\left(1-a_1/\lambda\right)\left(1-a_2/\lambda\right)\cdots 
\left(1-a_m/\lambda\right)}
=\left\{\begin{array}{ll}
1&,\,\, n=0\\
h_n\left(a_1,a_2,\ldots, a_m\right)&,\,\, n>0.
\end{array}\right.
\end{align} 
Therefore,
\begin{align}
\overset{\lambda}{\underset{=
}{\Omega}}\frac{e^{\lambda t}\lambda^{-n}}{\left(1-a_1/\lambda\right)\left(1-a_2/\lambda\right)\cdots 
\left(1-a_n/\lambda\right)}
&=\sum_{m\geq n}\frac{t^m}{m!}h_{m-n}\left(a_0,\ldots,a_n\right),\nonumber\\
&=\sum_{m\geq 0}\frac{t^{m+n}}{\left(m+n\right)!}h_m\left(a_0,\ldots,a_n\right),\nonumber\\
&=e^{[a_0,\ldots,a_n]t}.
\end{align} The last line here follows from the definition of the divided-difference exponential, Eq.~(\ref{expdivideddifference}).

\textbullet~Identity 3 (Scaling in a divided-difference exponential) It is sometimes convenient to extract a common factor in the arguments of a divided-difference exponential, e.g. for dimensional analysis. This is done as follows
\begin{align}\label{Id1}
e^{b[a_0,\ldots,a_n]t}=\overset{\lambda}{\underset{=
}{\Omega}} \frac{e^{b\lambda t}}{\lambda^n\prod_{k=0}^n\left(1-\frac{a_k}{\lambda}\right)}
=b^n\overset{\lambda}{\underset{=
}{\Omega}} \frac{e^{\lambda t}}{\lambda^n\prod_{k=0}^n\left(1-\frac{b a_k}{\lambda}\right)}=b^ne^{[ba_0,\ldots,ba_n]t},
\end{align} 
with the right-hand side coming from Identity~2. We highlight that the factor $b$ appears $n+1$ times in $[ba_0,\ldots,ba_n]$, but it is only raised to the power $n$ in front of the divided-difference exponential.

\textbullet~Identity 4 (Exchanging integrals and the Omega operator, $\int \Omega=\Omega \int$) Without loss of generality, by linearity, one may restrict our attention to $F(\lambda)=e^{\lambda t}/\lambda^n$. Then we have
\begin{align}\label{ExchangeProperty}
\int \underbrace{\overset{\lambda}{\underset{=
}{\Omega}}\,\frac{e^{\lambda t}}{\lambda^n}}_{=t^n/n!}dt
=\frac{t^{n+1}}{(n+1)!}
=\overset{\lambda}{\underset{=
}{\Omega}}\,\underbrace{\int \frac{e^{\lambda t}}{\lambda^n}dt}_{=e^{\lambda t}/\lambda^{n+1}}.
\end{align}

\textbullet~Identity 5 (Integrating the divided-difference exponentials). We have
\begin{align}\label{Intdd-exp}
\int e^{a[\pmb{b}]t}dt&=\int\overset{\lambda}{\underset{=
}{\Omega}} \frac{e^{a\lambda t}}{\lambda^m\prod_{i=0}^m(1-b_i/\lambda)},\nonumber\\
&=\overset{\lambda}{\underset{=
}{\Omega}} \int \frac{e^{a\lambda t}}{\lambda^m\prod_{i=0}^m(1-b_i/\lambda)},\nonumber\\
&=\frac{1}{a}\overset{\lambda}{\underset{=
}{\Omega}}\frac{e^{\lambda t}}{\lambda^{m+1}\prod_{i=0}^m(1-b_i/\lambda)}=\frac{e^{[\pmb{b},0]t}}{a},
\end{align} using Eqs.~(\ref{hsp}) and (\ref{ExchangeProperty}).

\subsection{General treatment of \texorpdfstring{$U_{ij}$}{the evolution operator}}
The proof of the form for the solution relies on the $\star$-Neumann series expansion for the $\star$-resolvents appearing in any of the $U_{ij}$. First of all, we observe that in all cases the diagonal terms $U_{11}=\Theta\star(1_\star - K\Theta)^{\star -1}$ and $U_{22}$ are always produced by path-sum kernels $K$ of the form
\begin{equation}\label{kernelforms}
K(t,s)=\sum_{j=1}^N c_j\, e^{a _j t}e^{ [b_j,0] (t-s)}.
\end{equation}
It follows that evaluating $\star$-powers of the above general form exactly is sufficient to determine the diagonal entries while, per path-sum, the off-diagonal entries follow from them after a single additional $\star$-operation.
By linearity, $\star$-powers of the above kernel $K$ are given by
\begin{equation}
(K\Theta)^{\star n}=\sum_{\pmb{k}=\pmb{1}_n}^{N\pmb{1}_n}\bigstar_{j=1}^n\left(c_{k_j} e^{a_{k_j} t}e^{[b_{k_j},0] (t-s)}\Theta\right),
\end{equation}
where $\pmb{k}\in[\![1,N]\!]^n$, $\pmb{1}_n=(1,\ldots,1)\in \mathbb{C}^n$ and $k_j$ designate the $j$th entry of $\pmb{k}$.
It follows from these observations that in order to analytically determine the $\star$-resolvent of all the relevant kernels, one needs only the products of the form 
\begin{equation}
\bigstar_{j=1}^n(c_{j} e^{a_{j} t}e^{[b_{j},0] (t-s)}\Theta).
\end{equation}
Remark that to alleviate the notation, and since only the form of the terms in the products is relevant, here, we write $c_j, a_j$ and $b_j$ instead of $c_{k_j},a_{k_j}$ and $b_{k_j}$. We shall establish the following result by induction on $n$:
\begin{align}\label{starprodF_j}
\bigstar_{j=1}^n(c_{j} e^{a_{j} t}e^{[b_{j},0](t-s)}\Theta)=e^{A_ns}e^{[A_1,B_1,A_2,B_2,\ldots,A_n,B_n](t-s)}\,\prod_{j=1}^n c_j\,\Theta
\end{align} with $A_i:=\sum_{j=1}^ia_j$ and $B_i:=A_i+b_i$.
Although a direct calculation is sufficient to establish the base case, it is better to do it following the more general pattern at work in the subsequent induction. In particular, most of the technical steps will effectively be the same in both the base and induction steps but will be easier to assimilate from the simpler base case.

We begin by rearranging the form of the kernel term $c_{j} e^{a_{j} t}e^{[b_{j},0](t-s)}$. Since  \(
e^{a_jt}=e^{a_j(t-s)}e^{a_js},
\)
then
\(
e^{a_1t}e^{[0,b_1](t-s)}=e^{a_1(t-s)}
e^{[0,b_1](t-s)}e^{a_1 s}.
\)
Furthermore, by linearity, for any $a, b_0,\ldots, b_k$,
\begin{align}\label{Identity2}
e^{a (t-s)}e^{[b_0,\ldots,b_k](t-s)}=e^{[a+b_0,\ldots,a+b_k](t-s)}.
\end{align}
Combining this with the preceding observation, it follows that 
\begin{equation}
c_{j} e^{a_{j} t}e^{[b_{j},0](t-s)}=c_je^{[a_j,a_j+b_j](t-s)}e^{a_j s}.
\end{equation}
 Then the base case corresponds to
\begin{align}
&c_1 e^{a_1t}e^{[0,b_1](t-s)}\Theta\star c_2 e^{a_2t}e^{[0,b_2](t-s)}\Theta\nonumber\\
&\hspace{20mm}=c_1c_2\,e^{A_2 s}\int_s^te^{[A_1,B_1](t-\tau)}e^{[A_2,B_2]
(\tau-s)}d\tau\,\Theta.
\end{align}
Since divided-difference exponentials are rational functions in the Omega domain, Eq.~(\ref{hsp}), and exchanging the Omega operator and integral Eq.~(\ref{ExchangeProperty}), we have
\begin{align}
&c_1e^{a_1t}e^{[0,b_1](t-s)}\Theta\star c_2e^{a_2t}e^{[0,b_2](t-s)}\nonumber\\
&=c_1c_2\,e^{A_2 s}\overset{\lambda,\mu}{\underset{=
}{\Omega}}\int_s^t\frac{e^{\lambda(t-\tau)}}{\lambda\left(1-\frac{A_1}{\lambda}\right)\left(1-\frac{B_1}{\lambda}\right)}\frac{e^{\mu(\tau-s)}}{\mu\left(1-\frac{A_2}{\mu}\right)\left(1-\frac{B_2}{\mu}\right)}d\tau\,\Theta,\nonumber\\
&=c_1c_2\,e^{A_2 s}\overset{\lambda,\mu}{\underset{=
}{\Omega}}\frac{e^{\mu(t-s)}-e^{\lambda(t-s)}}{\mu-\lambda}\frac{1}{\lambda\left(1-\frac{A_1}{\lambda}\right)\left(1-\frac{B_1}{\lambda}\right)}\frac{1}{\mu\left(1-\frac{A_2}{\mu}\right)\left(1-\frac{B_2}{\mu}\right)}\,\Theta,\nonumber\\
&=c_1c_2\,e^{A_2 s}\overset{\lambda,\mu}{\underset{=
}{\Omega}}\frac{e^{\mu(t-s)}-e^{\lambda(t-s)}}{\mu\left(1-\frac{\lambda}{\mu}\right)}\frac{1}{\lambda\left(1-\frac{A_1}{\lambda}\right)\left(1-\frac{B_1}{\lambda}\right)}\frac{1}{\mu\left(1-\frac{A_2}{\mu}\right)\left(1-\frac{B_2}{\mu}\right)}\,\Theta.
\end{align} 
Technically, in writing $(1-\lambda/\mu)^{-1}$ we  assumed without loss of generality that $|\lambda/\mu|<1$ so that we can use the geometric series for $(1-\lambda/\mu)^{-1}$. This can be achieved using invariance under the Omega operator $\lambda \rightarrow z\lambda$ and $\mu \rightarrow w\mu$ and taking $|z/w|<1$ as follows from Eq.~(\ref{Id0}).

At this point, expanding the above, one notices that all terms involving $e^{\lambda (t-s)}$ will be sent to $0$ by eliminating the Omega variable $\mu$. Indeed, the denominator produces a Laurent series with only strictly negative powers of $\mu$ while the numerator $e^{\lambda (t-s)}$ has no positive powers of $\mu$. Thus, none of the terms with the prefactor $e^{\lambda (t-s)}$ contribute to the coefficient of $\mu^0$ and all in all we get $0$. Continuing with this observation, we get
\begin{align}\label{StepinBaseCase}
&e^{a_1t}e^{[0,b_1](t-s)}\Theta\star e^{a_2t}e^{[0,b_2](t-s)}\Theta\nonumber\\
&\hspace{10mm}=c_1c_2\,e^{A_2s}\overset{\lambda,\mu}{\underset{=
}{\Omega}}\frac{e^{\mu(t-s)}}{\mu\left(1-\frac{\lambda}{\mu}\right)}\frac{1}{\lambda\left(1-\frac{A_1}{\lambda}\right)\left(1-\frac{B_1}{\lambda}\right)}\frac{1}{\mu\left(1-\frac{A_2}{\mu}\right)\left(1-\frac{B_2}{\mu}\right)}\,\Theta,
\end{align} with $A_1=a_1$, $A_2=a_1+a_2$, $B_1=a_1+b_1$ and $B_2=a_1+a_2+b_2$. 
We now eliminate the Omega variable $\mu$ from the remaining terms. Indeed, since, for $n\in\mathbb{N}$,
\begin{equation}
\overset{\lambda,\mu}{\underset{=
}{\Omega}}\,\frac{\mu^n}{1-\frac{\lambda}{\mu}}=
\overset{\lambda}{\underset{=
}{\Omega}}\left(\overset{\mu}{\underset{=
}{\Omega}}\,\frac{\mu^n}{1-\frac{\lambda}{\mu}}\right)=\overset{\lambda}{\underset{=
}{\Omega}}\, \mu^n\Big(1+\frac{\lambda}{\mu}+\frac{\lambda^2}{\mu^2}+\cdots\Big)=\overset{\lambda}{\underset{=
}{\Omega}}\,\lambda^n,
\end{equation} 
it follows that
\begin{equation}
\overset{\lambda}{\underset{=
}{\Omega}}
\frac{e^{\mu(t-s)}}{\mu\left(1-\frac{\lambda}{\mu}\right)}\frac{1}{\mu\left(1-\frac{A_2}{\mu}\right)\left(1-\frac{B_2}{\mu}\right)}
=\frac{e^{\lambda(t-s)}}{\lambda^2\left(1-\frac{A_2}{\lambda}\right)\left(1-\frac{B_2}{\lambda}\right)}.
\end{equation} 
Coming back to Eq.~(\ref{StepinBaseCase}) gives  
\begin{align}
&e^{a_1t}e^{[0,b_1](t-s)}\Theta\star e^{a_2t}e^{[0,b_2](t-s)}\Theta\\
&=c_1c_2\,e^{(a_1+a_2)s}\overset{\lambda}{\underset{=
}{\Omega}}\frac{e^{\lambda(t-s)}}{\lambda^3\left(1-\frac{a_1}{\lambda}\right)\left(1-\frac{a_1+b_1}{\lambda}\right)\left(1-\frac{a_1+a_2}{\lambda}\right)\left(1-\frac{a_1+a_2+b_2}{\lambda}\right)}\,\Theta.\nonumber
\end{align} 
By Eq.~(\ref{Id4}), eliminating the Omega variable $\lambda$ produces a divided-difference exponential,
\begin{align}
&e^{a_1t}e^{[0,b_1](t-s)}\Theta\star e^{a_2t}e^{[0,b_2](t-s)}\Theta\nonumber\\
&\hspace{15mm}=c_1c_2\,e^{(a_1+a_2)s}e^{[a_1,a_1+b_1,a_1+a_2,a_1+a_2+b_2](t-s)}\,\Theta,\nonumber\\
&\hspace{15mm}=c_1c_2\,e^{A_2 s}e^{[A_1,B_1,A_2,B_2](t-s)}\,\Theta.
\end{align} 
This establishes the base case. The general induction step now follows the same pattern. The induction hypothesis is
\begin{equation}
\bigstar_{j=1}^{n-1}c_j\,e^{a_j t}e^{[0,b_j](t-s)}\Theta=\prod_{j=1}^{n-1}c_j\,e^{A_{n-1}s}e^{[A_1,B_1,A_2,B_2,\ldots,A_{n-1},B_{n-1}]t},
\end{equation} 
where $A_i:=\sum_{j=1}^ia_j$ and $B_i:=A_i+b_i$. Then
\begin{align}
&\bigstar_{j=1}^n c_j\,e^{a_jt}e^{[0,b_j](t-s)}\Theta\nonumber\\
&=\;\bigstar_{j=1}^{n-1}c_j\,e^{a_jt}e^{[0,b_j](t-s)}\Theta\star c_n\,e^{a_n t}e^{[0,b_n](t-s)}\Theta,\nonumber\\
&=\prod_{j=1}^n c_j\int_s^t e^{A_{n-1}\tau}e^{[A_1,B_1,A_2,B_2,\ldots,A_{n-1},B_{n-1}](t-\tau)}
e^{[a_n,a_n+b_n](\tau-s)}e^{a_n\tau}d\tau\Theta,\nonumber\\
&=\prod_{j=1}^n c_j\,e^{A_ns}\int_s^t e^{[A_1,B_1,A_2,B_2,\ldots,A_{n-1},B_{n-1}](t-\tau)}
e^{A_{n-1}(\tau-s)}e^{[a_n,a_n+b_n](\tau-s)}d\tau\Theta,\nonumber\\
&=\prod_{j=1}^n c_j\,e^{A_n s}\int_s^t e^{[A_1,B_1,A_2,B_2,\ldots,A_{n-1},B_{n-1}](t-\tau)}
e^{[A_n,B_n](\tau-s)}d\tau\Theta,\nonumber\\
&=\prod_{j=1}^n c_j\,e^{A_n s}\overset{\lambda,\mu}{\underset{=
}{\Omega}}\int_s^t\frac{e^{\lambda(t-\tau)}}{\lambda^{2n-2} \prod_{i=1}^{n-1}\left(1-\frac{A_i}{\lambda}\right)\left(1-\frac{B_i}{\lambda}\right)}\frac{e^{\mu(\tau-s)}}{\mu\left(1-\frac{A_n}{\mu}\right)\left(1-\frac{B_n}{\mu}\right)}d\tau\Theta,\nonumber\\
&=\prod_{j=1}^n c_j\,e^{A_n s}\overset{\lambda}{\underset{=
}{\Omega}}\frac{e^{\lambda(t-s)}}{\lambda^{2n-1} \prod_{i=1}^n\left(1-\frac{A_i}{\lambda}\right)\left(1-\frac{B_i}{\lambda}\right)}\Theta,\nonumber\\
&=\prod_{j=1}^n c_j\,e^{A_n s}e^{[A_1,B_1,A_2,B_2,\ldots,A_n,B_n](t-s)}\Theta,
\end{align} 
where we evaluated the integral following the exact same strategy as in the base case.  Collecting the aforementioned results establishes Eq.~(\ref{starprodF_j}). 

We now proceed with all details pertaining to Case 1, as Cases 2 through 4 work in the same way since the relevant kernels are always of the form in Eq.~(\ref{kernelforms}). To that end, we obtain $U_{11}$ and $U_{12}$ using Eqs.~(\ref{U_ii}) and (\ref{U_ij}), respectively. More precisely, we use 
\begin{align}
A_{m,n}(t,s)\Theta&=i|g|^2  J_n(e_0/\omega_0)
J_m(e_0/\omega_0)e^{i(n -m)\omega_0t}e^{i[\varepsilon_0+m\omega_0,0](t-s)}~\Theta,\nonumber\\
&=-|g|^2 J_n(e_0/\omega_0)
J_m(e_0/\omega_0)e^{i(n-m)\omega_0t}e^{[i(\varepsilon_0+m\omega_0),0](t-s)}~\Theta,
\end{align} using Eq.~(\ref{ExchangeProperty}). Establishing the identifications $j\rightarrow(m,n)$ so that
$c_j\rightarrow J_nJ_m$, $a_j\rightarrow i(n-m)\omega_0$, and $b_j\rightarrow i(\varepsilon_0+n\omega_0)$ in Eq.~(\ref{starprodF_j}) and setting $s=0$ the result follows. For instance, consider the following $\star$-product,
\begin{align}
&\Theta \star A_{m_1,n_1}(t,s)\Theta\star A_{m_2,n_2}(t,s)\Theta|_{s=0}\nonumber\\
&=|g|^2 J_{n_1}
J_{m_1}J_{n_2}J_{m_2}e^{[i(N_1-M_1),i(\varepsilon_0+N_1),i(N_2-M_2),i(\varepsilon_0+N_2-M_1),0]t}~\Theta,\nonumber\\
&=|g|^2 J_{n_1}
J_{m_1}J_{n_2}J_{m_2}e^{i[N_1-M_1,\varepsilon_0+N_1,N_2-M_2,\varepsilon_0+N_2-M_1,0]t}~\Theta,
\end{align} using Eqs.~(\ref{Intdd-exp}) and (\ref{ExchangeProperty}). The general case; that is, $\Theta \star \bigstar_{j=1}^k (A_{m_j,n_j}(t,s)\Theta)|_{s=0}$ follows similarly.

For $U_{12}$, as per the path-sum results Eq.~(\ref{ReversePSG12}) we have both $U_{12}=\Theta\star (-i)H_{12}U_{22}\Theta$, which is Eq.~(\ref{U_ij}), and $U_{12}=U_{11}\star (-i)H_{12}\Theta$. The former yields
\begin{align}
&\frac{dU_{12}}{dt}=-ig\sum_nJ_ne^{i(\varepsilon_0+n\omega_0)t}U_{22},\nonumber\\
&=-i\sum_{k\geq 0}g|g|^{2k}\sum_{n,\pmb{m}_k,\pmb{n}_k} J_{n,\pmb{n}_k,\pmb{m}_k}e^{i(\varepsilon_0+n\omega_0)t}\,\, 
e^{-i[A_1,\varepsilon_0+B_1,A_2,\varepsilon_0+B_2,\ldots,A_k,\varepsilon_0+B_k,0]t},
\nonumber\\
&=-i\sum_{k\geq 0}g|g|^{2k}\sum_{n,\pmb{m}_k,\pmb{n}_k} J_{n,\pmb{n}_k,\pmb{m}_k} \nonumber
\\&\hspace{15mm}\times e^{-i[A_1-(\varepsilon_0+n\omega_0),B_1-n\omega_0,\ldots,A_k-(\varepsilon_0+n\omega_0),B_k-n\omega_0,-(\varepsilon_0+n\omega_0)]t},
\end{align} where we used Eqs.~(\ref{Symcc}) and (\ref{U11Case1}) to obtain the second equality and Eq.~(\ref{Identity2}) to obtain the last equality. Finally, collecting the results above, we obtain
\begin{align}
U_{12}&=\sum_{k\geq 0}g|g|^{2k}\sum_{n,\pmb{m}_k,\pmb{n}_k} J_{n,\pmb{n}_k,\pmb{m}_k}\nonumber\\
&\hspace{15mm}\times e^{-i[A_1-(\varepsilon_0+n\omega_0),B_1-n\omega_0,\ldots,A_k-(\varepsilon_0+n\omega_0),B_k-n\omega_0,-(\varepsilon_0+n\omega_0),0]t},\nonumber\\
&=\sum_{k\geq 0}g|g|^{2k} \sum_{n,\pmb{m}_k,\pmb{n}_k}J_{n,\pmb{n}_k,\pmb{m}_k}\,e^{i(\varepsilon_0+n\omega_0)t}\, 
 e^{-i[A_1,\varepsilon_0+B_1,\ldots, A_k,\varepsilon_0+B_k, \varepsilon_0+n\omega_0,0]t},
\end{align} using Eqs.~(\ref{Intdd-exp}) and (\ref{Identity2}).

\section{Case 1: recovering the time-independent case \texorpdfstring{$e_0=0$}{}.}\label{FirstCaseTimeIndep}
When $e_0=0$ the first case Hamiltonian becomes  time-independent and the evolution operator is its ordinary matrix exponential. Since for $e_0=0$ we have $J_n(0)=\delta_{n,0}$, the general, exact solution simplifies to
\begin{align}
U_{11}(t)&=\sum_{k\geq 0}|g|^{2k}e^{i[\varepsilon_0\pmb{1}_k,\pmb{0}_{k+1}]t},\nonumber\\
&=\sum_{k\geq 0}|g|^{2k}\overset{\lambda}{\underset{=
}{\Omega}} \frac{e^{i\lambda t}}{\lambda^{2k}\left(1-\frac{\varepsilon_0}{\lambda}\right)^k},\nonumber\\
&=\overset{\lambda}{\underset{=
}{\Omega}} \frac{e^{i\lambda t}}{1-\frac{|g|^2}{\lambda^2}\frac{1}{1-\varepsilon_0/\lambda}},\nonumber\\
&=\overset{\lambda}{\underset{=
}{\Omega}} \frac{e^{i\lambda t}(1-\varepsilon_0/\lambda)}{1-\frac{\varepsilon_0}{\lambda}-\frac{|g|^2}{\lambda^2}},
\end{align} where $\pmb{1}_n=(1,\ldots,1)\in \mathbb{C}^n$ and we can now use partial fractions to write
\begin{equation}
\frac{1-\varepsilon_0/\lambda}{1-\frac{\varepsilon_0}{\lambda}-\frac{|g|^2}{\lambda^2}}
=\frac{1}{r_+-r_-}\left(\frac{r_+-\varepsilon_0}{1-\frac{r_+}{\lambda}}-\frac{r_--\varepsilon_0}{1-\frac{r_-}{\lambda}}\right),
\end{equation} where
\(
r_{\pm}:=\big(\varepsilon_0\pm \sqrt{\varepsilon_0^2+4|g|^2}\big)/2. 
\) From this it follows that 
\begin{equation} 
U_{11}(t)=\frac{1}{r_+-r_-}\left[(r_+-\varepsilon_0)e^{ir_+t}-(r_--\varepsilon_0)e^{ir_-t}\right].
\end{equation} 
As expected, this expression yields $U_{11}(t)=1$ if $g=0$ and $U_{11}(t)=\cos(|g|t)$ if $\varepsilon_0=0$. The result agrees with the ordinary exponential of the time-independent Hamiltonian
\begin{align}
\left(e^{
-i\mathsf{H}_{e_0=0}t}
\right)_{11}&=e^{-iS_0t}U_{11}(t),\nonumber\\
&=\frac{1}{r_+-r_-}\left[(r_+-\varepsilon_0)e^{i(r_+-S_0)t}-(r_--\varepsilon_0)e^{i(r_--S_0)t}\right],
\end{align} and
\begin{align}
\left(e^{
-i\mathsf{H}_{e_0=0}t}
\right)_{22}&=e^{-i(S_1+i\Gamma)t}U_{22}(t),\nonumber\\
&=\frac{1}{r_+-r_-}\left[(r_+-\varepsilon_0)e^{-i(r_++S_1+i\Gamma)t}-(r_--\varepsilon_0)e^{-i(r_-+S_1+i\Gamma)t}\right].
\end{align}
We also have
\begin{align}
U_{12}(t)&=\Theta\star (-i)H_{12} U_{22}\Theta,\nonumber\\
&=\Theta\star (-i)\underbrace{H_{12}(t)}_{=ge^{i\varepsilon_0t}}\left\{\frac{1}{r_+-r_-}\left[(r_+-\varepsilon_0)e^{-ir_+t}-(r_--\varepsilon_0)e^{-ir_-t}\right]\right\}\Theta,\nonumber\\
&=\frac{g}{r_+-r_-}\left(e^{-i(r_+-\varepsilon_0)t}-e^{-i(r_--\varepsilon_0)t}\right).
\end{align} 
This agrees again with the classical results
\begin{equation}
\left(e^{
-i\mathsf{H}_{e_0=0}t}
\right)_{12}=e^{-iS_0t}U_{12}(t)
=\frac{g}{r_+-r_-}\left(e^{-i(r_+-\varepsilon_0+S_0)t}-e^{-i(r_--\varepsilon_0+S_0)t}\right)
\end{equation} and
\begin{align}
\left(e^{
-i\mathsf{H}_{e_0=0}t}
\right)_{21}&=e^{-i(S_1+i\Gamma)t}U_{21}(t),\nonumber\\
&=\frac{\bar{g}}{r_--r_+}\left(e^{i(r_+-\varepsilon_0-S_1-i\Gamma)t}-e^{i(r_--\varepsilon_0-S_1-i\Gamma)t}\right).
\end{align}

\section{Case 1: Exact re-summations in approximations}\label{Approx}
A possible approximation strategy to simplify the exact expression of $U_{11}$ consists in truncating the kernel $k_1\Theta:=\sum_{m,n\in\mathbb{Z}}A_{m,n}\Theta$ before taking its $\star$-resolvent (see Eq.~(\ref{Amn}) for $A_{m,n}$). This is equivalent to exactly re-summing a family of terms from the $\star$-Neumann series $U_{11}=\Theta\star  \sum_n (k_{11}\Theta)^{\star n}$. An advantage of this approach versus hard truncations of the $\star$-Neumann series itself is that it does not lead to divergences at finite times, capturing a crucial property of the exact solution. Concretely, let us proceed with the simplest situation of the first case as an example. 
We aim to retain only the $A_{0,0}$ term of the kernel, thus yielding the approximation $U_{11}(t)\simeq\Theta\star (1_\star-A_{0,0}\Theta)^{\star-1}$. By Eq.~(\ref{Amn}), this is
\begin{align}
\Theta\star (1_\star-A_{0,0}\Theta)^{\star-1}&=\sum_{k\geq 0}J_0^{2k}|g|^{2k}e^{i[\varepsilon_0\pmb{1}_k,\pmb{0}_{k+1}]t},\nonumber\\
&=\sum_{k\geq 0}J_0^{2k}|g|^{2k}\,\,\overset{\lambda}{\underset{=
}{\Omega}}\, \frac{e^{i\lambda t}}{\lambda^{2k}\left(1-\frac{\varepsilon_0}{\lambda}\right)^k},\nonumber\\
&= \overset{\lambda}{\underset{=
}{\Omega}} \frac{e^{i\lambda t}}{1-\frac{J_0^2|g|^2}{\lambda^2}\frac{1}{1-\varepsilon_0/\lambda}},\nonumber\\
&=\, \overset{\lambda}{\underset{=
}{\Omega}} \,\frac{e^{i\lambda t}\left(1-\varepsilon_0/\lambda\right)}{1-\frac{\varepsilon_0}{\lambda}-\frac{J_0^2|g|^2}{\lambda^2}},
\end{align} 
and we can now use partial fraction expansion to write
\begin{equation}
\frac{1-\varepsilon_0/\lambda}{1-\frac{\varepsilon_0}{\lambda}-\frac{J_0^2|g|^2}{\lambda^2}}
=\frac{1}{r_{0+}-r_{0-}}\left(\frac{r_{0+}-\varepsilon_0}{1-\frac{r_{0+}}{\lambda}}-\frac{r_{0-}-\varepsilon_0}{1-\frac{r_{0-}}{\lambda}}\right),
\end{equation} where
\(
r_{0\pm}:=\big(\varepsilon_0\pm \sqrt{\varepsilon_0^2+4J_0^2|g|^2}\big)/2
\) and, finally,
\begin{align} 
U_{11}(t)&\simeq \frac{1}{r_{0+}-r_{0-}}\Big((r_{0+}-\varepsilon_0)e^{ir_{0+}t}-(r_{0-}-\varepsilon_0)e^{ir_{0-}t}\Big),\nonumber\\
&\simeq e^{\frac{i\varepsilon_0 t}{2}} \left(\cos( \Omega t/2 )-\frac{i \varepsilon_0}{\Omega} \sin (\Omega t/2)\right).
\end{align} 
This indicates a simple rescaling $|g|\rightarrow |J_0g|$ of the present approximation with respect to the time-independent case as claimed in the main text. Similarly, we have 
\begin{align}
U_{12}(t)&\simeq e^{i\varepsilon_0t}\sum_{k\geq 0}J_0^{2k+1}g|g|^{2k}e^{-i[\varepsilon_0\pmb{1}_{k+1},\pmb{0}_{k+1}]t},\nonumber\\
&\simeq e^{i\varepsilon_0t}\sum_{k\geq 0}J_0^{2k+1}g|g|^{2k}\overset{\lambda}{\underset{=
}{\Omega}} \frac{e^{-i\lambda t}}{\lambda^{2k+1}\left(1-\frac{\varepsilon_0}{\lambda}\right)^{k+1}},\nonumber\\
&\simeq J_0ge^{i\varepsilon_0t}\overset{\lambda}{\underset{=
}{\Omega}} \frac{e^{-i\lambda t}}{\lambda\left(1-\frac{\varepsilon_0}{\lambda}-\frac{J_0^2|g|^2}{\lambda^2}\right)},
\end{align} 
and we can now use partial fractions to write
\begin{equation}
\frac{1/\lambda}{1-\frac{\varepsilon_0}{\lambda}+\frac{|g|^2}{\lambda^2}}
=\frac{1}{r_{0+}-r_{0-}}\left(\frac{1}{1-\frac{r_{0+}}{\lambda}}-\frac{1}{1-\frac{r_{0-}}{\lambda}}\right).
\end{equation} Finally, we obtain
\begin{equation} 
U_{12}(t)\simeq \frac{J_0g}{r_{0+}-r_{0-}}\left(e^{-i(r_{0+}-\varepsilon_0)t}-e^{-i(r_{0-}-\varepsilon_0)t}\right)=-\frac{2 i J_0 g}{\Omega
   } e^{\frac{i \varepsilon_0 t}{2}} \sin \left(\Omega t/2\right).
\end{equation}
This procedure is not limited to retaining only $A_{0,0}$ but rather extends straightforwardly to approximations based on wider families of kernel terms. For example, one could retain some or all `diagonal' kernel terms $A_{n,n}$, leading to an expression of $U_{11}$ and $U_{12}$ in terms of finite divided-difference exponentials whose frequencies are given by the roots of higher-order polynomials.
For example, with
\begin{equation}
U_{11}\simeq \Theta \star (1_\star - A_{0,0}\Theta-A_{1,1}\Theta-A_{-1,-1}\Theta)^{\star-1},
\end{equation}
we obtain
\begin{equation}
U_{11}\simeq (a_0\partial_{it}^2+a_1\partial_{it}+a_2)e^{i[\pmb{r}_4,0](t-s)},
\end{equation}
with $\textbf{r}_4=(r_1,r_2,r_3,r_4)$ and $r_i$ are the roots of 
\begin{align}
&P(X):=\\
&\hspace{5mm}(X-\varepsilon_0)^2 \left[X (X-\varepsilon_0)-|g|^2 \left(J_0^2+2
   J_1^2\right)\right]+\omega_0^2
  \left[|g|^2 J_0^2+X (\varepsilon_0-X)\right],\nonumber
\end{align}  
and $a_0=-|g|^2 (J_0^2+2J_1^2)$, $a_1=2|g|^2\varepsilon_0(J_0^2+2J_1^2)$, $a_2=-|g|^2\big[J_0^2(\varepsilon_0^2-\omega_0^2)+2J_1^2\varepsilon_0^2\big]$. This improves upon previous approximation and typically goes beyond the accuracy of the first orders of the high-frequency expansion. Inclusion of non-diagonal terms $A_{m, n\neq m}$ proceed in a similar way and can be evaluated using Omega-calculus. In this case, however, all expressions become more involved and cannot be obtained using Laplace or Fourier transforms. This is because non-diagonal kernel terms genuinely depend on both $t$ and $s$ and not solely on $t-s$, as diagonal terms do. This indicates that they break time-translation invariance, a characteristic of truly non-autonomous differential systems.

\section{Case 2: Resonances of the Bloch-Siegert model \texorpdfstring{$e_0=0$}{}}\label{AltKBS}
We begin by detailing Omega domain calculations to turn the path-sum kernel of the Bloch-Siegert model, $K_{\text{BS}}:=(-i)^{2}(H_{12}\Theta\star H_{21}\Theta)|_{e_0=0}$, into a single divided-difference function. 
The naive expression for the kernel as described by the general solution formula is
\begin{align}\label{KBSAppendix}
K_{\text{BS}}(t,s)&=\Big(e^{i[\varepsilon_0+\omega,0](t-s)}+e^{2i\omega t}e^{i[\varepsilon_0-\omega,0](t-s)}\nonumber\\
&\hspace{15mm} +e^{-2i\omega t}e^{i[\varepsilon_0+\omega,0](t-s)}+e^{i[\varepsilon_0-\omega,0](t-s)}\Big)\Theta.
\end{align}
At a glance, this expression seems to reveal a single resonance in $\varepsilon_0=\pm\omega$ (this is a single resonance if we fix the sign of $\varepsilon_0$, $\omega$ being positive). We shall reveal another resonance by basic algebraic manipulations of the above and prove that all subsequent resonances are secondary replications of these two through $\star$-powers of $K_{BS}$ because of its $e^{\pm 2i \omega t}$ and $e^{\pm 2i \omega s}$ dependencies.
Firstly, observe that we can exchange the dependencies in the two-time variables by the properties of the divided-difference exponentials. Since in physics we usually set $s$ to 0 (after evaluating all $\star$-products!), this profitably simplifies final expressions. Here the kernel is also
\begin{align}
K_{\text{BS}}(t,s)&=\Big(e^{i[\varepsilon_0+\omega,0](t-s)}+e^{i[\varepsilon_0-\omega,-2\omega](t-s)}e^{-2i\omega s}\nonumber\\
&\hspace{15mm}+e^{i[\varepsilon_0+\omega ,2\omega](t-s)}e^{2i\omega s}+e^{i[\varepsilon_0-\omega,0](t-s)}\Big)\,\Theta,\nonumber\\
&=\Bigg(\overset{\lambda}{\underset{=
}{\Omega}}\frac{e^{i\lambda (t-s)}}{\lambda(1-\frac{\varepsilon_0+\omega}{\lambda})}
+\overset{\lambda}{\underset{=
}{\Omega}}\frac{e^{-2i\omega s}e^{i\lambda (t-s)}}{\lambda(1-\frac{\varepsilon_0-\omega}{\lambda})(1+\frac{2\omega}{\lambda})}\nonumber\\
&\hspace{15mm}+\overset{\lambda}{\underset{=
}{\Omega}}\frac{e^{2i\omega s}e^{i\lambda (t-s)}}{\lambda(1-\frac{\varepsilon_0+\omega}{\lambda})(1-\frac{2\omega}{\lambda})}
+\overset{\lambda}{\underset{=
}{\Omega}}\frac{e^{i\lambda (t-s)}}{\lambda(1-\frac{\varepsilon_0-\omega}{\lambda})}\Bigg)\,\Theta,\nonumber\\
&=\overset{\lambda}{\underset{=
}{\Omega}}\frac{A(\lambda,e^{i \omega s})e^{i\lambda (t-s)}}{\lambda^4\left(1-\frac{\varepsilon_0+\omega}{\lambda}\right)\left(1-\frac{\varepsilon_0-\omega}{\lambda}\right)\left(1-\frac{4\omega^2}{\lambda^2}\right)}\,\Theta,\nonumber\\
&=\Big(A(\bullet,e^{i\omega s})e^{i\bullet(t-s)}\Big)[\varepsilon_0+\omega,\varepsilon_0-\omega,2\omega,-2\omega,0]\,\Theta,\label{ddKernelAppendix}
\end{align} where $A$ is given by Eq.~(\ref{A}).
The crucial step to reduce a sum of divided-difference exponentials to a single divided-difference function, i.e., passing from the second to the third line, is a simple partial fraction reduction in the Omega domain. Observe that in the form of Eq.~(\ref{ddKernelAppendix}), the kernel is revealed to have not one but two resonances: repeated arguments in the divided-difference occur if and only if $\varepsilon_0\pm\omega=2\omega$ that is $\varepsilon_0=\omega,3\omega$. 

These two resonances replicate themselves when taking $\star$-powers of the kernel, giving rise to all secondary resonances when $\varepsilon_0$ is an odd-integer multiple of $\omega$. Mathematically, this process is already visible with the $\star$-square of the kernel $K^{\star2}_{\text{BS}}:=(-i)^4(H_{12}\Theta\star H_{21}\Theta)^{\star 2}|_{e_0=0}$.
\begin{align}
K_{\text{BS}}^{\star 2}(t,s)&
=\Big(A(\bullet,e^{i\omega s})e^{i\bullet(t-s)}\Big)[\pmb{a}]\,\Theta\star\Big(A(\bullet,e^{i\omega s})e^{i\bullet(t-s)}\Big)[\pmb{a}]\,\Theta,
\end{align} 
where $[\pmb{a}]$ is a shorthand notation for $[\varepsilon_0+\omega,\varepsilon_0-\omega,2\omega,-2\omega,0]$. To calculate the above, it is again advantageous to exchange the dependencies of the two time variables in the left term so as to simplify the $\star$-product. 
Let $A_0(X):=2(X-\varepsilon_0)(X^2-4\omega^2)$ and $A_{\pm}(X)=X(X-\varepsilon_0\mp\omega)(X\mp2\omega)$ so that
\(
A(X,Y)=A_0(X)+A_+(X)Y^{-2}+A_-(X)Y^2
\). We may write
\begin{align}
&\Big(A(\bullet,e^{i\omega s})e^{i\bullet(t-s)}\Big)[\pmb{a}]\nonumber\\
&=\Big(A_0(\bullet)+A_+(\bullet)e^{-i2\omega s}+A_-(\bullet)e^{i2\omega s}\Big)e^{i\bullet(t-s)}[\pmb{a}],\nonumber\\
&=\Big(A_0(\bullet)+e^{-i2\omega t}A_+(\bullet)e^{i2\omega (t-s)}+e^{i2\omega t}A_-(\bullet)e^{-i2\omega (t-s)}\Big)e^{i\bullet(t-s)}[\pmb{a}],\nonumber\\
&=A_0(\bullet)e^{i\bullet(t-s)}[\pmb{a}]+e^{-i2\omega t}A_+(\bullet)e^{i\bullet(t-s)}[\pmb{a}_+]+e^{i2\omega t}A_-(\bullet)e^{i\bullet(t-s)}[\pmb{a}_-],\nonumber\\
&=\Big(B(e^{i\omega t},\bullet)e^{i\bullet (t-s)}\Big)[\pmb{b}],
\end{align} where $[\pmb{b}]:=[\pmb{a},\pmb{a}_+,\pmb{a}_-]$, \(
[\pmb{a}_{\pm}]:=(\varepsilon_0\pm 3\omega,\varepsilon_0\pm \omega,\pm 4\omega,\pm 2\omega,0)\) and
\begin{align}
B(X,Y)&=
A_0([\pmb{a}])\prod_{i=0}^4(X-a_{+i})\prod_{i=0}^4(X-a_{-i})\nonumber\\
&\hspace{15mm}+A_+([\pmb{a}_+])\prod_{i=0}^4(X-a_{-i})\prod_{i=0}^4(X-a_i)Y^{-2}\nonumber\\
&\hspace{25mm}+A_-([\pmb{a}_-])\prod_{i=0}^4(X-a_{+i})\prod_{i=0}^4(X-a_i)Y^2.
\end{align}
With this we now have the much simpler $\star$-product,
\begin{align}
K_{\text{BS}}^{\star 2}
&=\Big(B(e^{i\omega t},\bullet)e^{i\bullet (t-s)}\Big)[\pmb{b}] \star\Big(A(\circ,e^{i\omega s})e^{i\circ(t-s)}\Big)[\pmb{a}],\nonumber\\
&=\Big(\Big(B(e^{i\omega t},\bullet)e^{i[\bullet,\circ](t-s)}A(\circ,e^{i\omega s})\Big)[\pmb{b}]\Big)[\pmb{a}].
\end{align} This should be understood as a divided-difference function of two variables, here symbolically denoted $\bullet$ and $\circ$, $\bullet$ being evaluated in $[\pmb{b}]$ and $\circ$ in $[\pmb{a}]$. These evaluations commute so can be treated as two consecutive standard divided-differences.
As promised we can now see two novel resonances from the above divided-difference. Indeed, 
\begin{align}
[\pmb{a},\pmb{b}]=[\varepsilon_0+3\omega,\,\varepsilon_0-3\omega,\,(\varepsilon_0+\omega)\pmb{1}_3,\,(\varepsilon_0-\omega)\pmb{1}_3,\,4\omega,\,-4\omega,\,2\omega\,\pmb{1}_3,\,-2\omega\,\pmb{1}_3,\,\pmb{0}_4],
\end{align}
with $\pmb{1}_3=(1,1,1)$ and $\pmb{0}_4=(0,0,0,0)$. Repeated arguments now occur if and only if
$\varepsilon_0+3\omega=4\omega$, $\varepsilon_0-3\omega=4\omega,\pm2\omega$, and $\varepsilon_0\pm\omega=4\omega,2\omega$. This implies $\varepsilon_0=\omega,3\omega,5\omega,7\omega$, revealing two novel resonances. Similarly, by induction we prove that $K_{BS}^{\star n}$ has resonances at all $\omega,\ldots ,(2 n + 1)\omega$.

\section{Case 2, Bloch-Siegert model: Effective Hamiltonians}\label{Monodromy}
Since the exact evolution operator $\mathsf{U}$ is available, it is sufficient to evaluate it in $\mathrm{T}=2\pi/\omega$ to obtain $\mathsf{U}_{\text{eff}}$. The omission of order $0$ in the formulas further provides $\mathsf{U}_{\text{eff}}-\mathsf{Id}$. Since, furthermore,
$
\mathsf{U}_{\text{eff}}=\exp(-i \mathsf{H}_{\text{eff}}\mathrm{T})
$ it follows that 
$
-(\mathsf{U}_{\text{eff}}-\mathsf{Id})=i \mathrm{T}\,\mathsf{H}_{\text{eff}}+\cdots.
$
Hence, we have
\begin{equation}
-\frac{1}{i\mathrm{T}}(\mathsf{U}_{\text{eff}}-\mathsf{Id})=\mathsf{H}_{\text{eff}}+\mathcal{O}(\mathrm{T}^2).
\end{equation} The removal of the additional terms produced by the formula is achieved symbolically by setting all instances of $\mathrm{T}\to 0$ as the final step. Note that the formula, nonetheless, requires one to use $\exp(n\,i\omega \mathrm{T})=1$ for any integer $n$ to be validated. Note that it follows that $(\mathsf{H}_{\text{eff}})_{11}+(\mathsf{H}_{\text{eff}})_{22}=0$. Indeed, we have
\begin{align}
(\mathsf{H}_{\text{eff}})_{11}&=\overline{(\mathsf{H}_{\text{eff}})}_{11},\nonumber\\
&=\lim_{\mathrm{T}\rightarrow 0}\frac{1}{i\mathrm{T}}\left(\overline{(\mathsf{U}_{\text{eff}})}_{11}-1\right),\nonumber\\
&=\lim_{\mathrm{T}\rightarrow 0}\frac{1}{i\mathrm{T}}\left((\mathsf{U}_{\text{eff}})_{22}-1\right)
=-(\mathsf{H}_{\text{eff}})_{22},
\end{align} using  $(\mathsf{H}_{\text{eff}})_{ij}=\overline{(\mathsf{H}_{\text{eff}})}_{ji}$ and Eq.~(\ref{Symcc}).

\section{Recovering the Floquet-Magnus expansion in the $\star$-algebra}\label{FMFromStar}
The standard Floquet-Magnus expansion for $\mathsf{U}_{\text{eff}}$ is a consequence of $\star$-algebraic manipulations, with no path-sum nor Omega calculus involved. In the spirit of the main text, we demonstrate this explicitly in the case of the Bloch-Siegert Hamiltonian, but the method employed here is generally applicable to all periodic Hamiltonians. Recall that, in the $\star$-algebra we have $\mathsf{U}_{\text{rot}}\Theta = \Theta \star (\mathsf{Id}_\star - (-i) \mathsf{H}_{\text{rot}}\Theta)^{\star -1}$. Then 
\begin{equation}
\mathsf{U}_{\text{eff}}(\mathrm{T})=\lim_{\varepsilon_0\rightarrow \omega}\mathsf{U}_{\text{rot}}(\mathrm{T})=\Theta\star (\mathsf{Id}_{\star}+i\lim_{\varepsilon_0\rightarrow \omega}\mathsf{H}_{\text{rot}}(t,s)\Theta)^{\star-1}|_{t=\mathrm{T} \atop s=0},
\end{equation} and, on resonance, we have $\lim_{\varepsilon_0\rightarrow \omega}\mathsf{H}_{\text{rot}}(t,s)=(g/2)\sigma_x+(g/2)e^{2i\omega t}\sigma_++(g/2)e^{-2i\omega t}\sigma_-$.
To illustrate the capabilities of the $\star$-algebraic framework, we demonstrate a low order calculation  highlighting how the commutator structure at the heart of Floquet-Magnus expansions appears within the present framework. We have
\begin{equation}
\mathsf{U}_{\text{eff}}^{(1)}(\mathrm{T})=-i\Theta\star \mathsf{H}_{\text{rot}}(t,s)\Theta
|_{t=\mathrm{T} \atop s=0}=\frac{g}{2}\sigma_x(-i\mathrm{T})=\mathsf{H}_0(-i\mathrm{T}),
\end{equation} using $\mathsf{H}_0\equiv (g/2)\sigma_x$ and
\begin{equation}
\mathsf{U}_{\text{eff}}^{(2)}(\mathrm{T})=(-i)^2\Theta\star \mathsf{H}_{\text{rot}}(t,s)\Theta
\star \mathsf{H}_{\text{rot}}(t,s)\Theta|_{t=\mathrm{T} \atop s=0}.
\end{equation} In what follows, we use $[\mathsf{A},\mathsf{B}]:=\mathsf{A}\mathsf{B}-\mathsf{B}\mathsf{A}$ and
\begin{align}
&e^{i0(t-s)}\Theta \star e^{\alpha 2i\omega t}\Theta\star e^{\beta2i\omega t}\nonumber\Theta\nonumber\\
&=e^{i0(t-s)}\Theta \star e^{\alpha 2i\omega (t-s)}e^{\alpha 2i\omega s}\Theta\star e^{\beta2i\omega t}\Theta,\nonumber\\
&=e^{i0(t-s)}\Theta \star e^{\alpha 2i\omega (t-s)}\Theta\star e^{(\alpha+\beta)2i\omega t}\Theta,\nonumber\\
&=e^{i0(t-s)}\Theta \star e^{\alpha 2i\omega (t-s)}\Theta\star e^{(\alpha+\beta)2i\omega (t-s)}\Theta e^{(\alpha+\beta)2i\omega s},\nonumber\\
&=e^{i[0,\alpha 2\omega,(\alpha+\beta) 2\omega](t-s)}e^{(\alpha+\beta)2i\omega s}\Theta,
\end{align} with $\alpha,\beta\in \{\pm 1\}$ to obtain
\begin{align}
&\left(\frac{g}{2}\right)^2\left(\Theta\star e^{2i\omega t}\Theta \star e^{-2i\omega t}\Theta\sigma_+\sigma_-+\Theta\star e^{-2i\omega t}\Theta \star e^{2i\omega t}\Theta\sigma_-\sigma_+\right)|_{t=\mathrm{T} \atop s=0}\nonumber\\
&\hspace{15mm}=\left(\frac{g}{2}\right)^2\left(e^{i[0,0,2\omega]\mathrm{T}}\sigma_+\sigma_-+e^{i[0,0,-2\omega]\mathrm{T}}\sigma_-\sigma_+\right)\Theta,\nonumber\\
&\hspace{15mm}=\left(\frac{g}{2}\right)^2\frac{[\sigma_+,\sigma_-]}{2\omega}(-i\mathrm{T}),\nonumber\\
&\hspace{15mm}=\left(\frac{g}{2}\right)^2\left(\frac{[\sigma_+,\sigma_-]}{4\omega}+\frac{[\sigma_-,\sigma_+]}{-4\omega}\right)(-i\mathrm{T}),\nonumber\\
&\hspace{15mm}=\sum_{m=\pm2}\frac{[\mathsf{H}_{-m},\mathsf{H}_m]}{2m\omega}(-i\mathrm{T}),
\end{align}
\begin{align}
&\left(\frac{g}{2}\right)^2\left(\Theta\star \Theta \star e^{\pm 2i\omega t}\Theta\sigma_x\sigma_{\pm}+\Theta\star e^{\pm 2i\omega t}\Theta \star \Theta\sigma_{\pm}\sigma_x\right)|_{t=\mathrm{T} \atop s=0}\nonumber\\
&\hspace{15mm}=\left(\frac{g}{2}\right)^2\left(e^{i[0,0,\pm 2\omega]\mathrm{T}}\sigma_x\sigma_\pm+e^{i[0,\pm 2\omega,\pm 2\omega]\mathrm{T}}\sigma_\pm\sigma_x\right)\Theta,\nonumber\\
&\hspace{15mm}=\left(\frac{g}{2}\right)^2\frac{[\sigma_x,\sigma_\pm]}{\pm 2\omega}(-i\mathrm{T}),\nonumber\\
&\hspace{15mm}=\frac{[\mathsf{H}_{\mp2},\mathsf{H}_0]}{\mp2\omega}(-i\mathrm{T}),
\end{align} using $\mathsf{H}_{\mp 2}\equiv(g/2)\sigma_{\pm}$ and these agree with \cite[Eqs.~(31b,~31c)]{Mikami2016} (recall that here we take $s=0=t_0$). In a similar way, we can show
\begin{align}
&e^{i0(t-s)}\Theta \star e^{\alpha 2i\omega t}\Theta\star e^{\beta2i\omega t}\Theta \star e^{\gamma2i\omega t}\Theta\nonumber\\
&\hspace{15mm}=e^{i[0,\alpha 2\omega,(\alpha+\beta) 2\omega,(\alpha+\beta+\gamma)2\omega](t-s)}e^{(\alpha+\beta+\gamma)2i\omega s}\Theta,
\end{align}
with $\alpha,\beta,\gamma\in \{\pm 1\}$ and proceeding as above, we recover \cite[Eq.~(31d)]{Mikami2016} by computing
\begin{equation}
\mathsf{U}_{\text{eff}}^{(3)}(\mathrm{T})=(-i)^3\Theta\star \left(\mathsf{H}_{\text{rot}}(t,s)\Theta\right)^{\star 3}|_{t=\mathrm{T} \atop s=0}.
\end{equation} More generally, we can write
\begin{align}\label{HeffMagnus}
&\mathsf{H}_{\text{eff}}=-\frac{1}{i\mathrm{T}}(\mathsf{U}_{\text{eff}}-\mathsf{Id}\Theta)|_{\mathrm{T}=0}\nonumber,\\
&\hspace{2mm}=-\frac{1}{i\mathrm{T}}\left(\Theta\star \sum_{n \geq 0} (-i)^n(\mathsf{H}_{\text{rot}}\Theta)^{\star n}-\mathsf{Id}\Theta\right)\bigg|_{\mathrm{T}=0},\nonumber\\
&\hspace{2mm}=\left(\frac{1}{i\mathrm{T}}\sum_{n\geq 1}\sum_{m_1,\ldots,m_n=0,\pm 2}i^ne^{i\mathrm{T}[0,m_1\omega,(m_1+m_2)\omega,\ldots,(m_1+\cdots +m_n)\omega]}\right)\bigg|_{\mathrm{T}=0}\mathsf{H}_{\pmb{m}},\nonumber\\
&\hspace{2mm}=\left(\frac{1}{i\mathrm{T}}\sum_{n\geq 1}\sum_{m_1,\ldots,m_n=0,\pm 2}\frac{i^n}{\omega^n}e^{i\omega\mathrm{T}[0,m_1,m_1+m_2,\ldots,m_1+\cdots +m_n]}\right)\bigg|_{\mathrm{T}=0}\mathsf{H}_{\pmb{m}},
\end{align} using $\mathsf{H}_{\text{rot}}(t,s)=\sum_{m=0,\pm2}e^{-im\omega t}\mathsf{H}_{m}$ and defining $\mathsf{H}_{\pmb{m}}:=\mathsf{H}_{m_1}\mathsf{H}_{m_2}\cdots \mathsf{H}_{m_n}$. Contributions to $\mathsf{H}_{\text{eff}}$ occur whenever two arguments of the divided-difference exponential are equal. Although we worked out a particular example, Eq.~(\ref{HeffMagnus}) provides a divided-difference-based expansion of the generator of evolution, which is also compatible with the Magnus expansion generator obtained in \cite{Burum1981}. In particular, we have an explicit formula for $\mathsf{H}_j^k$ in \cite[Eq.~(18)]{Burum1981}, see also \cite{Haeberlen1968}. Eq.~(\ref{HeffMagnus}) can be seen as an instance of a  mould--in reference to mould calculus as introduced by \'Ecalle \cite[Section~2]{Li2019}. The removal of the non-trivial walk-like  contribution $\mathsf{H}_{m_1}\mathsf{H}_{m_2}\cdots \mathsf{H}_{m_n}$ to the above is possible only using path-sum in conjunction with the $\star$-algebra and Omega calculus, leading to the explicit solutions provided in this work.   

\section{Cases 3 \& 4: Karhunen-Lo\`eve Expansion}\label{Exactsmall}

The Karhunen-Lo\`eve theorem states that, on a bounded interval, a stochastic process can be expanded in terms of an infinite linear combination of orthogonal functions, similar to a Fourier series. Specifically, for any zero mean second order process, $X(t)$ such that $t \in [0,\mathcal{T}]$, with a covariance function, $K(t,s)$, continuous between time $t$ and $s$, then Mercer's theorem guarantees the existence of an orthonormal basis of eigenfunctions, $\phi_k(t)$, and eigenvalues, $\lambda_k$, such that 
\begin{equation}
K(t,s) = \sum_{j=1}^\infty \lambda_j \phi_j(s) \phi_j(t).
\end{equation}
Then, the process, $X(t)$, may be expanded on the basis of these eigenfunctions,
\begin{equation}\label{X(t)}
X(t) = \sum_{k=1}^\infty \sqrt{\lambda_k} Z_k \phi_k(t),
\end{equation}
where each $Z_k$ is an independent Gaussian random variable with zero mean and variance one. This expansion is known as the Karhunen-Lo\`eve expansion. Truncating the infinite series provides an approximation with the smallest mean squared error over all expansions.  

For the investigation in the main text, the primary stochastic process of interest is the Brownian motion.  The covariance function of the Brownian motion is $K(t,s)=\text{min}(s,t)$. In the interval $[0,\mathcal{T}]$, the integral equation used to determine the eigenfunctions is
\begin{equation}\label{intKlambdaphi}
\int_0^\mathcal{T} K(t,s)\phi(s)ds=\int_0^t s\phi(s)ds + t\int_t^\mathcal{T} \phi(s)ds = \lambda \phi(t).
\end{equation}
Evaluating and differentiating results in the derivative equations
\begin{equation}\label{intphilambdaphi'}
\int_t^\mathcal{T} \phi(s)ds = \lambda \frac{d}{dt}\phi(t),
\end{equation} and
\begin{equation}\label{ODE}
-\phi(t) = \lambda \frac{d^2}{dt^2} \phi(t).
\end{equation}
Solving the linear second order ODE (\ref{ODE}) and using a particular Robin boundary condition $\phi(0)=0=\phi'(\mathcal{T})$ which follows directly from the equations above by setting $t=0$ and $t=\mathcal{T}$ in (\ref{intKlambdaphi}) and (\ref{intphilambdaphi'}), respectively, one finds the eigenvalues
\begin{equation}
\lambda_i = \frac{4\mathcal{T}^2}{(2i-1)^2\pi^2},
\end{equation}
with eigenfunctions $\phi(t) = A \sin(t/\sqrt{\lambda})$. The orthonormality condition sets $A=\sqrt{2/\mathcal{T}}$. Therefore, the Karhunen-Lo\`eve expansion for the Brownian motion is represented as
\begin{equation}
W(t) = \sum_{i=1}^\infty\sqrt{\lambda_i}Z_iA\sin \left(\frac{t}{\sqrt{\lambda_i}}\right)=\sum_{i=1}^\infty \frac{2 \sqrt{2\mathcal{T}}Z_i}{(2i-1)\pi}\sin\left({\frac{(2i-1)\pi t}{2\mathcal{T}}}\right),
\end{equation} using Eq.~(\ref{X(t)}).

\section{Case 4: $H_{\text{eff}}(t)$ with time-dependent driving}\label{AppTDHeff}
We now show that our approach implies \cite[Eq.~15]{Zeuch2018}. To exemplify, we take
\begin{equation}
H_{12}(t)=e^{at}\cos(\omega t),
\end{equation} since all the examples considered can be recast in a Fourier type expansion. The first order; that is, the terms containing $e^{at}$ is non-zero only for $U_{12}$, yielding
\begin{align}
U_{12}^{(1)}(\mathrm{T})&=\lim_{\varepsilon_0 \rightarrow \omega}(-i)\Theta\star H_{12}|_{e_0=0\atop s=0},\nonumber\\
&=\frac{1}{2}\lim_{\varepsilon_0 \rightarrow \omega}\overset{\lambda}{\underset{=
}{\Omega}}\frac{(\lambda+2(\varepsilon_0-ia))e^{-i\lambda \mathrm{T}}}{\lambda^2 \left(1+\frac{\varepsilon_0-ia+\omega}{\lambda}\right)\left(1+\frac{\varepsilon_0-ia-\omega}{\lambda}\right)},\nonumber\\
&=\frac{1}{2}\overset{\lambda}{\underset{=
}{\Omega}}\frac{(\lambda+2(\omega-ia))e^{-i\lambda \mathrm{T}}}{\lambda^2 \left(1+\frac{2\omega-ia}{\lambda}\right)\left(1-\frac{ia}{\lambda}\right)},\nonumber\\
&=\frac{1}{2}(\bullet+2(\omega-ia))e^{-i\bullet \mathrm{T}}[-2\omega+ia,ia,0],\nonumber\\
&=\frac{(-ia) e^{a \mathrm{T}}}{4\omega(2\omega-ia)}+\frac{(2\omega-ia) e^{a \mathrm{T}}}{4i\omega a}-\frac{2(\omega-ia)}{2ia(2\omega-ia)},\nonumber\\
&=(-i)\frac{e^{a\mathrm{T}}-1}{2a}-\frac{e^{a\mathrm{T}}-1}{4\omega}+(-i)\frac{ae^{a\mathrm{T}}-a}{8\omega^2}+\mathcal{O}(\omega^{-3}),\nonumber\\
&=(-i)\int_0^T\left(\frac{e^{at}}{2}+(-i)\frac{(d/dt)e^{at}}{4\omega}+\frac{(d/dt)^2e^{at}}{8\omega^2}\right)dt+\mathcal{O}(\omega^{-3}),\nonumber\\
&=(-i)\int_0^T (\mathsf{H}_{\text{eff}}(t))_{12}dt+\mathcal{O}(\omega^{-3}).
\end{align} After considering the remaining entries of $\mathsf{H}_{\text{eff}}(t)$ we obtain
\begin{equation}
\mathsf{H}_{\text{eff}}(t)=\frac{e^{at}}{2}\sigma_x-\frac{e^{2at}}{4\omega}\sigma_z+\frac{(d/dt)e^{at}}{4\omega}\sigma_y-\frac{e^{3at}}{4\omega^2}\sigma_x+\frac{(d/dt)^2e^{at}}{8\omega^2}\sigma_x+\mathcal{O}(\omega^{-3}),
\end{equation} which agrees with \cite[Eq.~15]{Zeuch2018}.

\section{Self-commuting approximation}\label{ChapterToulouse}
We consider the evolution operator associated with the simple generator of evolution Hamiltonian
\begin{equation}
\mathsf{H}_1(t) = 
\begin{pmatrix}
0& g f(t)\\
\bar{g} \bar{f}(t)&0
\end{pmatrix},
\end{equation} with $f(t),g\in \mathbb{R}$. In what follows, we use the identity (assuming $t>0$ so that $\Theta(t)=1$)
\begin{equation}\label{Idstarintpower}
k!\Theta \star (f(t)\Theta)^{\star k}=\left(\int_0^tf(\tau)d\tau\right)^k,
\end{equation} as is readily apparent from taking the derivative with respect to $t$ on both sides of Eq.~(\ref{Idstarintpower}) to show that they satisfy the same first order ordinary linear differential system: $(d/dt)\Phi_k(t)=kf(t)\Phi_{k-1}(t)$ with $k\in \mathbb{Z}_{\geq 2}$ and $(d/dt)\Phi_1(t)=f(t)$ along with the initial condition: $\Phi_k(0)=0$ with $k\in \mathbb{Z}_{\geq 1}$. Eq.~(\ref{Idstarintpower}) follows from the aforementioned observation along with the uniqueness of the solution. Using only the $\star$-algebra, we get
\begin{align}
\mathsf{U}_1(t)= \Theta \mathsf{Id}\star \left(\mathsf{Id}_{\star}+i\mathsf{H}_1(t)\Theta\right)^{\star -1}=\exp\left(-i\int_0^tf(\tau)d\tau
~\sigma_x\right).
\end{align}  
Now suppose that we consider the dynamics generated by the following Hamiltonian
\begin{equation}
\mathsf{H}_2(t) = 
\begin{pmatrix}
S_0(t)& igf(t)\\
-igf(t)&S_1(t)
\end{pmatrix}.
\end{equation} After a phase change; that is, $\mathsf{U}_3(t)=\exp((i/2)\int_0^t(S_0(\tau)+S_1(\tau))d\tau)\mathsf{U}_2(t)$ we consider the simplified generator
\begin{equation}
\mathsf{H}_3(t) = 
\begin{pmatrix}
cf(t)& igf(t)\\
-igf(t)&-cf(t)
\end{pmatrix},
\end{equation} where we adopt the simplification $cf(t):=(S_0(t)-S_1(t))/2$. Under this simplified assumption, the operator $\mathsf{H}_3(t)$ commutes for different times; that is, $[\mathsf{H}_3(t),\mathsf{H}_3(s)]=\mathsf{O}$ resulting in a simplified dynamics, since there is no need to introduce the time-ordering operator. A direct calculation gives 
\begin{equation}
\mathsf{U}_3(t)=\begin{pmatrix}
\cos(F(t))-i\frac{c}{\sqrt{c^2+g^2}}\sin(F(t))&\frac{g}{\sqrt{c^2+g^2}}\sin(F(t))\\
-\frac{g}{\sqrt{c^2+g^2}}\sin(F(t))&\cos(F(t))+i\frac{c}{\sqrt{c^2+g^2}}\sin(F(t))
\end{pmatrix},
\end{equation} where 
\begin{align}
F(t)&=\sqrt{c^2+g^2}\int_0^tf(\tau)d\tau,\nonumber\\
&=\int_0^t\sqrt{c^2+g^2}f(\tau)d\tau,\nonumber\\
&=\int_0^t\sqrt{c^2f^2(\tau)+g^2f^2(\tau)}d\tau,\nonumber\\
&=\int_0^t\sqrt{\Delta^2(\tau)/4+g^2f^2(\tau)}d\tau,
\end{align} with $c^2f^2(\tau)=\Delta^2(\tau)/4$.
\bibliographystyle{plain}

\end{document}